\documentclass{jfm}

\usepackage{graphicx}
\usepackage{tabularx}
\usepackage{subcaption}
\usepackage{placeins} 
\usepackage{newtxtext}
\usepackage{newtxmath}
\usepackage{natbib}
\usepackage{capt-of}
\usepackage{hyperref}
\usepackage{float}
\hypersetup{
    colorlinks = true,
    linkcolor  = blue, 
    citecolor  = blue, 
    urlcolor   = gray, 
}

\newcommand{\RomanNumeralCaps}[1]
\linenumbers
\usepackage{makecell}
\usepackage{xcolor} 

\title{Translation dynamics of evaporating sessile binary-mixture droplet populations}

\author{Debarshi Debnath\aff{1}, Anna Malachtari \aff{2}, George Karapetsas\aff{2}\corresp{\email{gkarapetsas@auth.gr}},  Daniel Orejon\aff{1}, Khellil Sefiane\aff{1}, Alidad Amirfazli\aff{3}, Omar K. Matar\aff{4}, \and Prashant Valluri\aff{1}\corresp{\email{Prashant.Valluri@ed.ac.uk}
}
}
\affiliation{\aff{1}Institute for Multiscale Thermofluids, School of Engineering, University of Edinburgh, Edinburgh EH9 3FB, United Kingdom
\aff{2}Department of Chemical Engineering, Aristotle University of Thessaloniki, Thessaloniki 54124, Greece
\aff{3} Department of Mechanical Engineering, York University, Toronto, Ontario M3J 1P3, Canada
\aff{4} Department of Chemical Engineering, Imperial College London, London SW7 2AZ, United Kingdom 
}

\begin{document}

\maketitle

\begin{abstract}
 The translation dynamics of two binary mixture droplets is investigated theoretically and is corroborated with experiments. The proposed model accounts for the effects of Marangoni stresses generated by evaporative cooling and concentration gradients, as well as vapour diffusion, for both components of the binary mixture. We consider thin droplets, allowing us to use the lubrication theory to derive the evolution equation for the droplet profiles. We numerically solve the evolution equations using the finite element method and examine various cases of pure and binary droplet pairs exhibiting translational behaviours like attraction, repulsion, and 'chasing'. The results show that the combined effect of solutal Marangoni, capillary effect, and thermal Marangoni determines the movement of the droplets. The non-uniform evaporation generated from 'vapour shielding'  creates such effects. We observe that for droplets with the same initial composition, solutal Marangoni and capillary forces induce droplet attraction, while thermal Marangoni effects drive their repulsion. For droplets with different initial compositions, the drop with a higher concentration of the more volatile component pushes, or `chases', the drop with a lower initial concentration of this component, completely driven by the solutal Marangoni. We carried out experiments involving water-morpholine binary mixture droplets to validate the results predicted by our model.
\end{abstract}


\section{Introduction}
The evaporation of sessile droplets is an intriguing and complex phenomenon paramount to a wide range of applications such as inkjet printing \citep{singh_inkjet_2010, siregar_numerical_2013}, spray cooling of microelectronics \citep{bar-cohen_direct_2006, kim_spray_2007}, DNA control \citep{jing_automated_1998,dugas_droplet_2005}, disease diagnosis \citep{sefiane_formation_2010,brutin_pattern_2011}, among others.
Due to its widespread occurrence and numerous applications, droplet evaporation has been extensively studied for many years, uncovering several governing factors such as liquid volatility \citep{sefiane_self-excited_2008, karapetsas_convective_2012,wang_role_2024}, interfacial tension \citep{starov_evaporation_2009}, surface wettability, surfactants \citep{karapetsas_evaporation_2016}, shape of droplets \citep{Saenz2015,Saenz2017}, surface roughness \citep{hideo_nakae_effects_1998}, thermal conductivity \citep{Ristenpart2007,Dunn2009,wang_role_2024}, relative humidity \citep{fukatani_effect_2016}, atmospheric pressure and other vapour properties \citep{shahidzadeh-bonn_evaporating_2006}. 

The evaporation dynamics become more complex when droplets comprise miscible and/or immiscible mixtures \citep{Christy2011,Bennacer2014, WANG2022}. In the pioneering works of \cite{deegan_capillary_1997,deegan_contact_2000}, the effects of capillary flow in droplets during evaporation are illustrated in order to explain the coffee ring and salt deposit formations. Subsequently, \cite{Hu_and_Larson_2005} observed the effects of strong thermal Marangoni stresses on the evaporating droplets. Thermal Marangoni also generates strong convective flows within droplets, also termed hydrothermal waves \citep{sefiane_self-excited_2008, karapetsas_convective_2012} for volatile liquids. Marangoni effects may result from changes in surface tension caused by differences in not just temperature but also liquid composition. Surface tension forces due to changes in the liquid composition are commonly referred to as solutal Marangoni effects. 
\cite{Christy2011} showed the dominant effect of solutal Marangoni, which creates a chaotic inhomogeneous flow inside a pinned ethanol-water droplet. Subsequently, when the contact line is free to move, depending on the surface tension ratios between the components of the binary mixtures, the droplets tend to spread or contract. Binary mixture droplets such as ethanol-water, where the more volatile component (ethanol) has a lower surface tension compared to the low volatile component (water), show a super-spreading behaviour \citep{Williams2021} due to the action of solutal Marangoni stresses in the contact line region. On the other hand, according to \cite{karpitschka_marangoni_2017}, binary droplets contract when the more volatile component has a higher surface tension (e.g. water and propylene glycol mixture). 

Many studies have focused on the evaporation dynamics of individual droplets. However, in practical situations, droplets are rarely found in isolation. Recent observations have elucidated that droplets engage in mutual interactions via the vapour phase, a phenomenon termed as "vapor shielding", since typically the evaporation rates reduce for all involved droplets \citep{Wray2020,Masoud2021,Wray2021,Hatte2019, azzam_modeling_2024}. With the help of mathematical models \cite{Wray2020, Wray2021} and \cite{Masoud2021} could predict the asymmetric evaporation rates of pure droplets placed near each other, resulting in their extended lifetime. Recently, \cite{iqtidar_drying_2023} conducted experiments to predict the behaviour of evaporating droplet arrays. They found that droplets surrounded by other droplets had the longest lifetime, while those located at the edges of the array had the shortest lifetime due to reduced evaporation caused by vapour shielding. Furthermore, \cite{wen_vapor-induced_2019} observed that when highly volatile droplets, such as ethyl acetate, are placed on a super hydrophilic plasma-cleaned glass surface, not only do their lifetime increase, but they also approach each other and coalesce. Such motion of the droplets is attributed to the non-uniform evaporation rate, where droplets tend to move from the high-evaporation to the low-evaporation side, described by the minimum energy dissipation principle. Subsequently, \cite{sadafi_vapor-mediated_2019} proposed a surface-mediated and vapour-mediated interaction between two pure evaporating drops, causing them to move toward each other. More recently, the interaction of two pure volatile droplets residing on a soft substrate was examined by \cite{malachtari_dynamics_2024}, showing that the combined effect of non-uniform evaporation and elastocapillary phenomena determines whether the drop–drop interaction is attractive or repulsive. Moreover, it was demonstrated that, in the limit of rigid substrates, the nature of the interactions depends on the strength of thermal Marangoni stresses. 

Droplet interactions can be more prominent for binary mixture droplets \citep{Cira2015,Majhy2020,Man2017,zhao_long-range_2025}. One of the first observations of this behaviour of binary drops is reported in \cite{Cira2015}. It was discovered in their experiments that binary mixture drops (water and PG) placed near each other show attraction, repulsion, and chasing behaviour depending on their initial compositions and the pinning/depinning of the contact line. These behaviours are attributed to the solutal Marangoni effect generated from the differential concentrations of water in the droplet mixture and to the suppressed evaporation arising from the vapour shielding effect. To gain further insight into the translation of pure and binary droplets, \cite{Man2017} developed a theoretical model that takes into account the effects of capillary, Marangoni, and vapour concentration gradient.

In recent years, there has been a significant effort to develop accurate models for evaporating binary mixture droplets \citep{Diddens2017,Diddens2017a, Williams2021, Wang2021,karpitschka_marangoni_2017, baumgartner_marangoni_2022}. \cite{Diddens2017} developed a diffusion-limited model for an isolated multi-component evaporating drop using the lubrication approximation for both pinned and moving contact line conditions. They considered a water-ethanol mixture drop and observed strong solutal Marangoni flow during the rapid evaporation of ethanol and strong thermal Marangoni after the ethanol was fully evaporated from the mixture. Subsequently, \cite{Diddens2017a} also simulated droplets with higher contact angles, relaxing the lubrication approximation and observing the breaking of axial symmetry of such binary and even ternary mixture droplets. In recent years, \cite{Williams2021} developed a one-sided evaporation model for ethanol-water binary droplets with a moving contact line by introducing a precursor film. Their model was able to recreate the spreading and retraction dynamics observed in the experiments. Similar spreading and retraction were observed by \cite{baumgartner_marangoni_2022} using a Taylor dispersion model for the concentration of volatile components in the droplets of the binary and ternary mixture. 

In addition to these models for a single isolated evaporating drop, \cite{Pradhan2016} developed a numerical model for the diffusion-limited evaporation of pinned binary droplets placed adjacent to each other. They observed asymmetric Rayleigh convection flow inside the drop due to the concentration gradient generated as a result of the suppressed evaporation near the contact points of the drops facing each other. Although they have accurately predicted Rayleigh convection, the effect of Marangoni is completely neglected (since NaCl solution is considered), which has a strong influence on the flow field for mixtures such as ethanol-water or water-PG \citep{Cira2015, Diddens2021, karpitschka_marangoni_2017}. 
Recently, \cite{charlier_waterpropylene_2022} proposed a model for a water–PG droplet based on Taylor-dispersion-driven mixing. In their formulation, non-uniform evaporation generates a humidity gradient, which in turn produces a spatial variation in the apparent contact angle, similar to the mechanism proposed by \cite{sadafi_vapor-mediated_2019} and \cite{barrio2024droplet}. The resulting contact-angle variation acts analogously to a wettability gradient and drives the droplet toward the region of higher relative humidity.

In the short review above, it is worth mentioning that the phenomenon of multiple droplet evaporation for both pure and binary drops, which results in their motion, is not fully understood. In this paper, building on the works of \cite{Williams2021,Wang2021, malachtari_dynamics_2024,malachtari_evaporation_2025},  
we present a comprehensive lubrication theory-based diffusion-limited model for two pure and binary droplets placed adjacent to each other. 
It should be underscored that diffusion-limited evaporation from a two-dimensional droplet is inherently delicate to analyse. First, the associated vapour-diffusion field can produce a singular evaporative flux near the contact line, unless an appropriate microscopic regularisation is introduced \citep{fabrikant_potential_1985}.  Moreover, the two-dimensional setting represents a filament rather than a genuinely three-dimensional droplet, so the diffusion field and the resulting evaporation rate differ fundamentally from those of a finite sessile drop \citep{Wray2020}. Furthermore, the quasi-steady diffusion-limited formulation omits potential gas-phase convection and other transport mechanisms that may become significant under certain conditions \citep{colinet2001nonlinear,haut2005surface,cazabat2010evaporation}. In this study, these difficulties are addressed within a regularised two-dimensional thin-film model, in which a precursor-film/disjoining-pressure formulation smooths the contact-line region and yields a self-consistent description of the evaporating configuration \citep{SULTAN2005, malachtari_dynamics_2024}.

The proposed model considers the interaction of vapour from two droplets, along with thermo-capillary and soluto-capillary phenomena.
We focus on the translation of binary droplets as well as pure droplets and elucidate the intricate roles of thermal, solutal Marangoni, and capillary forces. Utilizing the model, we seek to comprehend the fundamental process behind the evaporation of various binary droplets, such as ethanol-water and water-morpholine, both in isolation and when positioned adjacent to each other. Depending on several physical parameters, we predict various translational motions of droplet pairs, such as attraction, repulsion, and chasing. Using our model, we strive to resolve the complex interactions of the Marangoni solutal and thermal effects and their impact on pairs of evaporating droplets. Our model’s results are further supported by corresponding experimental evidence.

The rest of the paper is organised as follows: Section \ref{s2} covers the problem formulation and governing equations. Sections \ref{s3} and \ref{s4} detail the scaling and evolution equations, respectively. The experimental methodology is outlined in Section \ref{s5}. The results and their discussion are provided in Section \ref{RD}, and the final remarks are given in Section \ref{s7}.

\section{Model formulation\label{s2}}

\subsection{Description of the problem}

\begin{figure}
\centering
    \includegraphics[width=13.5cm]{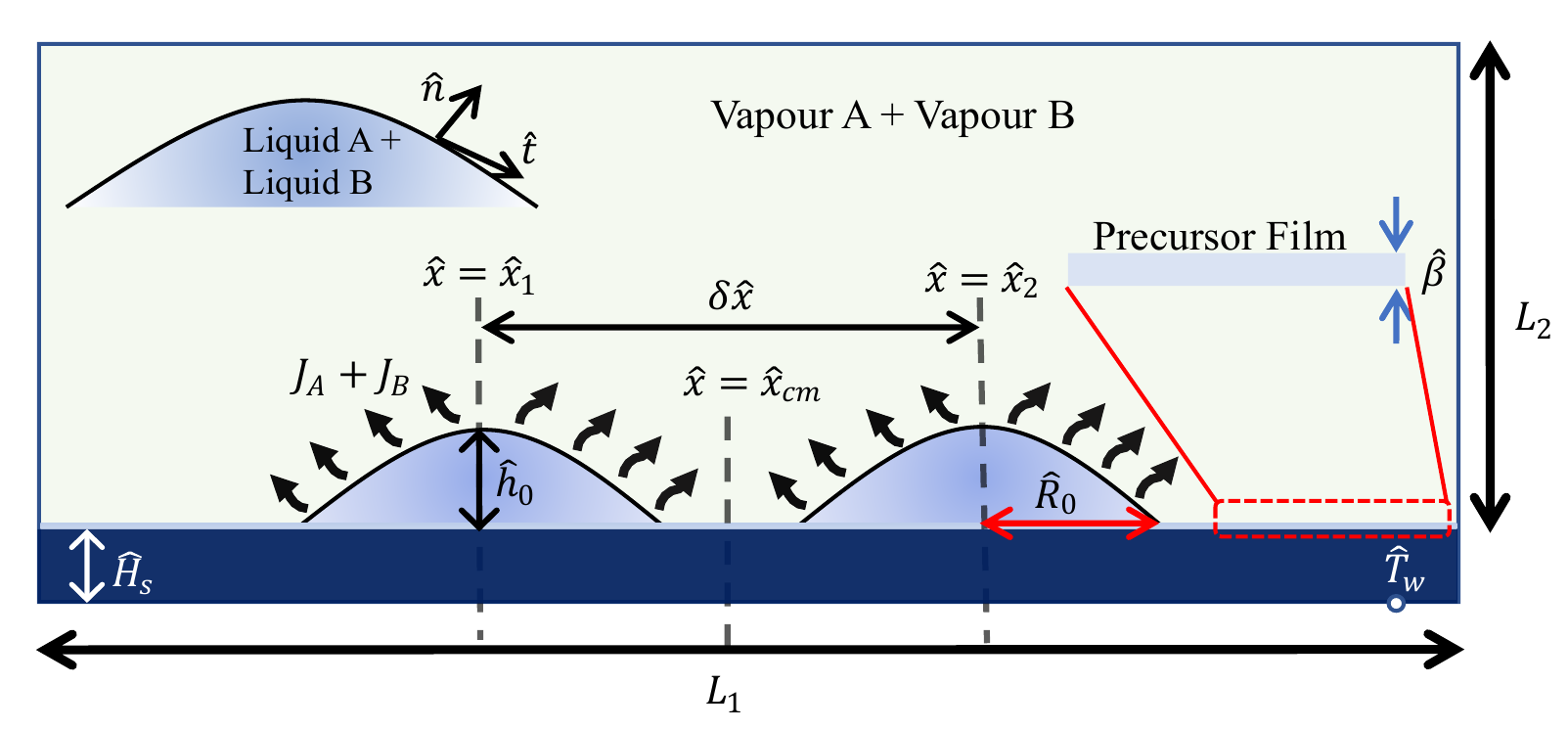}
    \caption{Initial configuration of the binary mixture droplets having initial height $\hat{h}_0$ and contact radius $\hat{R}_0$ placed with a separating distance of $\delta x$. The droplets are placed on a heated surface. The centroid locations of the droplets are $\hat{x}_1$ and $\hat{x}_2$. The droplets have a precursor film of thickness $\beta$. The gaseous phase is comprised of vapours of liquid A and liquid B.}
    \label{f1}
\end{figure}

We consider a system of two adjacent droplets placed on a rigid substrate, composed of a mixture of two volatile, miscible liquids (see figure~\ref{f1}).
Liquid A is the more volatile component (MVC), and liquid B is the less volatile component (LVC). The local mass fraction of liquid A and liquid B in the mixture are denoted as $\mathcal{X}_A$ and $\mathcal{X}_B$, respectively. The properties of the droplets, such as viscosity ($\hat{\mu}$), thermal conductivity ($\hat{k}$), and specific heat capacity ($\hat{c}_p$) are considered as the function of the local concentration and are defined using the following equations:
\begin{eqnarray}
\hat{\mu} = \mathcal{X}_A \hat{\mu}_A + (1-\mathcal{X}_A)\hat{\mu}_B, \label{2.1}\\
\hat{k} = \mathcal{X}_A \hat{k}_A+ (1-\mathcal{X}_A)\hat{k}_B, \label{2.2}\\
\hat{c}_p = \mathcal{X}_A \hat{c}_{p,A}+ (1-\mathcal{X}_A)\hat{c}_{p,B}, \label{2.3}
\end{eqnarray}
here, $\hat{\mu}_i$, $\hat{k}_i$ and $\hat{c}_{p,i}$ ($i=A,B$) denote the viscosities, thermal conductivities, and specific heats of liquid A and liquid B, respectively. Gravitational effects are considered to be negligible, and the densities of both components (A and B) are assumed to be equal ($\hat{\rho} = \hat{\rho}_A=\hat{\rho}_B$). The surface tension ($\hat{\sigma}$) of the liquid mixture is assumed to have a linear dependence on the local concentration of the liquid components and also the local temperature ($\hat{T}$), expressed using the following equation:
 \begin{equation}
     \hat{\sigma} = \mathcal{X}_A \left(\hat{\sigma}_A + \hat{\gamma}_A (\hat{T} -\hat{T}_\mathrm{ref})\right) + (1-\mathcal{X}_A) \left(\hat{\sigma}_B + \hat{\gamma}_B (\hat{T}-\hat{T}_\mathrm{ref})\right). \label{2.4}
 \end{equation}
 Here, $\hat{\sigma}_A$ and $\hat{\sigma}_B$ are the surface tensions of liquids A and B at the reference temperature $\hat{T}_\mathrm{ref}$, respectively; the reference temperature is assumed to be the bulk temperature of the surrounding gas phase ($\hat{T}_g$);
 $\hat{\gamma}_{A} $ and $\hat{\gamma}_{B}$ are the temperature gradients of the surface tension for liquid A $\left(\frac{\partial \hat{\sigma}_A }{\partial \hat{T}}\right)$ and liquid B $\left(\frac{\partial \hat{\sigma}_B }{\partial \hat{T}}\right)$, respectively. 
 The droplets are placed on a rigid solid surface of thickness $\hat{H}_s$ and thermal conductivity of $\hat{k}_s$. The liquid -air interface is located at $\hat{z} = \hat{h}(\hat{x},\hat{t})$. The outward normal ($\hat{\mathbf{n}}$) and tangential ($\hat{\mathbf{t}}$) unit vectors of the interface are $ \hat{\mathbf{n}}  = \frac{\left(-\frac{\partial \hat{h}}{\partial \hat{x}}, 1\right)}{ \sqrt{1+\left(\frac{\partial \hat{h}}{\partial \hat{x}}\right)^2} }$ and $   \hat{\mathbf{t}}  = \frac{\left(1, \frac{\partial \hat{h}}{\partial \hat{x}}\right)}{ \sqrt{1+\left(\frac{\partial \hat{h}}{\partial \hat{x}}\right)^2} }$, respectively. The length of the domain is $L_1$. In the domain, the centre of mass of each droplet is $\hat{x}=\hat{x}_1$ and 
$\hat{x} =\hat{x}_2$, respectively. The initial separation between the centers of mass of the two droplets is represented by $\delta \hat{x} = \hat{x}_2 - \hat{x}_1$.. The droplets are considered to be very thin, having aspect ratio $\epsilon = \hat{h}_0/\hat{R_0} < 1$ ($\hat{h}_0$ is the initial droplet height and $\hat{R}_0$ is the initial droplet contact radius).

We assume that the droplets are released into a thin precursor film of thickness $\hat{\beta}$ (see figure~\ref{f1}) with disjoining pressure that accounts for the intermolecular van der Waals interactions. The precursor film and disjoining pressure approach is a well-established method for regularising the moving contact line problem in lubrication models \citep{deGennes1985}, and has been widely employed in studies of evaporating sessile drops \citep{Moosman1980, Ajaev2005, SULTAN2005, Charitatos2021, Williams2021, malachtari_dynamics_2024, kavuri_evaporation-driven_2024}. Inclusion of the precursor film removes the stress singularity that arises at the moving contact line \citep{huh1971hydrodynamic,davis1974motion}, circumventing the well-known incompatibility between the no-slip boundary condition and the moving contact line without the need to introduce an empirical slip length. A key advantage of this approach over alternative contact line models, such as prescribed slip conditions or imposed contact angle boundary conditions, is that both the constant contact radius (CCR) and constant contact angle (CCA) modes of evaporation emerge naturally from the dynamics rather than being imposed as external constraints \citep{Charitatos2021, pham2017drying}. The approach is thermodynamically consistent, with the disjoining pressure derived from the intermolecular potential accounting for van der Waals interactions between the liquid-gas and liquid-solid interfaces \citep{oron1997long, israelachvili2011intermolecular}. The precursor film thickness $\hat{\beta}$ is not a free parameter but is determined self-consistently by the equilibrium condition balancing the disjoining pressure against the ambient vapour, ensuring that the precursor film is in thermodynamic equilibrium with the environment far from the drops \citep{Moosman1980, SULTAN2005}. This approach also allows us to naturally account for the evaporation of multiple droplets and their interactions through the connecting precursor film, without requiring explicit tracking of the contact line positions.
Under this consideration, we apply the lubrication theory to derive reduced-order governing equations in a two-dimensional Cartesian coordinate system ($\hat{x}, \hat{z}$).

 \subsection{Mathematical model}
 
\subsubsection{Liquid phase}
 The mass, momentum, energy, and concentration equations in the liquid phase are as follows:

\begin{eqnarray}
 \nabla \cdot \mathbf{\hat{u}} &=& 0, \label{2.7}\\
  \hat{\rho} \left(\frac{\partial \mathbf{\hat{u}}}{\partial \hat{t}} + \mathbf{\hat{u}} \cdot \hat{\nabla} \mathbf{\hat{u}}\right) &=& \hat{\nabla} \cdot \mathbf{\hat{E}}, \label{2.8}\\
\hat{\rho} \hat{c}_p \left(\frac{\partial \hat{T}}{\partial \hat{t}} + \mathbf{\hat{u}} \cdot \hat{\nabla} \hat{T} \right)
&=&  \hat{\nabla} \cdot (\hat{k}\hat{\nabla}\hat{T}), \label{2.9}\\
   \frac{\partial \mathcal{X}_A}{\partial \hat{t}} + \mathbf{\hat{u}} \cdot  \hat{\nabla}\mathcal{X}_A&=&\hat{\mathcal{D}}_A \hat{\nabla}^2\mathcal{X}_A. \label{2.10}
\end{eqnarray}
In equation~\ref{2.8}, $\hat{\mathbf{E}}$ is the total stress tensor which is expressed as:
\begin{equation}
    \hat{\mathbf{E}} = -\hat{p}\mathbf{I} + \hat{\mu}(\hat{\nabla}\mathbf{\hat{u}} + \hat{\nabla} \mathbf{\hat{u}}^T). \label{2.11}
\end{equation}
Here, $\mathbf{I}$ is the identity sensor. In equation~\ref{2.10}, $\hat{\mathcal{D}}_A$ is the diffusion coefficient of component A in the droplet.
Along the interface, the mass balance can be written with the help of the velocity of the liquid mixture ($\hat{\mathbf{u}}$) and the velocity of the interface ($\hat{\mathbf{u}}_s$) of the drop as
 \begin{equation}
          \hat{J} = \hat{\rho}(\hat{\mathbf{u}} - \hat{\mathbf{u}}_s)\cdot \hat{\mathbf{n}} \label{2.12}
  \end{equation}
 Here, $\hat{J}$ denotes the total evaporating mass flux, which is the sum of the evaporation fluxes of component A ($\hat{J}_A$) and component B ($\hat{J}_B$). The energy balance across the interface can be expressed as 

\begin{equation}
       \hat{J}_A \hat{L}_{v,A} + \hat{J}_B \hat{L}_{v,B} + \hat{k}\hat{\nabla}\hat{T} \cdot \hat{\mathbf{n}} = 0 \label{2.13}
\end{equation}
assuming that the thermal conductivity of the gas phase is much smaller than that of the liquid phase; $\hat{L}_{v,A}$, and $\hat{L}_{v,B}$ denote the latent heat of vaporisation for liquids A and B, respectively.

The normal and tangential stress boundary conditions at the interface can be expressed as
\begin{eqnarray}
    \hat{\mathbf{n}} \cdot \hat{\mathbf{E}} \cdot \hat{\mathbf{n}}  = 2\hat{\kappa} \hat{\sigma} + \hat{\mathrm{\Pi}}-\hat{p}_g \label{2.14}\\
    \hat{\mathbf{n}} \cdot \hat{\mathbf{E}} \cdot \hat{\mathbf{t}} = \hat{\nabla}_s\hat{\sigma} \cdot \hat{\mathbf{t}} \label{2.15}
\end{eqnarray}
Here, we have considered that viscous effects are negligible in the gas phase; $\kappa$ denotes the mean curvature of the liquid-vapour interface ($2\kappa = -\hat{\nabla}_s \cdot \hat{\mathbf{n}}$), $\hat{\nabla}_s$ is the surface gradient operator, and  $\hat{\Pi}$ represents the disjoining pressure at the contact line,
\begin{equation}
    \hat{\Pi} = \mathcal{\hat{A}} \left[\left( \frac{\hat{A}}{\hat{h}} \right)^m -   \left(\frac{\hat{A}}{\hat{h}} \right)^n \right], \label{2.16}
\end{equation}
where $\hat{\mathcal{A}} = \frac{\mathcal{\hat{A}}_{Ham}}{\hat{A}^3}$ is a constant, which is attributed to the interaction between liquid-vapour and liquid-solid interfaces and $\hat{A}$ is of the same order of magnitude as the precursor film thickness ($\beta$). In equation~\ref{2.16}, the first and second term describe the repulsion and attraction between liquid-solid and liquid-vapour interfaces, respectively. For a completely wetting droplet, the second term in equation~\ref{2.16} can be neglected \citep{Charitatos2021, Williams2021}, resulting in the following expression for the disjoining pressure containing only the repulsion term:
\begin{equation}
    \hat{\Pi} = \mathcal{\hat{A}} \left( \frac{\hat{A}}{\hat{h}} \right)^m.
\end{equation}
%
The kinematic boundary condition along the interface can be expressed as:
\begin{equation}
    \frac{\partial \hat{h}}{\partial \hat{t}} + \hat{u}_s \frac{\partial \hat{h}}{\partial \hat{x}} = \hat{v}_s.
\end{equation}
The generalized species mass jump condition \citep{whitaker1992species} from the liquid side to gas side over the interface is represented as 
\begin{equation}
    \left[\hat{\rho}_i\left(\hat{\mathbf{u}}_i - 
    \hat{\mathbf{u}}_s\right)\cdot\hat{\mathbf{n}} - 
    \hat{\mathbf{j}}_i\cdot\hat{\mathbf{n}}\right] = 0,
    \label{2.17a}
\end{equation}
 The concentration balance over the interface for component $i$ is defined as,
\begin{equation}
    \mathcal{X}_i(\hat{\mathbf{u}}-\hat{\mathbf{u}}_s) \cdot \hat{\mathbf{n}} -  \hat{\mathcal{D}_i}(\hat{\mathbf{n}} \cdot \hat{\nabla}\mathcal{X}_i)_{\hat{z} = \hat{h}} = \frac{\hat{J}_i}{\hat{\rho}}. \label{2.17}
\end{equation}
Using the mass balance across the interface, given by equation~\ref{2.12}, the concentration balance (for component A) is modified as, 
\begin{equation}
   \hat{\mathcal{D}_A}(\hat{\mathbf{n}} \cdot \hat{\nabla}\mathcal{X}_A)_{\hat{z} = \hat{h}} = \frac{\mathcal{X}_A \hat{J} - \hat{J}_A}{\hat{\rho}}. \label{2.18}
\end{equation}

\subsubsection{Gas phase}
The gas phase is assumed to consist of air and vapours of MVC and LVC (i.e. species A and B, respectively). Droplet evaporation is modelled using the generalised diffusion-limited model \citep{SULTAN2005,malachtari_dynamics_2024,malachtari_evaporation_2025} 
The quasi-steady diffusion equation for both species
in the gas phase are:
\begin{equation}
\hat{\nabla}^2 \hat{\rho}^{v, i} =0.
\end{equation}
Here, $\rho^{v,i}$ is the local vapour concentration for the component $i$ ($i=A, B$). Considering that the evaporation is limited by diffusion, the evaporative flux can be evaluated at the interface using Fick's law of diffusion:
   \begin{equation}
   \hat{J}_i = -\mathcal{D}_{v,i}(\hat{\textbf{n}} \cdot \hat{\nabla} \hat{\mathcal{\rho}}^{v,i}). \label{24} 
   \end{equation}
To predict the mass flux across the droplet interface induced by evaporation or condensation, we begin with the Hertz--Knudsen relation \citep{Plesset1976,Moosman1980}:
\begin{equation}
\hat{J}_{i} =
\left(
\frac{\hat{R}_g \hat{T}_{int}}{2 \pi \hat{M}_i}
\right)^{1/2}
\left(
\alpha_i \hat{\rho}^{ve,i} - \beta_i \hat{\rho}^{v,i}
\right),
\label{2.20}
\end{equation}
where \(\hat{\rho}^{ve,i}\) is the equilibrium vapour concentration of component \(i\), \(\hat{\rho}^{v,i}\) is the vapour concentration adjacent to the interface, \(\alpha_i\) and \(\beta_i\) are the accommodation coefficients for evaporation and condensation, respectively, \(\hat{T}_{int}\) is the interface temperature, and \(\hat{M}_i\) is the molecular weight.
In the present work, we adopt the near-equilibrium approximation \(\alpha_i=\beta_i=1\). For a binary mixture, the equilibrium vapour concentration of component \(i\) at the interface is assumed to scale with its liquid-phase mass fraction \(\mathcal{X}_i\), so that the interfacial mass flux is written as
\begin{equation}
\hat{J}_i =
\left(
\frac{\hat{R}_g \hat{T}_{int}}{2 \pi \hat{M}_i}
\right)^{1/2}
\left(
\mathcal{X}_i \hat{\rho}^{ve,i} - \hat{\rho}^{v,i}|_{int}
\right).
\label{2.21}
\end{equation}
At thermodynamic equilibrium there is no net interfacial mass transfer, and hence \(\hat{J}_i=0\). Equation \eqref{2.21} then reduces to
\[
\hat{\rho}^{v,i}|_{int}=\mathcal{X}_i \hat{\rho}^{ve,i},
\]
which is Raoult's law for component \(i\).
The equilibrium vapour concentration ($\hat{\rho}^{ve,i}$) for each component ($i =A,B$) in the binary mixture can be evaluated by equating the chemical potential of liquid and vapour phase, following a similar procedure as described in \cite{Plesset1976} and \cite{Moosman1980}, which results in the following form:
 \begin{equation}
   \frac{\hat{\rho}^{ve,i}}{\hat{\rho}^{v,i}_\mathrm{ref}} = 1+ \frac{\hat{M}_i }{\rho_l \hat{R}_g \hat{T}_{g}} (\hat{p}-\hat{p}_{g}) + \frac{\hat{M}_i \hat{L}_{v,i} }{\hat{R}_g \hat{T}_{g}^2} (\hat{T}_{int}-\hat{T}_{g}).
\end{equation}

Here, $\hat{\rho}^{v,i}_\mathrm{ref}$ is the saturation concentration of component $i$ at reference temperature. The reference temperature is considered as the bulk temperature of the surrounding vapour ($\hat{T}_g$). 

Using equation~\ref{2.21}, the boundary condition on the liquid-vapour interface is written as:
  \begin{equation}
      \left( \frac{\hat{R}_g \hat{T}_{int}}{2 \pi \hat{M}_i} \right)^{1/2} (\mathcal{X}_i \hat{\rho}^{ve,i} - \hat{\rho}^{v,i}|_{int}) =  -\mathcal{D}_{v,i}(\hat{\textbf{n}} \cdot \hat{\nabla} \hat{\mathcal{\rho}}^{v,i}). \label{2.23}
  \end{equation}
The boundary condition far away from the interface is presented as:
\begin{eqnarray}
    \hat{\rho}^{v,A} = \hat{\rho}^{v,A}_i, \\
    \hat{\rho}^{v,B} = \hat{\rho}^{v,B}_i,
\end{eqnarray}
where $\hat{\rho}^{v,A}_i$ and $\hat{\rho}^{v,B}_i$ are the initial concentrations of vapours A and B, respectively. 
The relative importance of heat transport, interfacial
phase change and gas-phase vapour transport depends on
the physical conditions and the interfacial accommodation
coefficients
\citep{colinet2001nonlinear,haut2005surface,
cazabat2010evaporation, wilson2023evaporation}.
As reviewed by \citet{MURISIC2011}, reported
accommodation coefficients span $O(10^{-6})$--$O(1)$,
with substantial uncertainty particularly for water.
Experimentally inferred values depend on liquid purity,
surface contamination, experimental conditions and the
modelling assumptions used to interpret measurements.
For their sessile-drop experiments with moving contact
lines, \citet{MURISIC2011} inferred coefficients
of approximately $3.5\times10^{-4}$ for water and
$9.3\times10^{-4}$ for isopropanol. Their non-equilibrium
one-sided model better reproduced the water experiments,
whereas the diffusion-limited description gave better
agreement for isopropanol. These findings concern their
particular experimental configurations and do not
establish a universal evaporation regime for either liquid.
Conversely, \citet{haut2005surface} obtained an interface
close to local equilibrium in their ethanol--nitrogen
example using an accommodation coefficient of $0.01$,
illustrating that proximity to equilibrium depends on
the relative transport resistances rather than requiring
unit accommodation.

In the present model, gas-phase vapour diffusion is
coupled to a Hertz--Knudsen interfacial relation through
a Robin boundary condition, retaining both gas-phase
transport and interfacial kinetic resistance.
We adopt $\alpha_i=\beta_i=1$ as a simplifying assumption
of unit evaporation and condensation coefficients,
rather than as measured properties of the mixtures
or a necessary consequence of near-equilibrium conditions.
The small vapour Péclet numbers support the adopted
gas-phase transport approximation within the chosen
scaling, but do not alone establish that diffusion
limits the overall evaporation rate.
The formulation resolves the vapour fields responsible
for interactions between neighbouring droplets and is
used here to investigate their coupling with Marangoni
transport and droplet motion. The selection of an appropriate evaporation closure,
including the accommodation coefficients and the relative
importance of the competing transport resistances,
remains an open, condition-dependent question rather
than a universally resolved modelling problem
\citep{cazabat2010evaporation, wilson2023evaporation,MURISIC2011}. For the binary mixtures considered
here, these quantities have not been independently
established; quantitative predictions of evaporation
rates remain conditional on the adopted
modelling assumptions.

\subsection{Scaling \label{s3}}
All the aforementioned equations and parameters ( $\hat{}$ designates dimensional quantities) are non-dimensionalized using the following scaling relations:
\begin{equation}
\left. \begin{array}{llll}  
\displaystyle  \hat{x} = \hat{R}_0 x,
  \quad \hat{z} = \hat{h}_0 z, \quad \hat{u} = \hat{U}^*u, \quad \hat{v} = \frac{ \hat{h}_0}{\hat{R}_0}\hat{U}^* v, \quad \hat{\rho}^{v,i} = \hat{\rho}^{v,i}_\mathrm{ref} \rho^{v,i},\\[12pt]
\displaystyle \hat{t} = \frac{\hat{R}_0}{\hat{U}^{*}}t,  \quad \hat{J}_i = \frac{\hat{k}_A \Delta \hat{T}}{\hat{L}_{v,A} \hat{h}_0} J_i, \quad \hat{c}_{p} = \hat{c}_{p,A} c_p, \quad \Delta \hat{T} = \hat{T}_w - \hat{T}_g,\\[12pt]
\displaystyle  \hat{p} = \frac{\hat{\mu}_A \hat{U}^*}{\epsilon^2 \hat{R}_0} p,\quad  \hat{T} = \hat{T}_{g} +T\Delta \hat{T}, \quad \hat{\mu} = \hat{\mu}_A \mu, \quad \hat{\sigma} = \hat{\sigma}_A \sigma, \quad \hat{k} = \hat{k}_A k. \\ \label{2.28} 
 \end{array} \right\} \
\end{equation}
A characteristic thermo-capillary velocity scale is defined as $\hat{U}^* = \epsilon \frac{\partial \hat{\sigma}_A }{\partial \hat{T}} \frac{\Delta \hat{T}}{\hat{\mu}_A}$.
The physical properties of the liquids are made non-dimensional using the properties of the more volatile component (A):
\begin{equation}
\left. \begin{array}{llll}
\displaystyle \sigma_R = \frac{\hat{\sigma}_B}{\hat{\sigma}_A}, \quad \gamma_R = \frac{\hat{\gamma}_B}{\hat{\gamma}_A}, \quad \mathcal{L}= \frac{\hat{L}_{v,B}}{\hat{L}_{v,A}}, \quad k_R = \frac{\hat{k}_B}{\hat{k}_A}  ; \\[12pt]
\displaystyle M_R = \frac{\hat{M}_B}{\hat{M}_A}, \quad \mu_R = \frac{\hat{\mu}_B}{\hat{\mu}_A},  \quad \rho^{v}_R = \frac{\hat{\rho}^{v,B}}{\hat{\rho}^{v,A}_\mathrm{ref}}.\\
\end{array} \right\} \
\end{equation}

\subsubsection{Liquid phase}
\label{liquid_phase}
 All the non-dimensional governing equations after applying lubrication approximation $(\epsilon << 1)$ and neglecting all the terms multiplied by $\epsilon$ are presented as: 
\begin{eqnarray}
      \frac{\partial u}{\partial x} + \frac{\partial v}{\partial z} = 0 \label{2.31}, \\
       -\frac{\partial p}{\partial x} + \frac{\partial }{\partial z}\left(\mu \frac{\partial u}{\partial z}\right) = 0,  \label{2.32} \\
      \frac{\partial p}{\partial z} = 0, \label{2.33}\\
      \frac{\partial}{\partial z} \left(k \frac{\partial T}{\partial z}\right) = 0. 
      \label{2.34}
\end{eqnarray}
  The concentration equation for component A in the liquid phase can be written as :
  \begin{equation}
     \frac{\partial \mathcal{X}_A}{\partial t} + u  \frac{\partial \mathcal{X}_A}{\partial x} + v  \frac{\partial \mathcal{X}_A}{\partial z} = \frac{1}{Pe} \left( \frac{\partial^2 \mathcal{X}_A}{\partial x^2}+ \frac{1}{\epsilon^2}\frac{\partial^2 \mathcal{X}_A}{\partial z^2}\right) \label{2.35}
  \end{equation}
  Here, $Pe= \frac{\hat{U} \hat{R}_0}{\hat{D}_A}$ is the Peclet number.
  To simplify the concentration equation and the concentration boundary condition at the interface, we assume the limit of weak diffusion $(Pe \approx O(\epsilon^{-2}))$ \citep{Matar2002, Williams2021, Wang2021} and we substitute $Pe = Pe' \epsilon^{-2}$ where $Pe'$ is the modified Peclet number $(Pe' \approx 0(1))$. An expression of $\mathcal{X}_A $ independent of $z$ by approximating a Galerkin expansion is considered: 
 \begin{equation}
     \mathcal{X}_A(x,z,t) = \mathcal{X}_{A0} (x,t) + \mathcal{X}_{A1}(x,t)\left(\frac{z^2}{h^2}- \frac{1}{3}\right), \label{3.8}
\end{equation}
for which transverse diffusion remains comparable to advection, while axial diffusion is \(O(\epsilon^2)\) and is therefore negligible at leading order. The concentration equation \ref{2.35} becomes
\begin{equation}
    \frac{\partial \mathcal{X}_A}{\partial t}
+ u \frac{\partial \mathcal{X}_A}{\partial x}
+ v \frac{\partial \mathcal{X}_A}{\partial z}
=
\frac{1}{Pe'}\frac{\partial^2 \mathcal{X}_A}{\partial z^2}. \label{2.36}
\end{equation}
Hence, to leading order in \(\epsilon\), axial diffusion is higher order, whereas transverse diffusion is retained.
The basis function in equation~\ref{3.8} is chosen so that it has zero depth average,
\[
\frac{1}{h}\int_0^h \left(\frac{z^2}{h^2}-\frac{1}{3}\right)\,dz = 0,
\]
and therefore \(\mathcal{X}_{A0}\) represents the depth-averaged concentration. In addition, the form \eqref{3.8} satisfies the substrate boundary condition
\[
\left.\frac{\partial \mathcal{X}_A}{\partial z}\right|_{z=0}=0
\]
automatically, while still allowing a non-zero concentration gradient at the free surface. A linear profile would not, in general, satisfy both the no-flux condition at the substrate and the evaporative flux condition at the interface simultaneously, and the quadratic term is therefore the lowest-order polynomial consistent with both.

Differentiating \ref{3.8} with respect to \(z\) gives
\begin{equation}
\frac{\partial \mathcal{X}_A}{\partial z}
=
\frac{2\mathcal{X}_{A1}z}{h^2},
\label{eq:dXAdz}
\end{equation}
so that
\begin{equation}
\left.\frac{\partial \mathcal{X}_A}{\partial z}\right|_{z=h}
=
\frac{2\mathcal{X}_{A1}}{h}.
\label{eq:dXAdz_interface}
\end{equation}
Substitution of \eqref{eq:dXAdz_interface} into the interfacial condition (equation~\ref{2.18})
\begin{equation}
    \left.\frac{\partial \mathcal{X}_A}{\partial z}\right|_{z=h}
=
Pe' E \left( \mathcal{X}_A J - J_A \right) \label{2.47}
\end{equation}
then yields the required relation between \(\mathcal{X}_{A1}\) and the interfacial evaporative flux.
Here, $E$ is the evaporation number, defined as $ E = \frac{\hat{k}_A \Delta \hat{T} \hat{R}_0}{\hat{h}^2_0 \hat{L}_{v,A}\hat{\rho}_A \hat{U}^*}$, which quantifies the strength of evaporation.

Further, equation~\ref{eq:dXAdz} is substituted into equation~\ref{2.47} for $z =h$, rearranged to obtain an expression for $\mathcal{X}_A$:
\begin{equation}
    \mathcal{X}_A = \frac{2\mathcal{X}_{A1}}{Pe'EJh}+\frac{J_A}{J}. \label{XA2}
\end{equation}
To obtain an expression of $\mathcal{X}_{A1}$, we evaluate equation~\ref{3.8} at $z=h$ and substitute the expression of $\mathcal{X}_{A}$ in equation~\ref{XA2}. The expression obtained for $\mathcal{X}_{A1}$ is then substituted in equation~\ref{eq:dXAdz} to arrive at the final form of the concentration balance over the interface,
\begin{equation}
        \left[\frac{\partial \mathcal{X}_A}{\partial z}\right]_{z=h} = \frac{(J_A - \mathcal{X}_{A0} J)}{h\left(\frac{J}{3} - \frac{1}{Pe'Eh}\right)}. \label{3.14}
\end{equation}

The liquid mixture properties mentioned in Eqs. \ref{2.1} - \ref{2.4} in dimensionless forms are as follows:
  \begin{eqnarray}
          \mu = \mathcal{X}_A + (1-\mathcal{X}_A)\mu_R, \\
          k = \mathcal{X}_A + (1-\mathcal{X}_A)k_R,\\
          c_p = \mathcal{X}_A + (1-\mathcal{X}_A)c_{pR}, \\
          \sigma  = \mathcal{X}_A (1-Ma T_{int}) + (1-\mathcal{X}_A) \sigma_R \left(1- \frac{Ma \gamma_R}{\sigma_R}T_{int}\right). \label{3}
  \end{eqnarray}
Here, $Ma$ is the thermal Marangoni number $\left(Ma = \frac{\hat{\gamma}_A \Delta\hat{T}}{\hat{\sigma}_A}\right)$.
Along the interface, we get for the mass, energy and force balances in the normal and tangential coordinate,
  \begin{align}
            EJ = - \frac{\partial h}{\partial x} (u-u_s) + (v-v_s),\\
            J_A + L_{v,R} J_B + k \frac{\partial T}{\partial Z} = 0, \label{2.42} \\
            p =  -\frac{\epsilon^2}{Ma}  2\kappa \sigma - \Pi,\label{2.40}
  \end{align}
\begin{equation}
     \frac{\partial u}{\partial z} = \left(\frac{1}{ \mu Ma}\right) \frac{\partial \sigma}{\partial x}.
\end{equation}
The kinematic boundary condition is given by
\begin{eqnarray}
     \frac{\partial h}{\partial t} +u \frac{\partial h}{\partial x} - v + EJ = 0 \label{3.16}.  
\end{eqnarray}
The following expressions give the scaled disjoining pressure for partial wetting case:
\begin{equation}
    \Pi = \mathcal{A} \left[\left( \frac{\mathcal{B}}{h}\right)^m -\left( \frac{\mathcal{B}}{h}\right)^n \right], \label{dsj_1}
\end{equation}
subsequently, for complete wetting, 
\begin{equation}
    \Pi = \mathcal{A} \left( \frac{\mathcal{B}}{h}\right)^m, \label{dsj_2}
\end{equation}
where, $\mathcal{B} = \frac{\hat{A}}{\hat{h}_0}$ and $\mathcal{A} = \frac{\hat{\mathcal{A}}_{ham} \epsilon \hat{h}_0}{\hat{A} \hat{\mu}_A \hat{U}^*}$ is the dimensionless Hamaker constant.
\subsubsection{Solid substrate}
Using a similar scaling (as shown in equation~\ref{2.28}), the energy equation in the solid  becomes:
\begin{equation}
    \frac{\partial^2 T_w}{\partial z^2} = 0 \label{2.48}
\end{equation}
The modified boundary conditions are: 
\begin{equation}
 k_w \frac{ \partial T_w}{\partial z} \Bigr \rvert_{z = 0} = k \frac{\partial T}{\partial z} \Bigr \rvert_{z = 0}, \quad T_w \Bigr \rvert_{z = 0} = T \Bigr \rvert_{z = 0}, \quad T_w \Bigr \rvert_{z = -H_s} = T_s, \label{2.49}
\end{equation}
\subsubsection{Gas phase}
\label{2.3.3}
Since the gas phase in the atmosphere may extend to large distances above the liquid phase, the scaling in the $z$ direction for this equation should be the same as the $x$ direction (i.e., $\hat{z} = \hat{R}_0 z^*$ and $ \hat{x} = \hat{R}_0 x)$. The
dimensionless conservation equation for the vapour concentration is then given by 
       \begin{eqnarray}
\frac{\partial^2 \rho^{v,i}}{\partial x^2} + \frac{\partial^2 \rho^{v,i}}{\partial z^{*2}} =0. \label{35}
      \end{eqnarray}
At the interface ($z = h$) and in the far field ($z = L_2$), the boundary conditions in non-dimensional form are
    \begin{eqnarray}
       \frac{Pe_{vA}}{Kn} (\mathcal{X}_A \rho^{ve,A} - \rho^{v,A}|_{int}) = - \frac{\partial \rho^{v,A}}{ \partial z^*} \Bigr\rvert_{z = h}, \label{flux_rho_A}\\
       \frac{Pe_{vB}}{Kn} \left(\frac{\rho^v_R }{\sqrt{M_R}}\right) \left[ (1-\mathcal{X}_A)\rho^{ve,B} - \rho^{v,B}|_{int}\right] = - \frac{\partial \rho^{v,B}}{ \partial z^*} \Bigr\rvert_{z = h} \label{flux_rho_B}, \\
\rho^{v,A}  \Bigr\rvert_{z= L_2} = \mathcal{H}_A, \quad \rho^{v,B}  \Bigr \rvert_{z= L_2} = \mathcal{H}_B \label{RH}
    \end{eqnarray}
where, $Pe_{v,A} = \frac{\hat{k}_A \Delta \hat{T} \hat{R}_0}{\hat{\mathcal{D}}_{v,A} \hat{\rho}^{v, A}_\mathrm{ref} \hat{L}_{v,A} \hat{h}_0}$ and $Pe_{v,B} = \frac{\hat{k}_A \Delta \hat{T} \hat{R}_0}{\hat{\mathcal{D}}_{v,B} \hat{\rho}^{v, B}_\mathrm{ref} \hat{L}_{v,A} \hat{h}_0}$ denote the vapour Peclet numbers, while $Kn = \frac{\hat{k}_A \Delta \hat{T}}{\hat{L}_{v,A} \hat{h}_0 \hat{\rho}_\mathrm{ref}^{v,A}}\sqrt{\frac{2 \pi \hat{M}_A}{\hat{R}\hat{T}_b}}$ represents the Knudsen number. In this context, $\mathcal{H}_A$ and $\mathcal{H}_B$ are the relative concentrations $\left(\frac{\hat{\rho}^{v,i}}{\hat{\rho}^{v,i}_\mathrm{ref}}\right)$ of the species corresponding to components A and B, respectively.

The dimensionless form of the equilibrium vapour concentrations are expressed as
    \begin{align}
    \rho^{ve,A} = 1+\delta p + \psi T_{int} \label{rhoveA}, \\
     \rho^{ve,B} = 1+ M_R\delta p + \mathcal{L} M_R\psi T_{int}. \label{rhoveB}
    \end{align}
Here, $\delta = \frac{\hat{\mu_A}\hat{U}^{*}\hat{M_A}}{\hat{\rho}_l \hat{R}_g\hat{T}_b \epsilon^2 \hat{R}_0}$, which signifies the measure of Kelvin effect and $\psi = \frac{\hat{L}_{v,A} \hat{M}_A \Delta \hat{T}}{\hat{R}\hat{T}_g^2}$, which is a quantification of the effect of local temperature on the mass flux. 

We solve equation~\ref{35} using Eqs. \ref{flux_rho_A} to \ref{RH} and compute the associated evaporation fluxes through the following expressions
    \begin{align}
    Kn ~J_A = (\mathcal{X}_A \rho^{ve,A} - \rho^{v,A}|_{int}),\label{2.54} \\
    Kn ~J_B = \left(\frac{\rho^v_R }{\sqrt{M_R}}\right) \left[ (1-\mathcal{X}_A)\rho^{ve,B} - \rho^{v,B}|_{int}\right]. \label{2.55}
    \end{align}
We note that the conventional diffusion-limited model is regularised through 
two physically motivated mechanisms 
that arise naturally from the model 
formulation. The first is the 
Hertz-Knudsen evaporation model, 
which imposes a Robin rather than 
Dirichlet boundary condition at the 
interface and introduces a finite 
kinetic resistance that prevents the 
square-root divergence in evaporation flux \citep{Saxton2017} and renders 
the total evaporation rate finite 
for all finite Knudsen numbers $Kn$ 
\citep{Saxton2017}.
The remaining limitations of the 
diffusion-limited framework, the 
quasi-steady vapour field, the neglect 
of gas-phase convection, and the logarithmic decay of 
droplet interactions are inherent 
to the lubrication-based thin-film 
approach and are shared by the 
majority of existing evaporating 
droplet models in the literature 
\citep{Craster2009, SULTAN2005}.


The dimensionless equilibrium thickness of the precursor film ($h = \beta$) is obtained under the assumptions that the film has zero mean curvature ($\kappa = 0$) and is sufficiently thin for attractive van der Waals interactions to suppress evaporation. Imposing $J_A = 0$ and $J_B = 0$ and combining Eqs. \ref{rhoveA} and \ref{rhoveB} yields
\begin{align}
    \left(1 - \frac{\rho^{v,A}}{1 - \delta \Pi_\infty}\right)\left(1 - \mu_R \delta \Pi_\infty \right) - \rho^{v,B} = 0,
\end{align}
where, for a purely wetting case, $ \Pi_\infty = \mathcal{A} \left( \frac{\mathcal{B}}{\beta}\right)^m$, and for partial wetting, $ \Pi_\infty = \mathcal{A} \left[\left( \frac{\mathcal{B}}{\beta}\right)^m - \left( \frac{\mathcal{B}}{\beta}\right)^n \right] $. At the initial state, prescribed values of $\rho^{v,A}$ and $\rho^{v,B}$ are used to determine $\beta$ at equilibrium. The initial values of $\rho^{v,A}$ and $\rho^{v,B}$ are $\mathcal{H}_A$ and $\mathcal{H}_B$, respectively. 

\subsection{Evolution equations\label{s4}}
In order to derive the evolution equations, the Eqs. \ref{2.31} and \ref{2.32} are integrated with respect to $z$ to obtain the expressions for $u$ and $v$: 
\begin{eqnarray}
     u = \frac{1}{\mu}\frac{\partial p}{\partial x} \left[\frac{z^2}{2} - hz\right] + \left(\frac{1}{\mu Ma}\right) \frac{\partial \sigma}{\partial x} z , \label{4.1}\\
    v = - \frac{\partial^2 p}{\partial x^2}\left[\frac{z^3}{6}- \frac{z^2h}{2}\right] - \left(\frac{1}{\mu Ma}\right) \frac{\partial^2 \sigma}{\partial x^2} \frac{z^2}{2}. \label{4.2}
\end{eqnarray}
To derive the evolution equation for $h(x, t)$, Eqs. \ref{4.1} and \ref{4.2} are combined with the kinematic boundary condition (equation~\ref{3.16}):
\begin{equation}
    \frac{\partial h}{\partial t}+ \frac{\partial q}{\partial x} = -EJ, \label{4.3}
\end{equation}
where $q$ represents the flow rate, given by
\begin{equation}
    q= \left[ -\frac{h^3}{3 \mu} \frac{\partial p}{\partial x} + \frac{h^2}{2}\left(\frac{1}{\mu Ma}\right) \frac{\partial \sigma}{\partial x} \right]. \label{eq:flowrate}
\end{equation}
The concentration equation (equation~\ref{2.36}) is integrated in the $z$-direction and combined with Eqs. \ref{2.47} and \ref{3.14} to obtain the evolution equation for $\mathcal{X}_{A0}$,
\begin{align}
    \frac{\partial}{\partial t}(h\mathcal{X}_{A0})
+
\frac{\partial}{\partial x}\left[
q\mathcal{X}_{A0}
+
\left(
-\frac{7}{180}\frac{h^4}{\mu}\frac{\partial p}{\partial x}
+\frac{1}{12\mu Ma}h^3\frac{\partial \sigma}{\partial x}
\right)\mathcal{X}_{A1}
\right] \nonumber\\
=
\frac{J_A-J\mathcal{X}_{A0}}
{Pe' h\left(\frac{J}{3}-\frac{1}{Pe'Eh}\right)}
+EJ\,\mathcal{X}_{A0}
-\frac{2}{3}EJh^3\mathcal{X}_{A1}
+\frac{\partial h}{\partial x}\,q^*\mathcal{X}_{A1}
\label{4.4}
\end{align}

Next, we integrate equation~\ref{2.34} and combine it with equation~\ref{2.42} to obtain the following expression for liquid temperature:

\begin{equation}
    T_{int} = - (J_A + \mathcal{L} J_B) \left(\frac{h}{k} +Bi \right) + Ts, \label{2.65}
\end{equation}
where $Bi$ is Biot number $\left( Bi = \frac{H_s}{k_s} = \frac{\hat{H}_s \hat{k}_A}{\hat{h}_0 \hat{k}_s}\right)$.
\begin{table}
    \centering
    \caption{Physical properties of Water and Morpholine used in the experiments.}
    \begin{tabular}{ccc}
    \hline
        Properties & Water (A)& Morpholine (B)\\
        \hline
        \makecell{Density\\(${\rm kg/m^3}$)} & 1000  & 1000 \\ [1.2em]
        \makecell{Viscosity \\ $({\rm Pa~s})$} & 0.001 & 0.0022 \\[1.2em]
        \makecell{Surface tension \\ $({\rm mN-m})$} & 72 & 37.5 \\[1.2em]
        \makecell{Latent heat \\ $({\rm kj/kg})$} & 2453 & 505 \\[1.2em]
        \makecell{Specific heat \\ $({\rm kj/Kg~K})$} & 4.1 & 0.173 \\[1.2em]
        \makecell{Vapour pressure\\ $({\rm kPa})$}& 3.2&  1.06 \\[1.2em]
        \makecell{Molecular weight \\ $({\rm kg/mol})$} & 0.018 & 0.0847\\[1.2em]
        \makecell{Thermal conductivity \\ ${\rm W/m~K}$} & 0.602 & 0.164 \\[1.2em]
    \end{tabular}
        \label{tab:1}
\end{table}
\subsection{Boundary conditions and numerical method}
The weak forms of Eqs. \ref{2.40}, \ref{2.54}, \ref{2.55}, \ref{4.3}, \ref{4.4} (Appendix \ref{appA}) 
are solved using the Finite Element Method subjected to the following conditions at $x = 0$ and $x = L_1$:
\begin{equation}
    \frac{\partial J_A}{\partial x} =  \frac{\partial J_B}{\partial x} =  \frac{\partial h}{\partial x} =  \frac{\partial p}{\partial x} =
     \frac{\partial \mathcal{X}_A}{\partial x}=0. \label{26}
\end{equation}
The diffusion equations for vapour concentrations for both components are solved in the $2$D domain ($0 < x< L_1$ and $0<z<L_2$) and subjected to the following boundary conditions:
\begin{equation}
 \begin{array}{lll}
\displaystyle \frac{\partial \rho^{v,A}}{\partial x} \Bigr \rvert_{x =0} = \frac{\partial \rho^{v,B}}{\partial x}\Bigr \rvert_{x =0} =0, \quad \rho^{v,A}  \Bigr\rvert_{z= L_2} = \mathcal{H}_A, \quad \rho^{v,B}  \Bigr \rvert_{z= L_2} = \mathcal{H}_B.
\end{array}
\end{equation}
This study examines two distinct types of binary mixtures: water-morpholine and ethanol-water. For the water-morpholine mixture, $\mathcal{H}_A$ represents the relative humidity $\mathcal{H}$, while $\mathcal{H}_B = 0$. Conversely, for the ethanol-water mixture, $\mathcal{H}_A = 0$ and $\mathcal{H}_B$ corresponds to the relative humidity ($\mathcal{H}$). For the initial droplet thickness, we used a fourth-order polynomial which satisfies $\frac{\partial h}{\partial x} = \frac{\partial^3 h}{\partial x^3} = 0$ at the centre of the drop, whereas, $\frac{\partial h}{\partial x} =0$ and $h=\beta$ at $R_0$ from the droplet centre ($R_0$ is the initial radius of the drop). In this study, we treat the substrate as effectively thin by taking $\hat{H}_s \to 0$, which yields $Bi = 0$.

The equations were solved using the finite element method in COMSOL Multiphysics. We used a fully implicit finite difference scheme and the PARDISO iterative solver. Typically, we used 10,000 elements for discretization and stopped the simulations when the system mass decreased by $90\%$. The influence of the domain size on the translation dynamics is illustrated in Appendix \ref{appB}.
\begin{table}
    \centering
    \caption{Order of magnitude for all the non-dimensional parameters for Water-morpholine mixture droplets. Here, $\hat{R}_0 = 2~\mathrm{mm}$, and $\hat{h}_0 = 0.4~\mathrm{mm}$.}
    \begin{tabular}{lcr}
        \hline
        No-dimensional number & Definition  & Order of magnitude \\
        \hline
         \makecell{Knudsen number \\ ($Kn$)} & $\frac{\hat{k}_A \Delta \hat{T}}{\hat{L}_{v,A} \hat{h}_0 \hat{\rho}_\mathrm{ref}^{v,A}}\sqrt{\frac{2 \pi \hat{M}_A}{\hat{R}\hat{T}_g}}$ &  $0.0001- 0.1$\\[1.2em]
         \makecell{Evaporation number \\ ($E$)}& $ \frac{\hat{k}_A \Delta \hat{T} \hat{R}_0}{\hat{h}^2_0 \hat{L}_{v,A}\hat{\rho}_A \hat{U}^*}$ &  $0.0001-0.001$\\[1.2em]
       \makecell{Marangoni number \\ $(Ma)$}  & $\frac{\hat{\gamma}_A \Delta\hat{T}}{\hat{\sigma}_A}$ &  $0.001-0.1$\\ [1.2em]
        \makecell{$\psi$} & $ \frac{\hat{L}_{v,A} \hat{M}_A \Delta \hat{T}}{\hat{R}_g\hat{T}_g^2}$ &  $0.1-1$\\[1.2em]
        \makecell{Peclet Number \\  $(Pe)$} & $ \frac{\hat{U}^* \hat{R}_0}{\hat{D}_A}$ &  $1-100$
\\[1.2em]
        \makecell{$\delta$} &$\frac{\hat{\mu_A}\hat{U}^{*}\hat{M_A}}{\hat{\rho}_l \hat{R}_g\hat{T}_g \epsilon^2 \hat{R}_0}$  &  $10^{-5} - 10^{-4}$\\[1.2em]
        \makecell{$Pe_{v,A}$} & $ \frac{\hat{k}_A \Delta \hat{T} \hat{R}_0}{\hat{\mathcal{D}}_{v,A} \hat{\rho}^{v, A}_\mathrm{ref} \hat{L}_{v,A} \hat{h}_0}$ &  $0-1$\\ [1.2em]
         \makecell{$Pe_{v,B}$}& $ \frac{\hat{k}_A \Delta \hat{T} \hat{R}_0}{\hat{\mathcal{D}}_{v,B} \hat{\rho}^{v, B}_\mathrm{ref} \hat{L}_{v,A} \hat{h}_0}$ &  $0-1$\\[1.2em]
    \end{tabular}
    \label{tab:placeholder_label}
\end{table}
\section{Results and Discussion}\label{RD}
The competitive evaporation of binary drops is extremely complex and depends on several factors. In order to gain a strong understanding of several physical phenomena involved, we begin our investigation with pure liquid drops. For pure droplets, the concentration of liquid A in the droplet is taken to be zero ($\mathcal{X}_{A,i} = 0$), and all other base values of the parameters are illustrated in Table \ref{tab1}. In this scenario, we have used a partial wetting condition for the disjoining pressure, as mentioned in equation~\ref{dsj_1}. 
\begin{table}
\caption{Base case parameters for pure droplets ($\mathcal{X}_{A,i} = 0$)}
    \begin{minipage}{0.3\linewidth}
        \centering
        \begin{tabular}{ll}
        
             $\mu_R$ & 1  \\ 
             $\sigma_R$& 1\\
             $M_R$& 1 \\
             $\mathcal{L}$& 1\\
             $k_R$& 1 \\ 
             $Pe_{v,B}$ & 0.1 \\
        \end{tabular}
    \end{minipage}  
    \hfill
    \begin{minipage}{0.32\linewidth}
        \centering
        \begin{tabular}{ll}
            $\gamma_R$ & 1 \\ 
            $\rho^v_{R}$& 1 \\ 
            $C_{p,R}$& 1 \\  
            $M_R$ & 1 \\  
            $\mathcal{A}$ & 150 \\ 
            $\mathcal{H}$ & 0.5 \\
        \end{tabular}
    \end{minipage}
    \hfill
    \begin{minipage}{0.32\linewidth}
        \centering
        \begin{tabular}{ll}
            $Kn$ &  $ 1 \times 10^{-3}$\\ 
            $E$ &  $2 \times 10^{-4}$\\   
            $Pe$ & - \\ 
            $\delta$ & $10^{-4}$ \\ 
            $\psi$ & 0.1 \\
            $\epsilon$ & 0.2
        \end{tabular}
    \end{minipage}
      \label{tab1}
\end{table}
\begin{figure}
\begin{subfigure}[t]{0.5\textwidth}
    \centering
    \includegraphics[width=6cm]{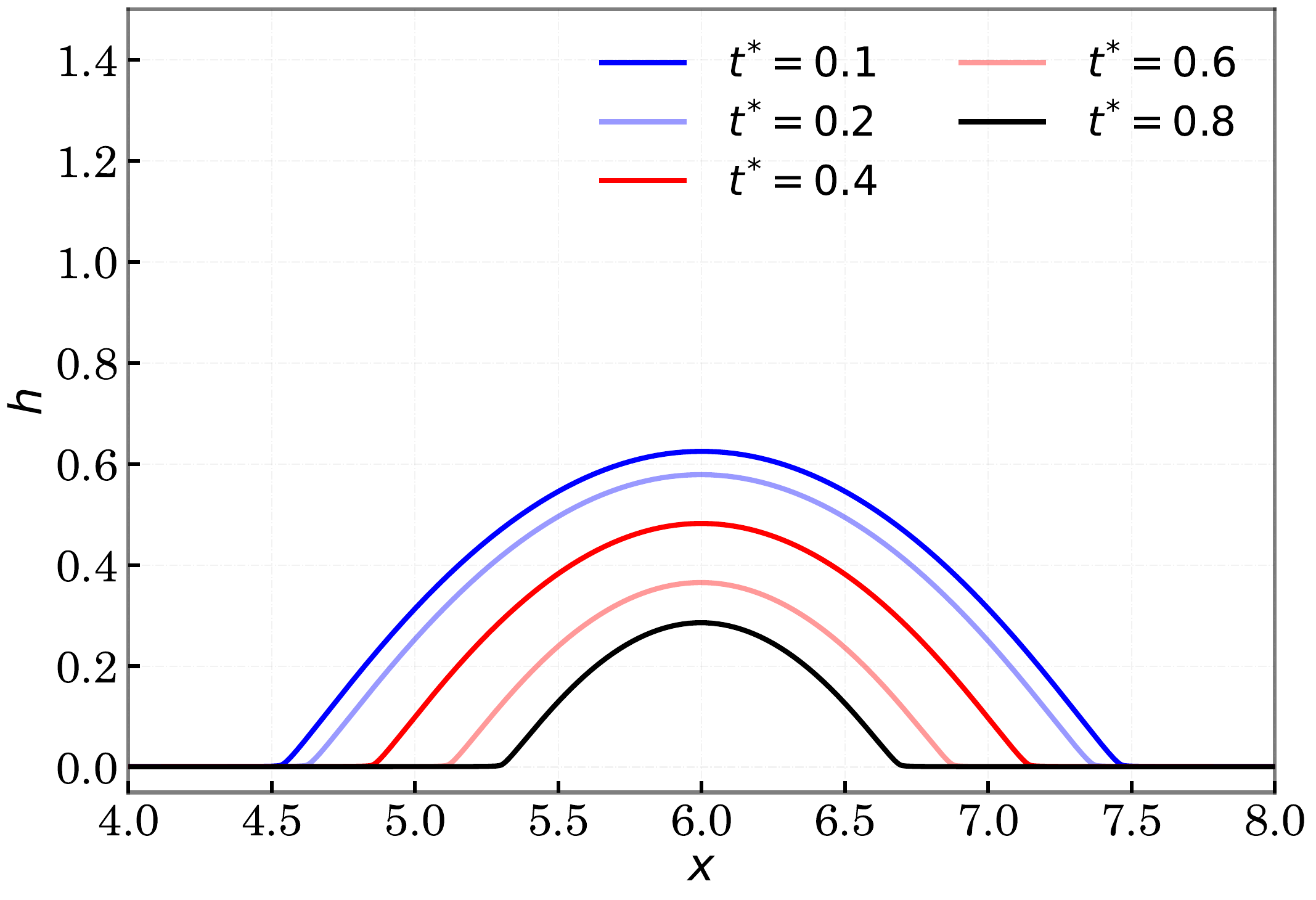}
    \subcaption{}
    \label{fig:enter-label}
    \end{subfigure}
    \begin{subfigure}[t]{0.5\textwidth}
    \centering
    \includegraphics[width=6cm]{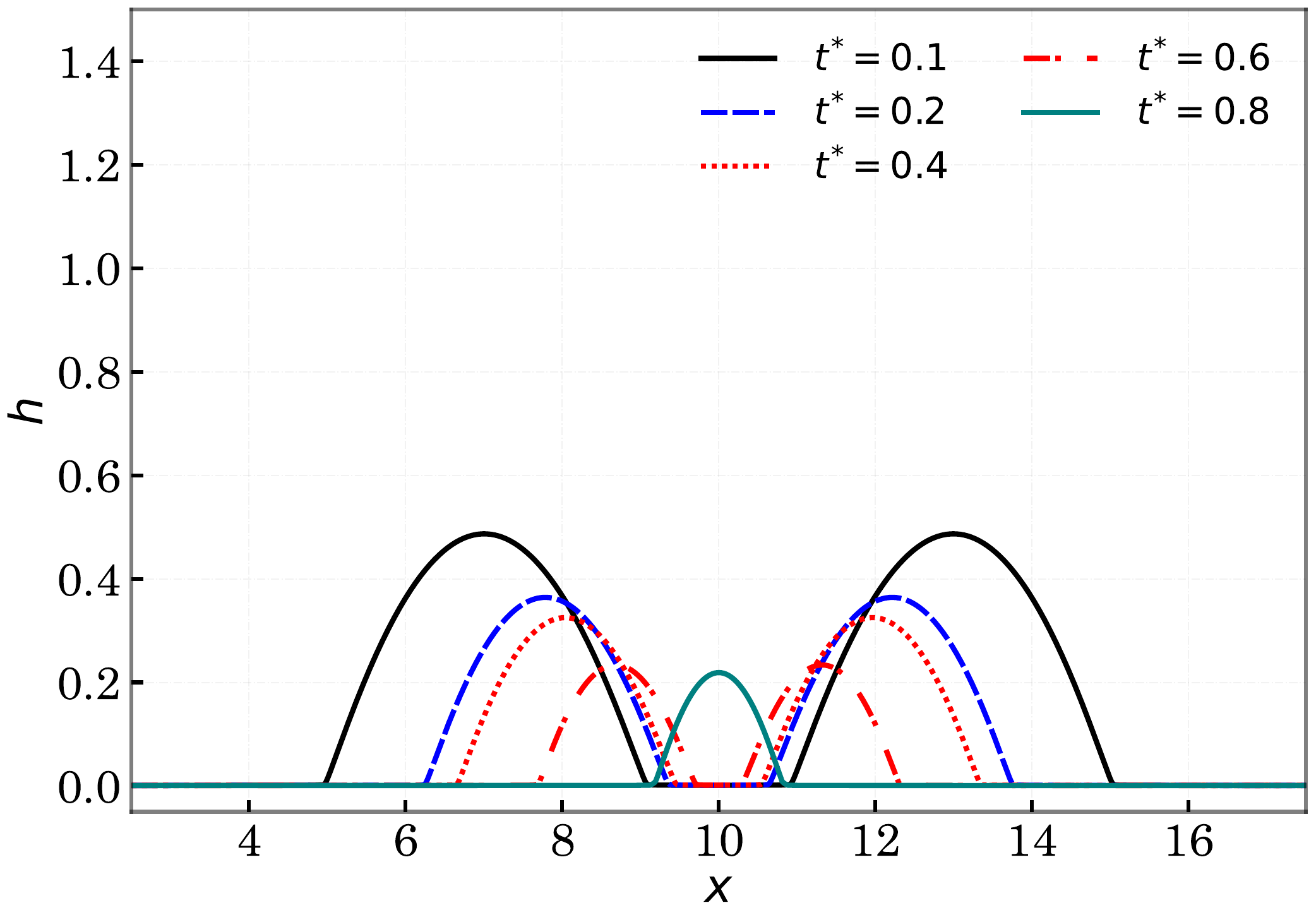}
    \subcaption{}
    \label{fig:enter-label}
    \end{subfigure}
    \begin{subfigure}[t]{0.5\textwidth}
    \centering
    \includegraphics[width=6cm]{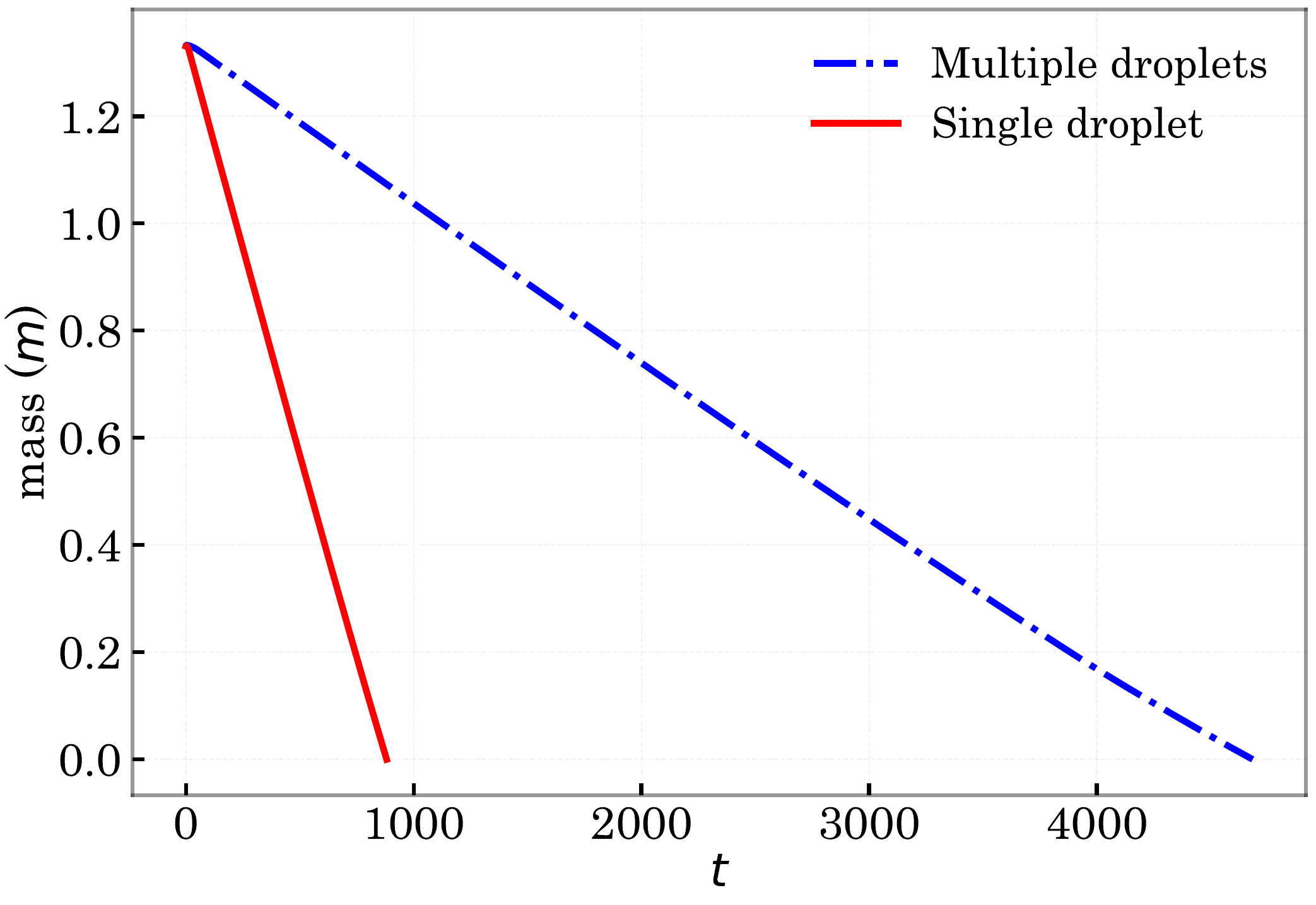}
    \subcaption{}
    \label{fig:enter-label}
    \end{subfigure}
    \begin{subfigure}[t]{0.5\textwidth}
    \centering
    \includegraphics[width=6cm]{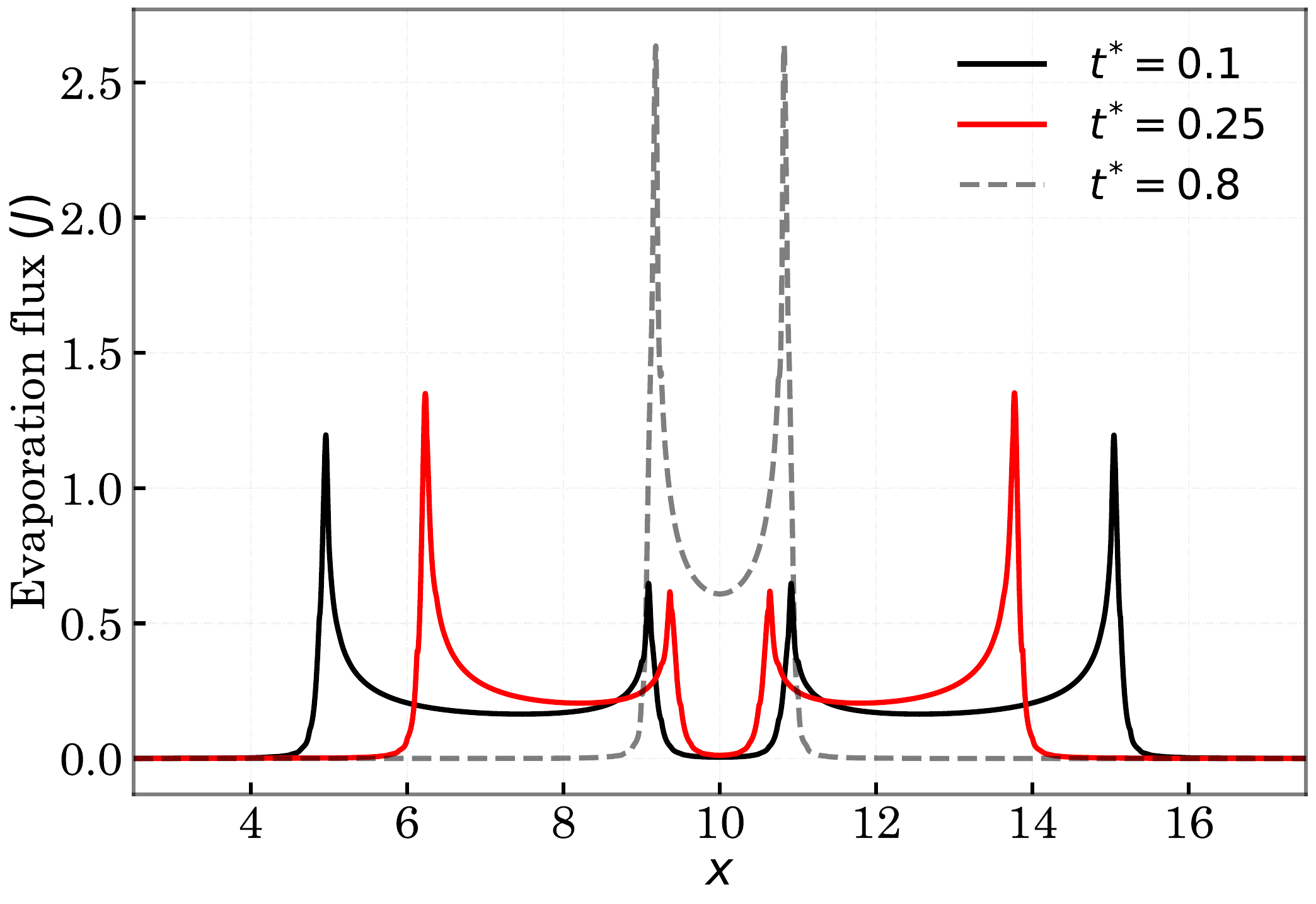}
    \subcaption{}
    \label{fig:enter-label}
    \end{subfigure}
    \caption{Time evolution of interface profile of (a) Isolated droplet, and (b) Two droplets placed near each other; (c) Comparison of 
    depletion with time between the isolated droplet and each of the droplets in the two droplet case, (d) Distribution of evaporative flux along the interface of the droplets at different times.}
    \label{F2}
\end{figure}
\begin{figure}
    \centering
    \includegraphics[width=0.8\textwidth]{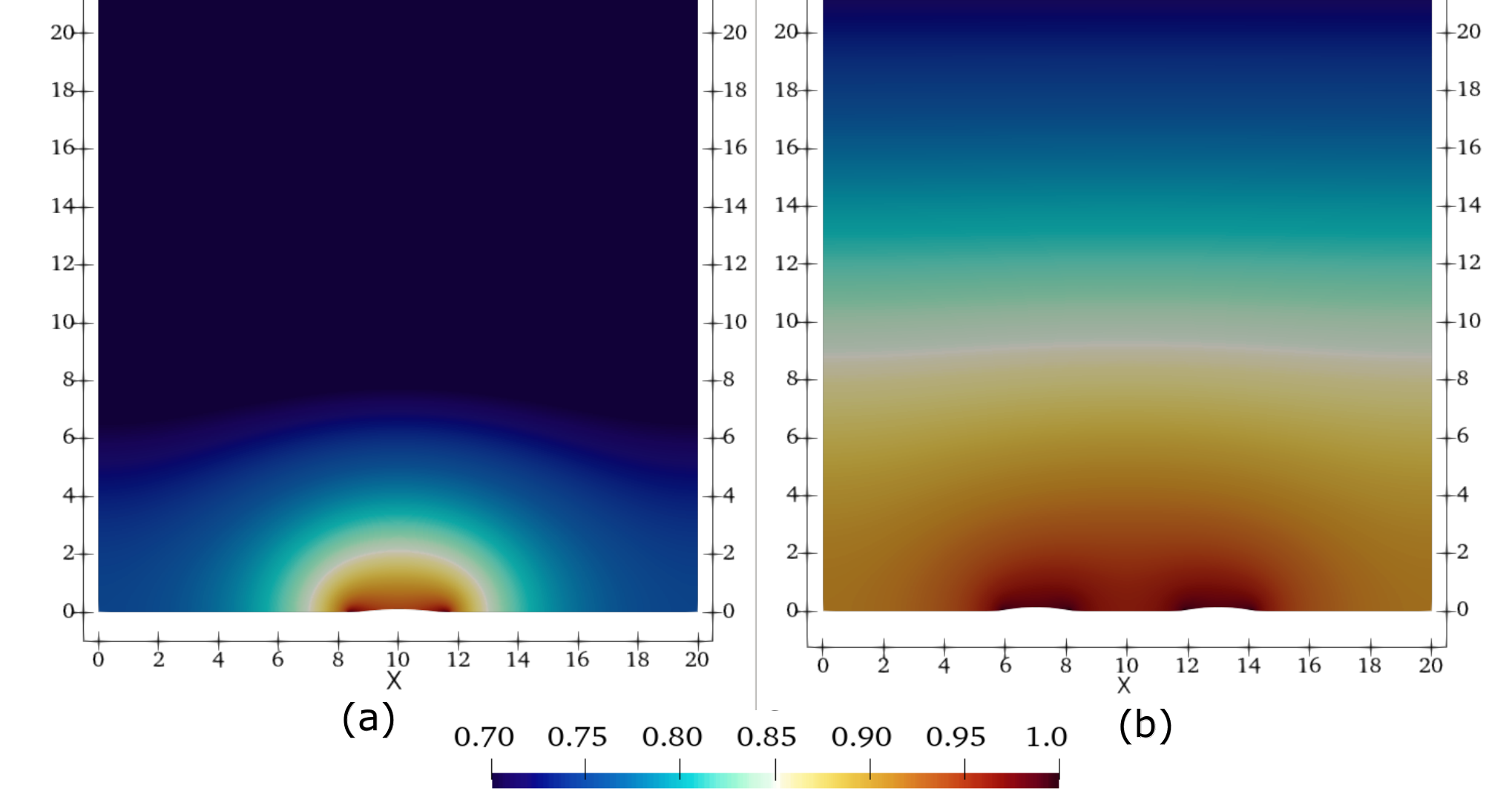}
    \caption{Vapour concentration for (a) isolated droplet, and (b) Two droplet case, at $t^* = 0.2$.}
    \label{F3}
\end{figure}
\subsection{Evaporation of a pair of pure droplets: Vapour shielding effect}
To understand the effect of adjacent droplets on their respective evaporation dynamics, figure~\ref{F2} represents a comparison between an isolated droplet and multiple adjacent droplets for the physical conditions mentioned in Table \ref{tab1}. In figure~\ref{F2}a, we present the time evolution of the droplet interface ($h$) evaporating in a constant contact angle mode. In figure~\ref{F2}b, it can be observed that two evaporating droplets translate toward each other and coalesce, in direct agreement with the experiments for highly volatile pure liquids (Ethyle Acetate) performed by \cite{sadafi_vapor-mediated_2019}, who reported evaporation-driven attraction of sessile droplets on a rigid surface.
The droplets interact because they are able to communicate through the gas phase and because the vapour concentration between two adjacent droplets is higher than that on their periphery, thereby inducing a lower evaporation flux in the middle and a higher one on the periphery. This is clearly shown in figure~\ref{F2}d where we plot the evaporation flux ($J$) at different time instants.
The evaporation flux $(J)$ on the proximal side (contact points of the droplets facing each other) of the droplet is less compared to the distal side (other side of the droplets), creating an asymmetry in the evaporation flux contrary to the symmetric evaporating flux of an isolated drop. 
As the droplets come towards each other and coalesce, the evaporation flux becomes 
symmetric 
with higher evaporation flux in the contact line region as would be expected for a thin isolated drop ($t^* = 0.8$ in figure~\ref{F2}d). Before coalescence, suppression of the evaporation flux between the two droplets occurs due to the accumulation of vapour in the area between the droplets, because the diffusion of the vapour is much slower than the rate of evaporation of the droplet. This phenomenon is reported in figure~\ref{F3}, where the vapour concentration is plotted for the case of isolated drop evaporation (figure~\ref{F3}a) and for multiple drop evaporation (figure~\ref{F3}b) at a time when $25\%$ of the drops have evaporated. For the case of two droplets evaporating, high vapour concentration can be observed in the proximal area between the droplets, resulting in reduced evaporation flux, as depicted in figure~\ref{F2}d. Our model very accurately predicts this phenomenon, also known as vapour shielding in the literature \citep{Wray2021}.

Due to the effect of vapour shielding, the evaporation time of the droplets is also significantly affected, i.e., leading to an increased droplet lifetime, as shown in figure~\ref{F2}c. The vapour shielding effect is also responsible for the translation of the droplets, as shown in figure~\ref{F2}b. The differential evaporation causes an asymmetric droplet profile, and as a result, the capillary stresses drive the two droplets to move toward each other. This effect can be enhanced or suppressed depending on the droplet volatility and the strength of thermo-capillary effects. 
This asymmetric evaporation arises only while the droplets remain distinct and interact with one another; after coalescence, the merged droplet behaves as an isolated droplet, and the asymmetry disappears.

\begin{figure}
    \begin{minipage}{1\textwidth}
    \centering
    \includegraphics[width=0.5
    \linewidth]{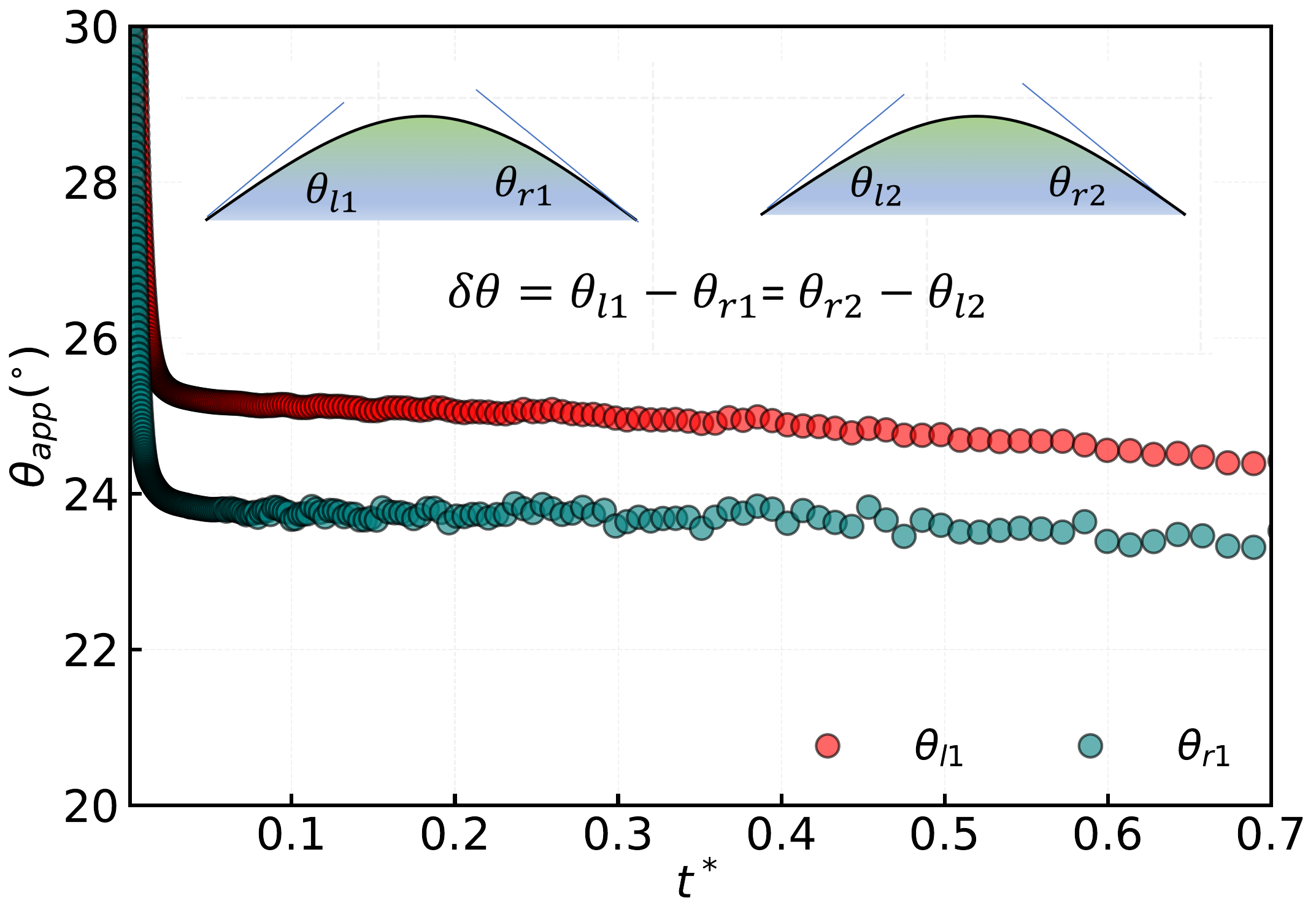} 
    \\ \footnotesize(a)
    \end{minipage}
    \begin{minipage}{0.45\textwidth}
    \centering
    \includegraphics[width=\linewidth]{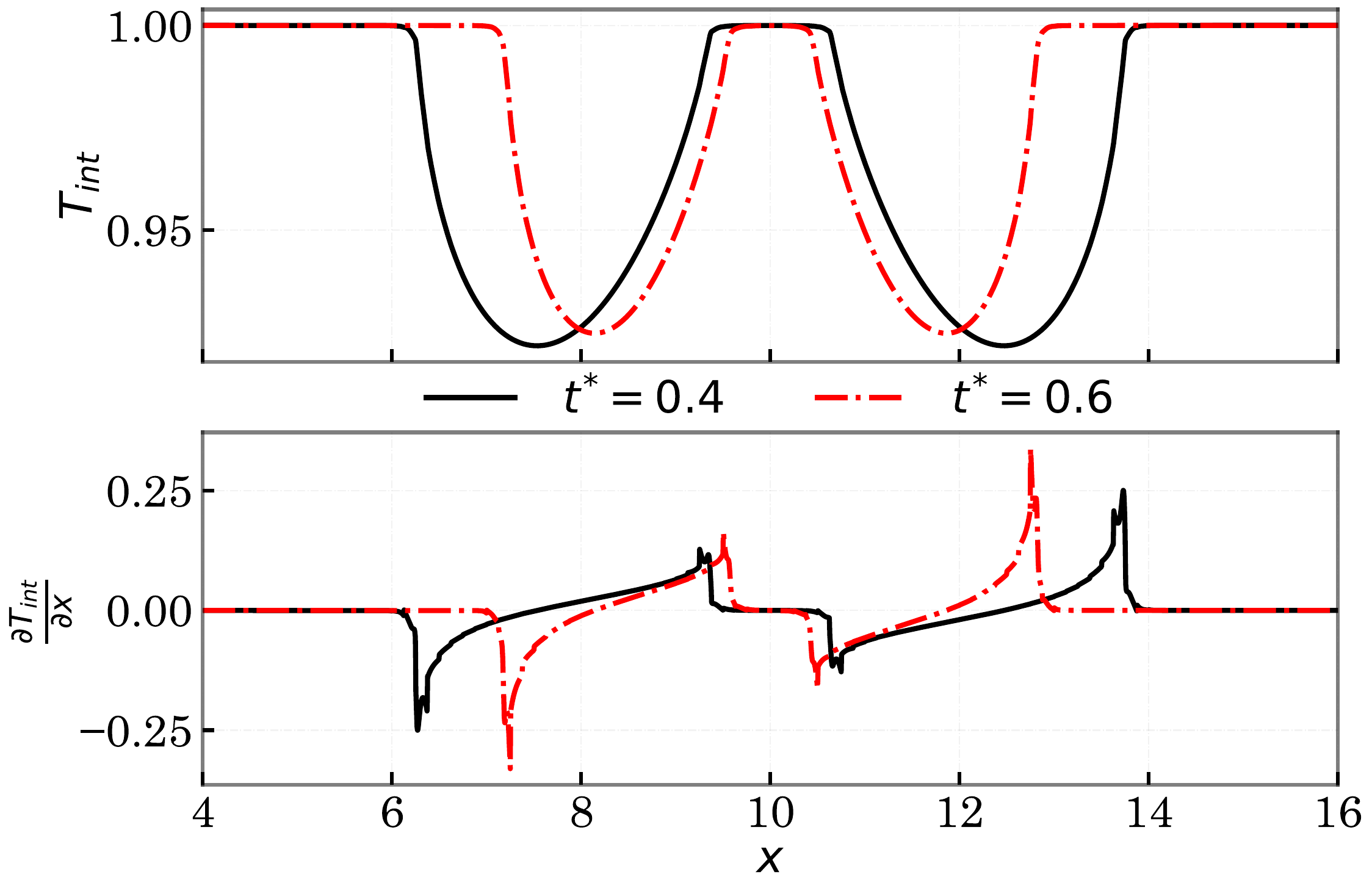}  
    \\ \footnotesize(b)
    \end{minipage}
    \begin{minipage}{0.45\textwidth}
    \centering
    \includegraphics[width=\linewidth]{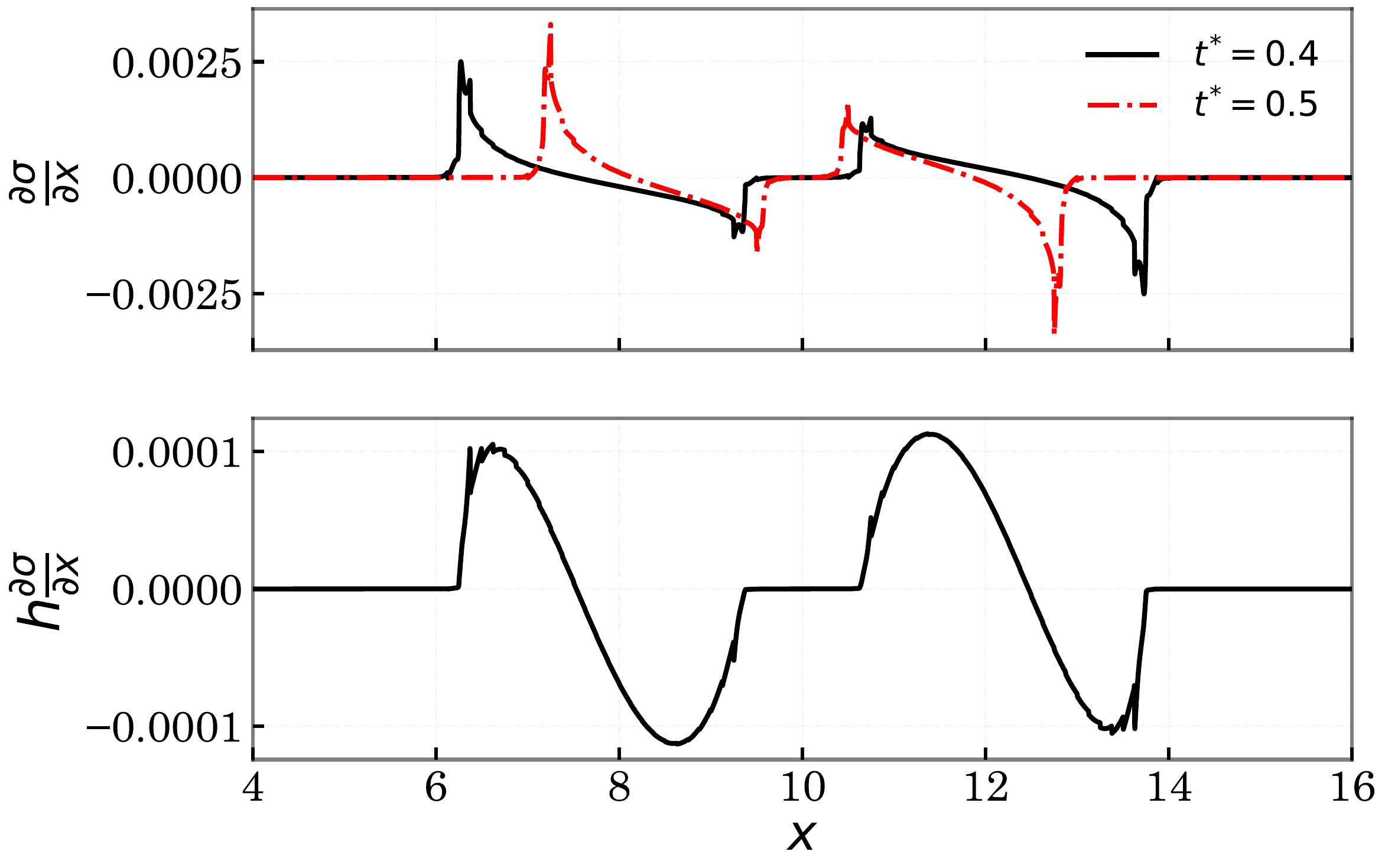}   
    \\ \footnotesize(c)
    \end{minipage}
    \captionof{figure}{(a) Apparent contact angel of the proximal ($\theta_{r1}, \theta_{l2}$) and distal ($\theta_{l1}, \theta_{r2}$) contact side of one of the drops, (b) Variation of droplet interface temperature($T_{int}$) and the gradient of interface temperature $\left(\frac{\partial T_{int}}{\partial x}\right)$, (c) Surface tension gradient and Marangoni stress $\left(h\frac{\partial \sigma}{\partial x}\right)$ due to the surface tension gradient.}
    \label{F5}
\end{figure}

 \begin{center}
    \begin{minipage}{0.45\textwidth}
    \centering
    \includegraphics[width=\linewidth]{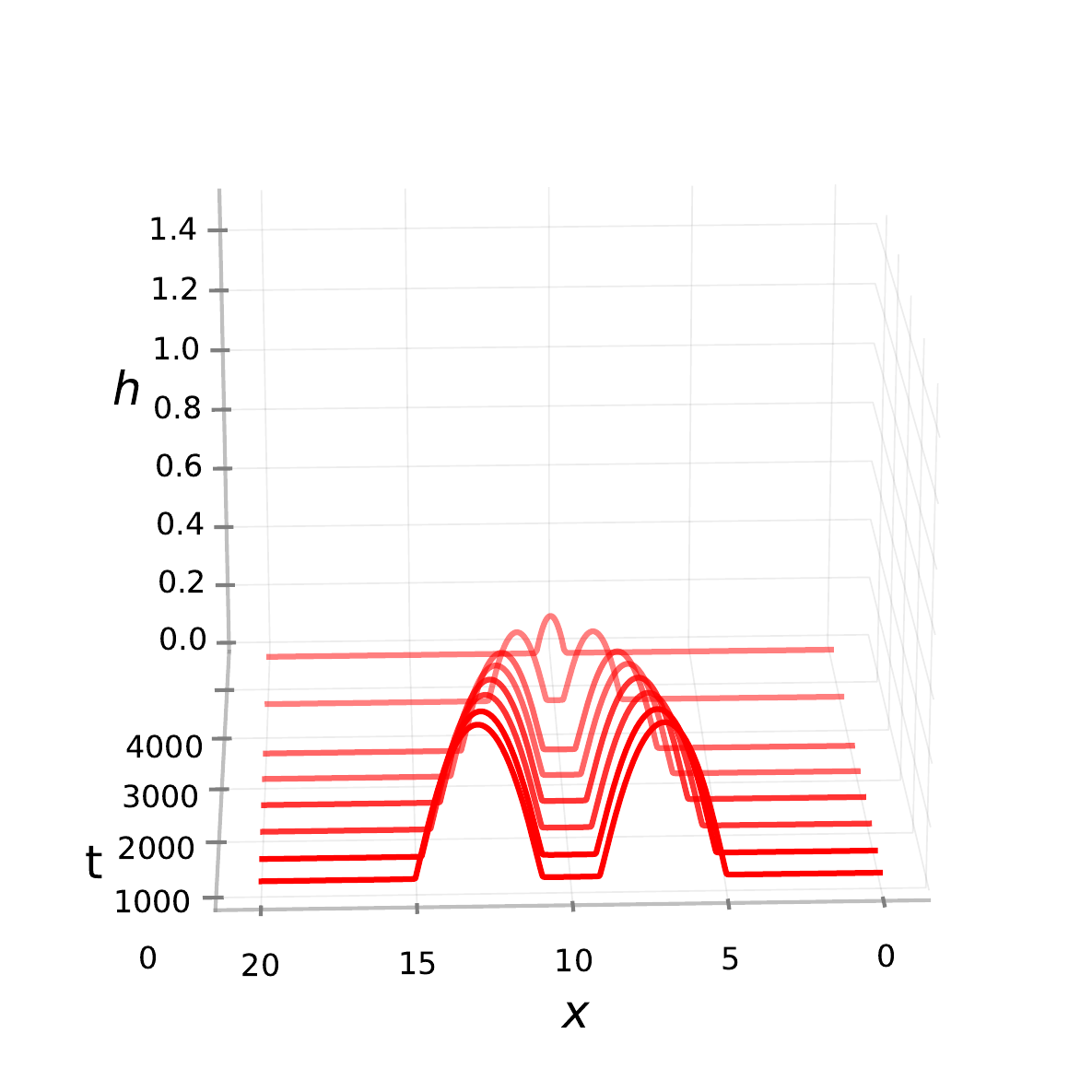} \\
    (a)
    \end{minipage}\hfill
    \begin{minipage}{0.45\textwidth}
    \centering
    \includegraphics[width=\linewidth]{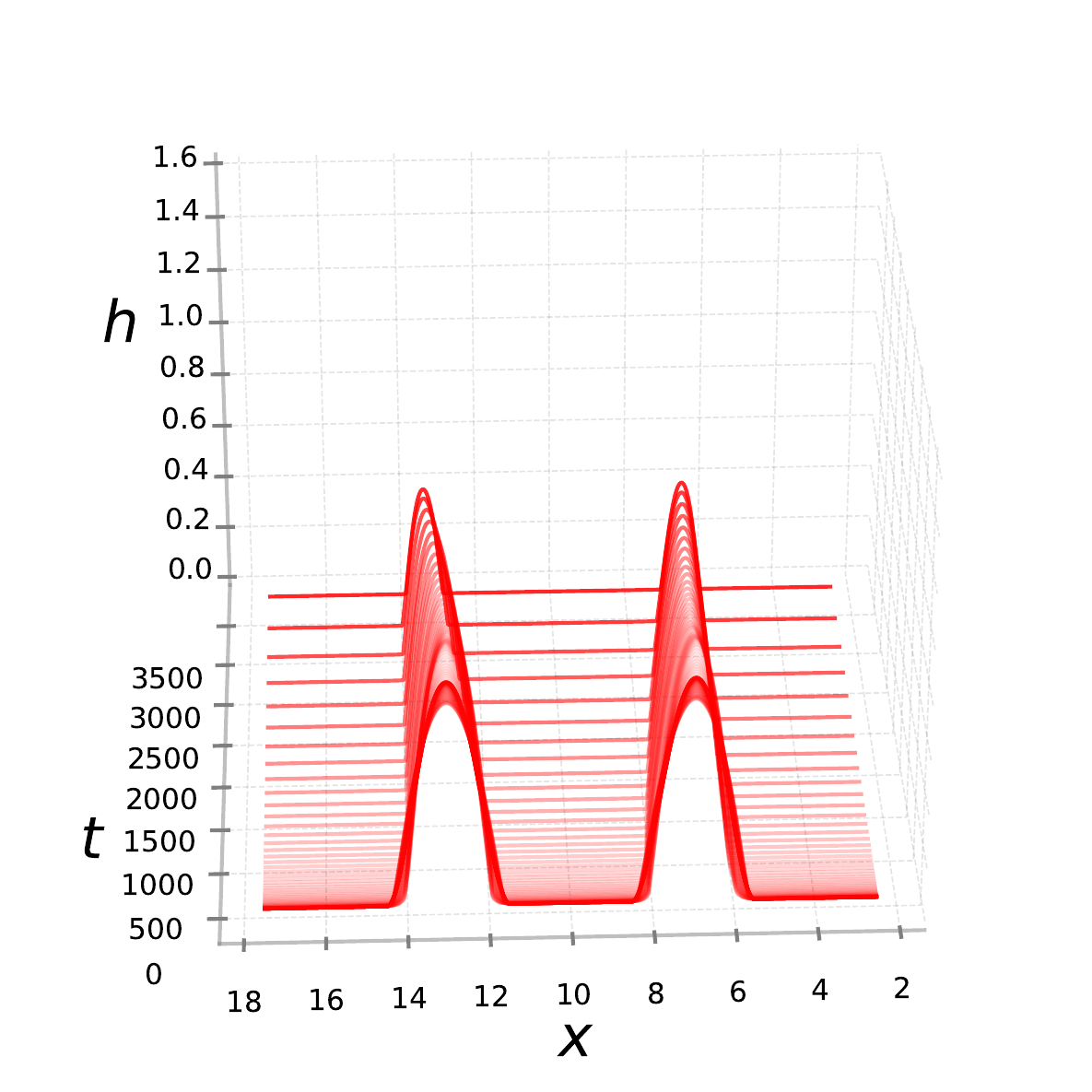}\\
    (b)
    \end{minipage}
    \captionof{figure}{Space-time plots for droplet shape ($h$) for (a) $Ma =0.001$, and (b) $Ma=0.01$ showing attraction and repulsive behaviour of the droplets. All other parameters are the same as table \ref{tab1}.}
    \label{F4}
    \end{center}
In the present model, the thermo-capillary effect is quantified by the thermal Marangoni number ($Ma$). Figure~\ref{F4} shows the space-time plots for two evaporating droplets for two different thermal Marangoni numbers, keeping all other properties the same. In the case of weak thermocapillarity (e.g., see figure~\ref{F4}a for $Ma=0.001$), 
the droplets tend to attract each other and coalesce.
On the other hand, for $Ma = 0.01$ (figure~\ref{F4}b), the droplets do not translate toward each other, thus indicating a counteracting repulsive behaviour due to the effect of thermal Marangoni stresses.
This effect has also been reported by \cite{malachtari_dynamics_2024} who showed that for rigid substrates and under sufficiently strong thermal Marangoni stresses, the droplets may even repulse; as shown by these authors an additional factor that may affect the nature of droplet interactions is also the rigidity of the substrate.
There are two distinct mechanisms at play here: capillary pressure, which drives attraction, and thermal Marangoni effects, which drive repulsion. Depending on the value of $Ma$, one of these mechanisms becomes dominant over the other. 
The differential 
evaporation makes the inner side of 
the droplet locally thicker and less 
curved (vapour shielding reduces 
evaporation there) and the outer side 
locally thinner and more curved. The 
resulting asymmetric curvature produces 
a higher capillary pressure 
on the outer side and lower on the 
inner side, giving $\partial p/\partial x 
< 0$ across the left droplet and hence directed toward the 
neighbouring droplet. Similarly, the right droplet experiences an equal 
and opposite velocity. This is shown in figure~\ref{F5}a where we depict the variation of the apparent contact angles of the drops on each contact side. We observe that, the contact angles on the left and right contact points of the droplets are different. The contact angles of the drops on the proximal side ($\theta_{r1}, \theta_{l2}$) are lower than the contact angles at the distal contact points ($\theta_{l1}, \theta_{r2}$), creating a resultant capillary flow from the distal side to the proximal side for the droplet on the left, promoting attraction. This differential contact angle of the droplets resulting from the evaporation gradient was also suggested as one of the mechanisms for pure droplet attraction by \cite{sadafi_vapor-mediated_2019}. 
\medskip
To understand the reason behind these 
observations and how thermocapillarity 
affects droplet interactions, the 
interfacial temperature $T_{int}$ and 
its axial gradient $\partial T_{int}/
\partial x$ are plotted in 
figure~\ref{F5}b. The physical picture 
has two distinct temporal stages that 
must be considered separately.
At early times, before significant 
vapour shielding has developed between 
the two droplets, the evaporation 
field is approximately symmetric about 
each droplet centre. The interfacial 
temperature is maximum at the contact 
lines, where the film is thin and 
evaporative cooling is strongest, and 
minimum at the apex where the film 
is thickest. Since 
$\partial{\sigma}/\partial{T} 
< 0$, the contact lines have higher 
surface tension than the apex, 
producing a surface tension gradient 
directed inward from both contact 
lines toward the apex. This drives 
a symmetric inward Marangoni flow 
that exerts a compressive stress on 
the droplet, reducing the droplet 
footprint as both contact lines are 
drawn inward. This compressive 
behaviour can be clearly observed at 
early times in figure~\ref{F5}c, where 
the surface tension gradient $h\partial\sigma/
\partial x$ is directed inward 
symmetrically from both contact lines. 
The net effect at this stage is a 
contraction of the droplet profile 
with no net translation, since the 
stress is symmetric and the two 
contact lines recede at equal rates.
As time progresses, vapour shielding 
develops on the inner facing sides 
of the two droplets, suppressing 
evaporation and hence evaporative 
cooling there. The temperature field 
becomes asymmetric: the inner contact 
line is warmer than the outer contact 
line, as the suppressed evaporation 
on the inner side reduces the local 
cooling. This asymmetry is clearly 
visible in the $\partial T_{int}/\partial x$ 
profile in figure~\ref{F5}b. The 
Marangoni stress now has two 
components: the symmetric compressive 
component that continues to reduce 
the droplet footprint, and a new 
asymmetric component arising from 
the temperature difference between 
the inner and outer sides. Since the 
inner side is warmer and hence at 
lower surface tension, the asymmetric 
component of the surface tension 
gradient $\partial\sigma/\partial x$ 
is directed outward (from inner low 
$\sigma$ to outer high $\sigma$), 
driving a net outward tangential 
Marangoni flow directed away from the neighbouring 
droplet. This outward flow carries 
liquid from the inner side toward 
the outer side of each droplet, 
draining liquid away from the inner 
contact line and depositing it at 
the outer contact line. For a sufficiently small Marangoni number, for example $Ma = 0.001$, the thermal Marangoni effect is too weak and is dominated by the induced capillary flow, which results in attraction.  
Conversely, for larger thermal Marangoni numbers, such as $Ma = 0.01$, the capillary forces are overwhelmed by the thermal Marangoni flow, and the droplets therefore do not attract each other, as already illustrated in figure~\ref{F4}a and figure~\ref{F4}b, respectively.
\begin{table}
    \caption{Non-dimensional parameter values for water-morpholine mixture.}
    \begin{minipage}{0.3\linewidth}
        \centering
        \begin{tabular}{ll}
             $\mu_R$ & 2  \\ 
             $\sigma_R$& 0.5 \\
             $M_R$& 4.7 \\
             $\mathcal{L}$& 0.2\\
             $k_R$& 0.6 \\ 
             $Pe_{v,A}$ & 0.1 \\
        \end{tabular}
    \end{minipage}
    \hfill
    \begin{minipage}{0.32\linewidth}
        \centering
        \begin{tabular}{ll}
            $\gamma_R$ & 0.58 \\ 
            $\rho^v_{R}$& 0.16 \\ 
            $C_{p,R}$& 0.16 \\    
            $\mathcal{A}$ & 200 \\ 
            $Pe_{v,B}$ & 0.3 \\
             $\epsilon$ & 0.2
        \end{tabular}
    \end{minipage}
    \hfill
    \begin{minipage}{0.32\linewidth}
        \centering
        \begin{tabular}{ll}
            $Kn$ &  0.001\\ 
            $E$ &  $1 \times 10^{-4}$\\   
            $Pe$ & 20 \\ 
            $\delta$ & $10^{-4}$ \\ 
            $\psi$ & 0.1 \\
           
        \end{tabular}
    \end{minipage}
    \label{tab:4}
\end{table}

\begin{figure}
    \centering

    \begin{subfigure}[t]{0.45\textwidth}
        \centering
        \includegraphics[width=\linewidth]{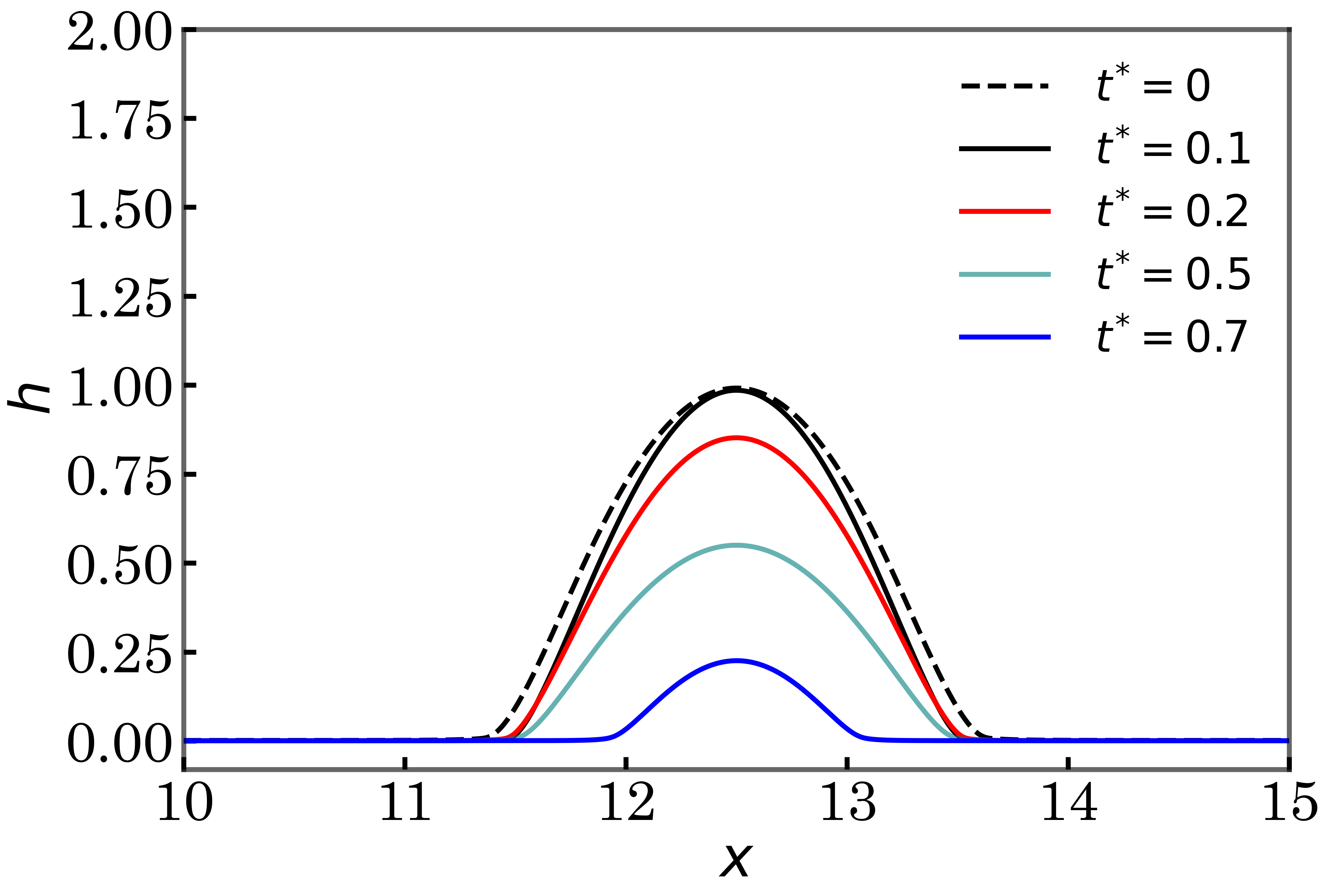}
        \caption{}
    \end{subfigure}\hfill
    \begin{subfigure}[t]{0.45\textwidth}
        \centering
        \includegraphics[width=\linewidth]{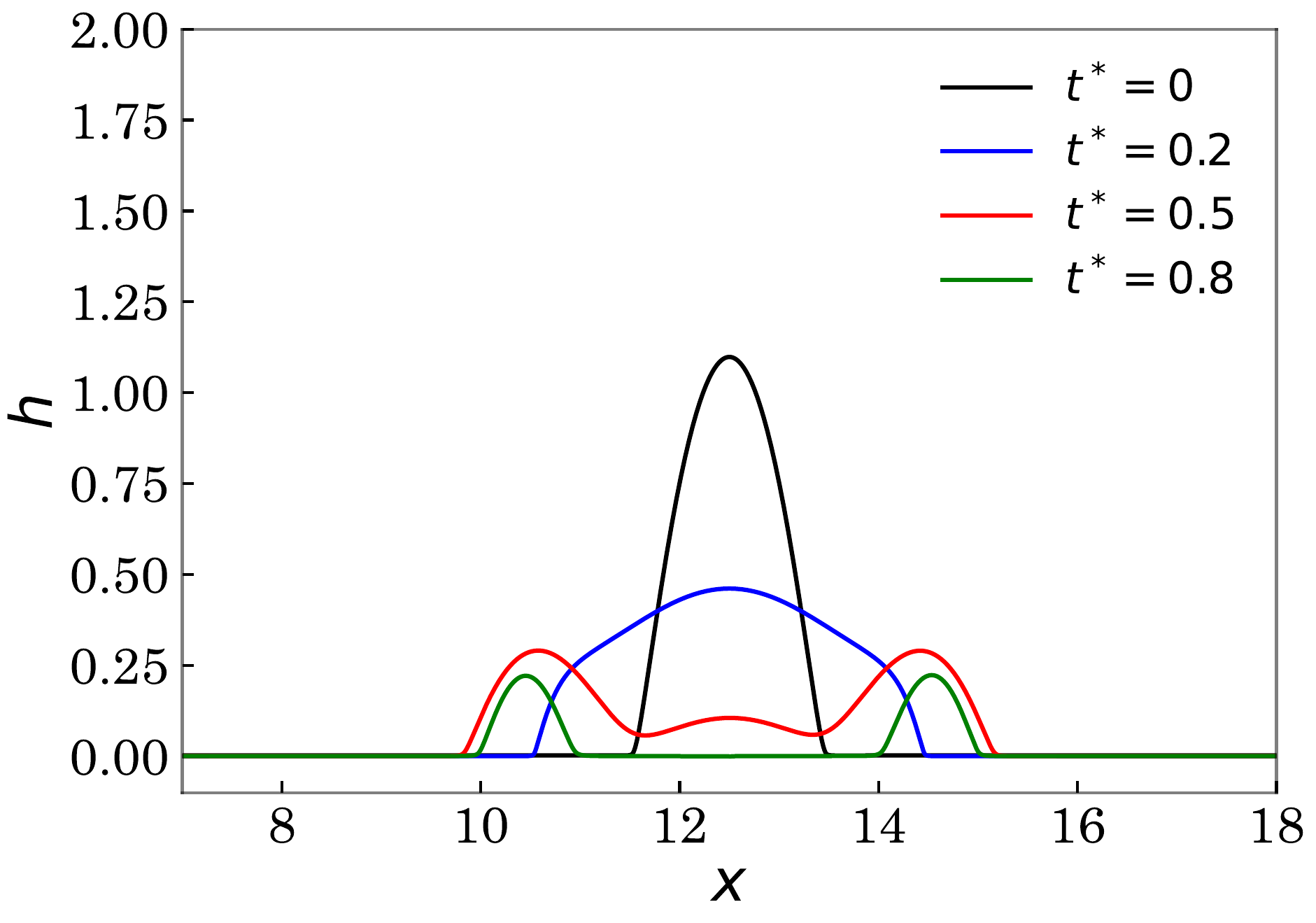}
        \caption{}
    \end{subfigure}

    \begin{subfigure}[t]{0.45\textwidth}
        \centering
        \includegraphics[width=\linewidth]{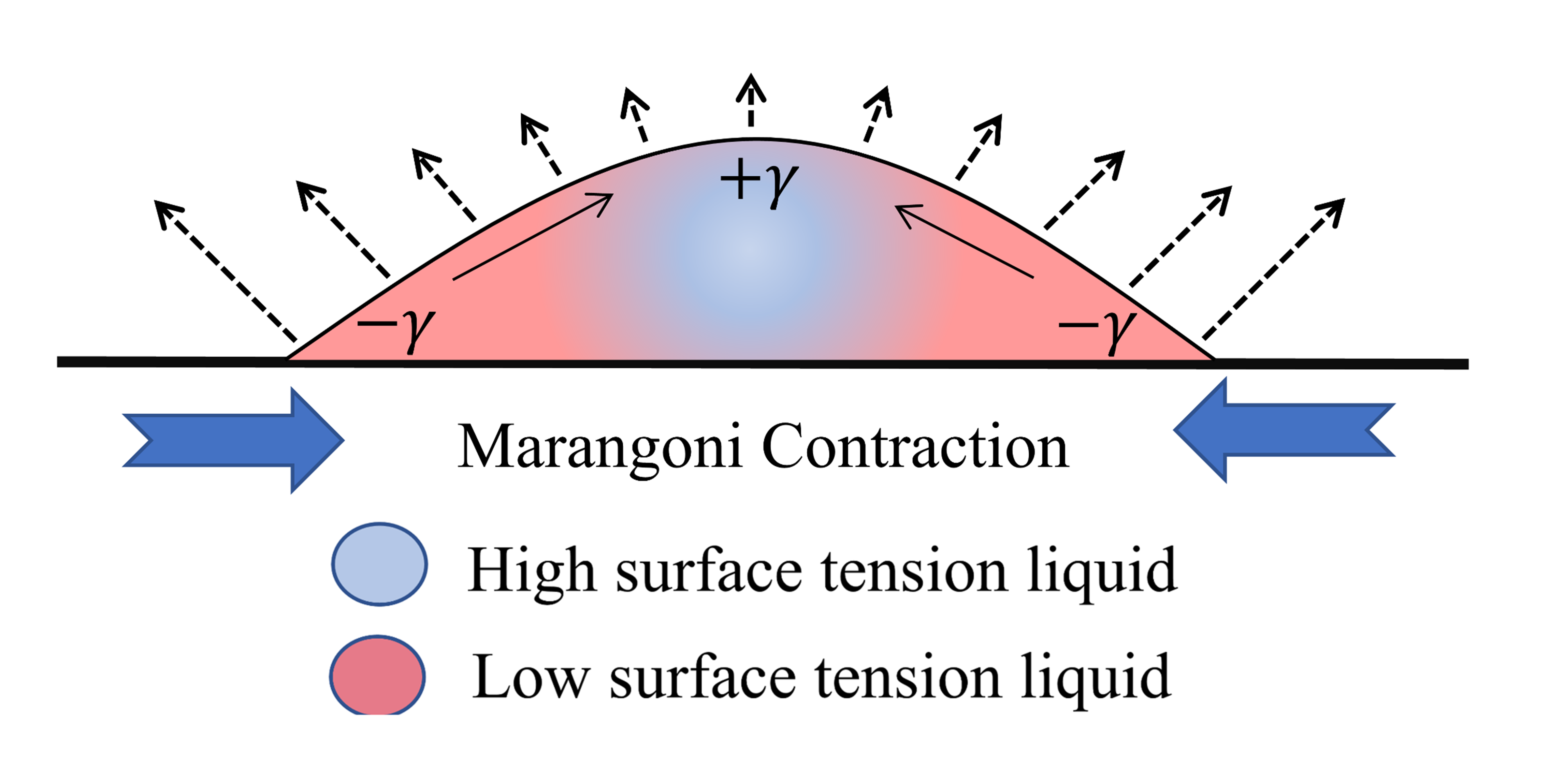}
        \caption{}
    \end{subfigure}\hfill
    \begin{subfigure}[t]{0.45\textwidth}
        \centering
        \includegraphics[width=\linewidth]{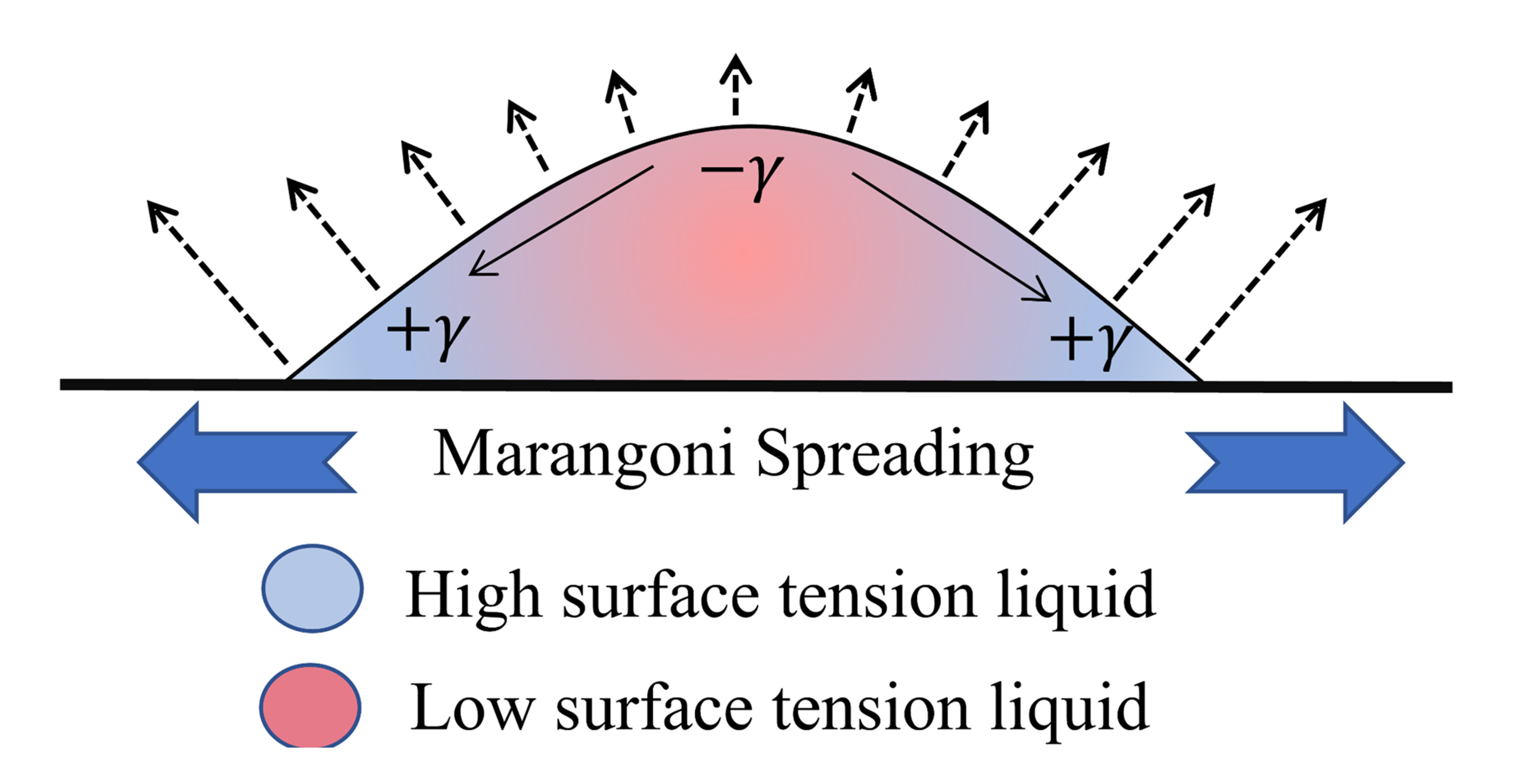}
        \caption{}
    \end{subfigure}

    \par\medskip
    \begin{subfigure}[t]{0.6\textwidth}
        \centering
        \includegraphics[width=\linewidth]{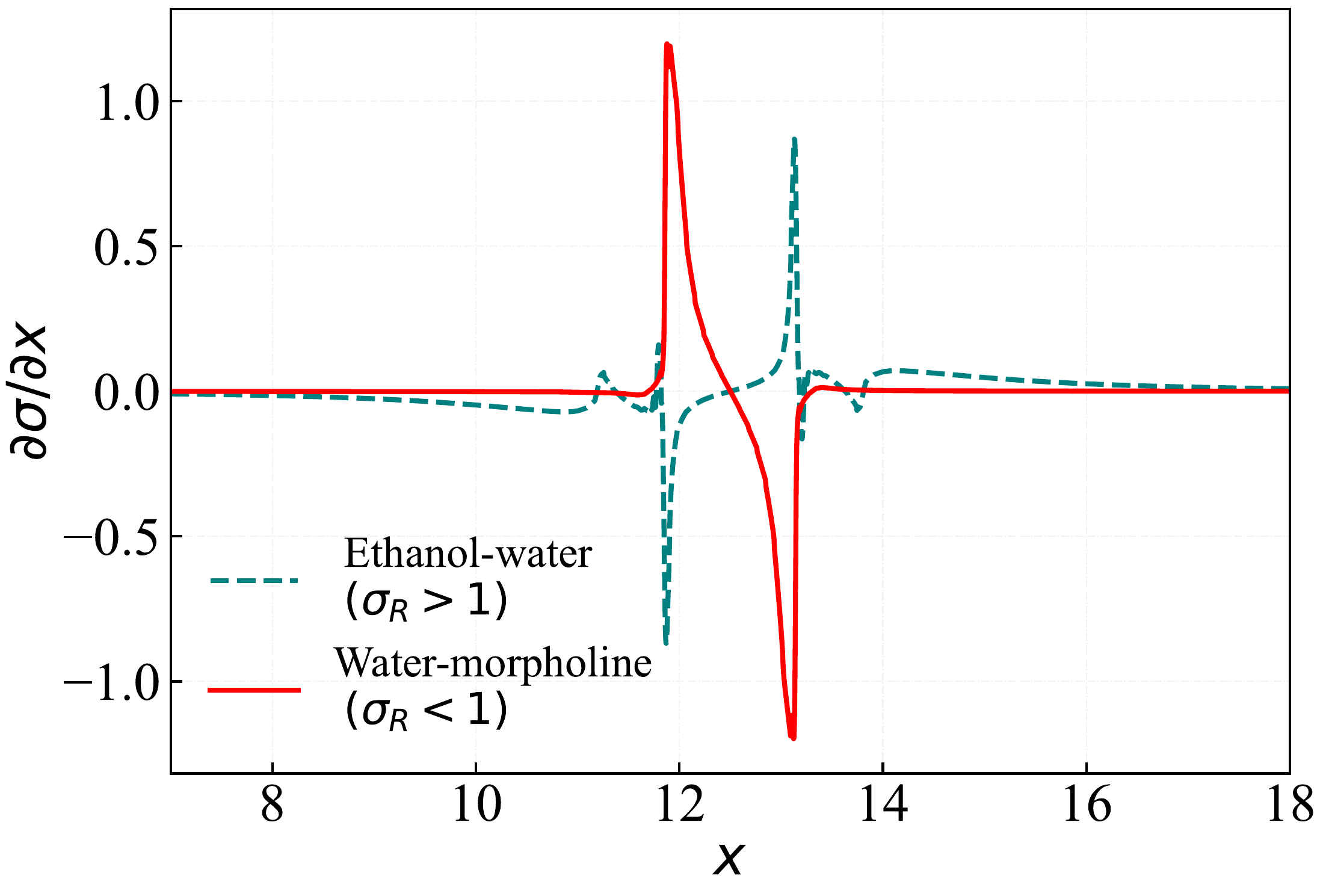}
        \caption{}
    \end{subfigure}

    \caption{Time evolution of droplet interface ($h$) for two different binary mixtures: (a) Water-morpholine ($\sigma_R < 1 $), and (b) water-ethanol ($\sigma_R > 1$) mixtures. Schematic of solutal Marangoni effects arising from the surface tension gradient for (c) water-morpholine mixture and (d) water-ethanol mixture, (e) Surface tension gradient for droplets for varying $\sigma_R$ at a time frame where $50 \%$ of the drops are evaporated.}
    \label{fig6}
\end{figure}
\begin{table}
    \caption{Non-dimensional parameter values for ethanol-water mixture.}
    \begin{minipage}{0.3\linewidth}
        \centering
        \begin{tabular}{ll}
             $\mu_R$ &0.84  \\ 
             $\sigma_R$& 3.2\\
             $M_R$& 0.39 \\
             $\mathcal{L}$& 2.3\\
             $k_R$& 0.6 \\ 
             $Pe_{v,A}$ & 0.02 \\
        \end{tabular}
    \end{minipage}
    \hfill
    \begin{minipage}{0.32\linewidth}
        \centering
        \begin{tabular}{ll}
            $\gamma_R$ & 1.81 \\ 
            $\rho^v_{R}$& 0.16 \\ 
            $C_{p,R}$& 0.16 \\  
            $M_R$ & 0.39 \\  
            $\mathcal{A}$ & 150 \\ 
            $Pe_{v,B}$ & 0.1 \\
        \end{tabular}
    \end{minipage}
    \hfill
    \begin{minipage}{0.32\linewidth}
        \centering
        \begin{tabular}{ll}
            $Kn$ &  $8.8 \times 10^{-4}$\\ 
            $E$ &  $2.6 \times 10^{-4}$\\   
            $Pe$ & 50 \\ 
            $\delta$ & $10^{-5}$ \\ 
            $\psi$ & 0.2 \\
            $\epsilon$ & 0.2
        \end{tabular}
    \end{minipage}

    \label{tab:5}
\end{table}

\subsection{Behaviour of isolated binary droplets}\label{s6.2}
We now include a second component in the liquid and thus consider a system that comprises a binary mixture.
A binary mixture droplet can behave differently depending on the composition of the mixture, exhibiting spreading or retraction \citep{Williams2021, Cira2015}, depending on the surface tension ratio of the binary mixture ($\sigma_R$). 
Unlike the pure droplet case, from this point 
onward for the binary mixture cases we employ the complete-wetting 
condition for the disjoining pressure, using equation~\ref{dsj_2}, 
so that the binary mixture itself determines the droplet's contact 
angle. We note that a precursor film is present in both the pure 
and binary drop formulations; the distinction lies solely in the 
form of the disjoining pressure, specifically whether both 
repulsive and attractive terms are retained (partial wetting, 
equation~\ref{dsj_1}) or only the repulsive term (complete wetting, 
equation~\ref{dsj_2}). The complete-wetting model is physically suitable for the binary systems examined here because each component in the water–morpholine and ethanol–water mixtures, on its own, fully wets the substrate \citep{Cira2015,karpitschka_marangoni_2017}. Moreover, this formulation permits the solutal Marangoni stresses to dictate freely whether the droplet footprint expands or shrinks, depending on the surface tension ratio $\sigma_R$, which is the key physical parameter that differentiates binary droplets from pure ones. Imposing a 
partial-wetting equilibrium contact angle would artificially 
constrain the footprint evolution and suppress the 
solutal-Marangoni-driven dynamics that are the focus of the 
present study. Furthermore, the primary objective of this work 
is to investigate the translation dynamics of evaporating droplet 
pairs; if the binary drops were to spread and form thin films, 
the asymmetric evaporation flux and Marangoni forces responsible 
for translation would become vanishingly small and the translation 
behaviour could not be meaningfully studied, as is observed in 
the ethanol-water case at low $Ma$ in later sections.
In figure~\ref{fig6}a, we present a case that corresponds to an isolated droplet of the water-morpholine binary mixture and depict the evolution of the interface shape over time.
For the water-morpholine mixture, the more volatile component, water, has a higher surface tension than the less volatile component, morpholine ($\sigma_R < 1$, all parameters are given in Table \ref{tab:4}) 
The droplet stays pinned until $t^* = 0.5$, this phenomenon is also known as Marangoni contraction, and a schematic representation of the mechanism is shown in figure~\ref{fig6}c.
The apparent pinning of the contact line 
until $t^* = 0.5$ is not a thermodynamic pinning but a 
dynamic balance between the inward solutal Marangoni flow, 
driven by the enrichment of the low-surface-tension morpholine 
at the contact line due to preferential evaporation of water (figure~\ref{fig6}c), 
and the outward spreading tendency of the complete-wetting 
film. During the interval of
approximately constant contact radius, this inward
transport opposes the spreading tendency of the
complete-wetting drop. Direct numerical evidence for this
mechanism is provided in Appendix \ref{app:no_marangoni}:
suppressing the Marangoni contribution to the liquid flux,
while retaining the other governing equations and initial
conditions, produces sustained spreading
(figure~\ref{fig:no_marangoni}). Thus, the numerical
complete-wetting formulation permits spreading, and its
suppression in the full model depends on Marangoni
transport. This interpretation is consistent with the
quasistationary competition between capillary and
Marangoni fluxes described by
\citet{karpitschka_marangoni_2017} for evaporating binary
droplets (water--diol).
The subsequent evaporation regimes, such as the very
short-lived CCA regime and the late stage in which both
the contact angle and the contact radius vary, arise
because the inward solutal Marangoni stress, the inward
thermal Marangoni stress, and the resisting capillary
pressure scale with the drop geometry in the same way as
the drop shrinks.

Furthermore, for the water--morpholine drops, the
characteristic timescales satisfy the strict ordering
$t_{sol} < t_{sp} < t_{ev}$. Here, $t_{sol}$ is the
solutal Marangoni timescale, $t_{sp}$ is the spreading
timescale driven by the repulsive disjoining pressure,
and $t_{ev}$ is the evaporation timescale.
These timescales, and especially $t_{sp}$, apply solely
to the present numerical model using a precursor-film
formulation and should not be interpreted as quantitative
experimental values. Since $t_{sol}$ is the shortest
timescale, solutal Marangoni flow can oppose and compete
with spreading early, before any substantial contact-line
advance occurs. Further details of these timescale
estimates are provided in Appendix \ref{app:no_marangoni}.

%
%
Comparable phenomena were observed in the current experimental study and corroborated by experimental observations in \cite{karpitschka_marangoni_2017, Cira2015} pertaining to binary mixture droplets, where the more volatile component has a higher surface tension.
On the other hand, if the surface tension ratio, $\sigma_R$, exceeds 1, meaning that the more volatile liquid possesses a lower surface tension, the droplets undergo Marangoni spreading. Figure~\ref{fig6}b depicts the spreading behaviour of the evaporating $50\%$ ethanol-water droplet (all values of nondimensional parameters for the ethanol-water mixture are presented in Table \ref{tab:5}). The droplet spreads rapidly and forms ridges as a result of the counteracting mechanisms of the solutal and thermal Marangoni stresses. The spreading behaviour can be described with figure~\ref{fig6}d. The more volatile component, with a lower surface tension, accumulates near the apex of the droplet, whereas the less volatile component, with a higher surface tension, concentrates in the vicinity of the contact line. This distribution engenders a surface tension gradient, thereby facilitating the spreading of the droplet. Such spreading phenomena of ethanol-water droplets and ridge formation are reported in \cite{Williams2021} with the help of experiments and simulations using a one-sided evaporation model. In contrast to the one-sided framework proposed by \cite{Williams2021}, the current model yields predictions that are considerably more aligned with experimental data. More details on the spreading of isolated ethanol-water droplets are depicted in Appendix \ref{appC}.

These predictions for isolated droplets within both types of binary mixture droplets from our model provide an ideal foundation to explore the mechanism of droplet translation when such droplets are positioned next to one another.


\subsection{Translation of Multiple binary drops}
\begin{figure}
\begin{subfigure}[t]{0.5\textwidth}
    \centering
        \includegraphics[width=6cm]{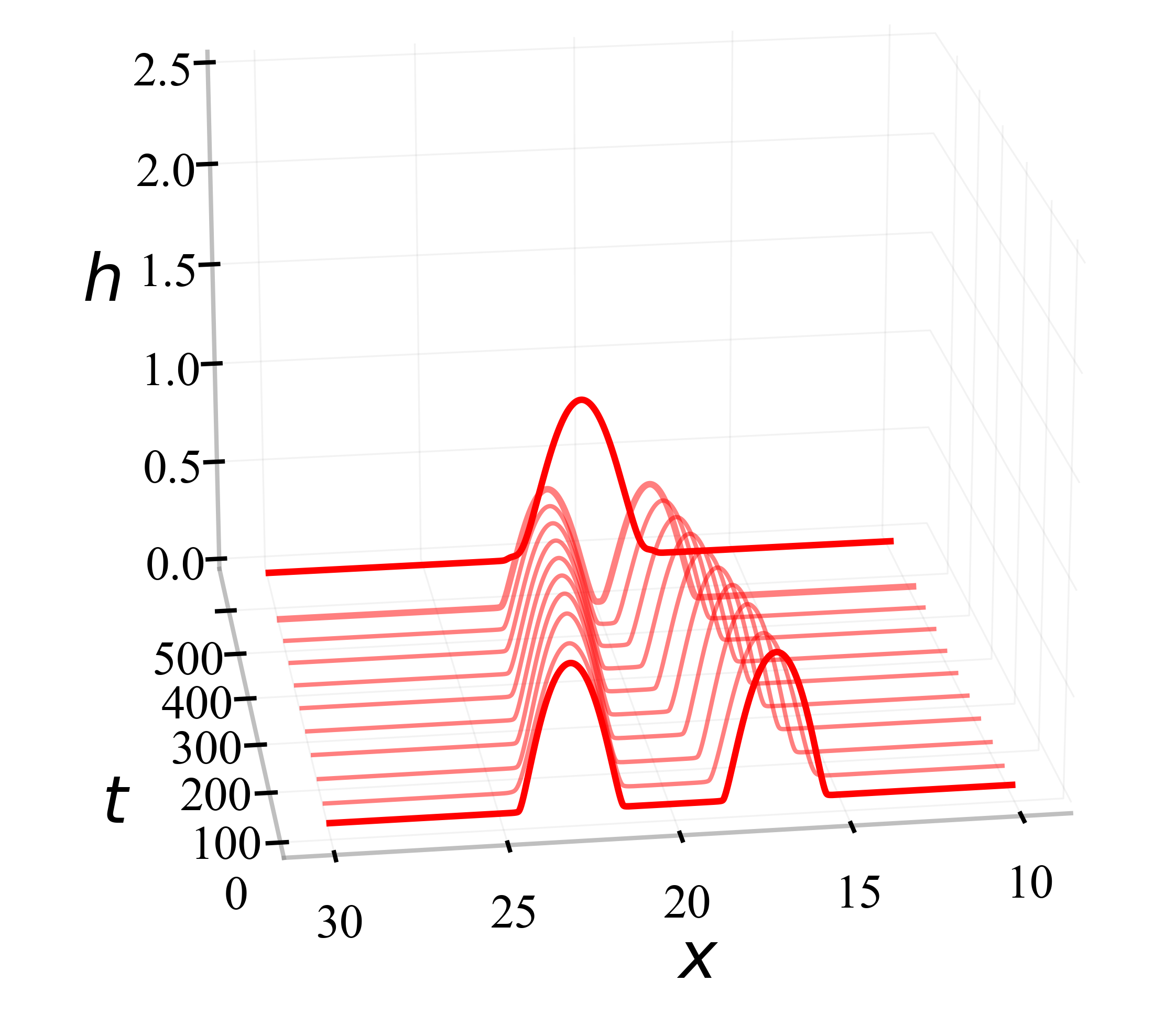}
        \subcaption{}
        \label{fig7a}
    \end{subfigure}
    \begin{subfigure}[t]{0.5\textwidth}
    \centering
        \includegraphics[width=6cm]{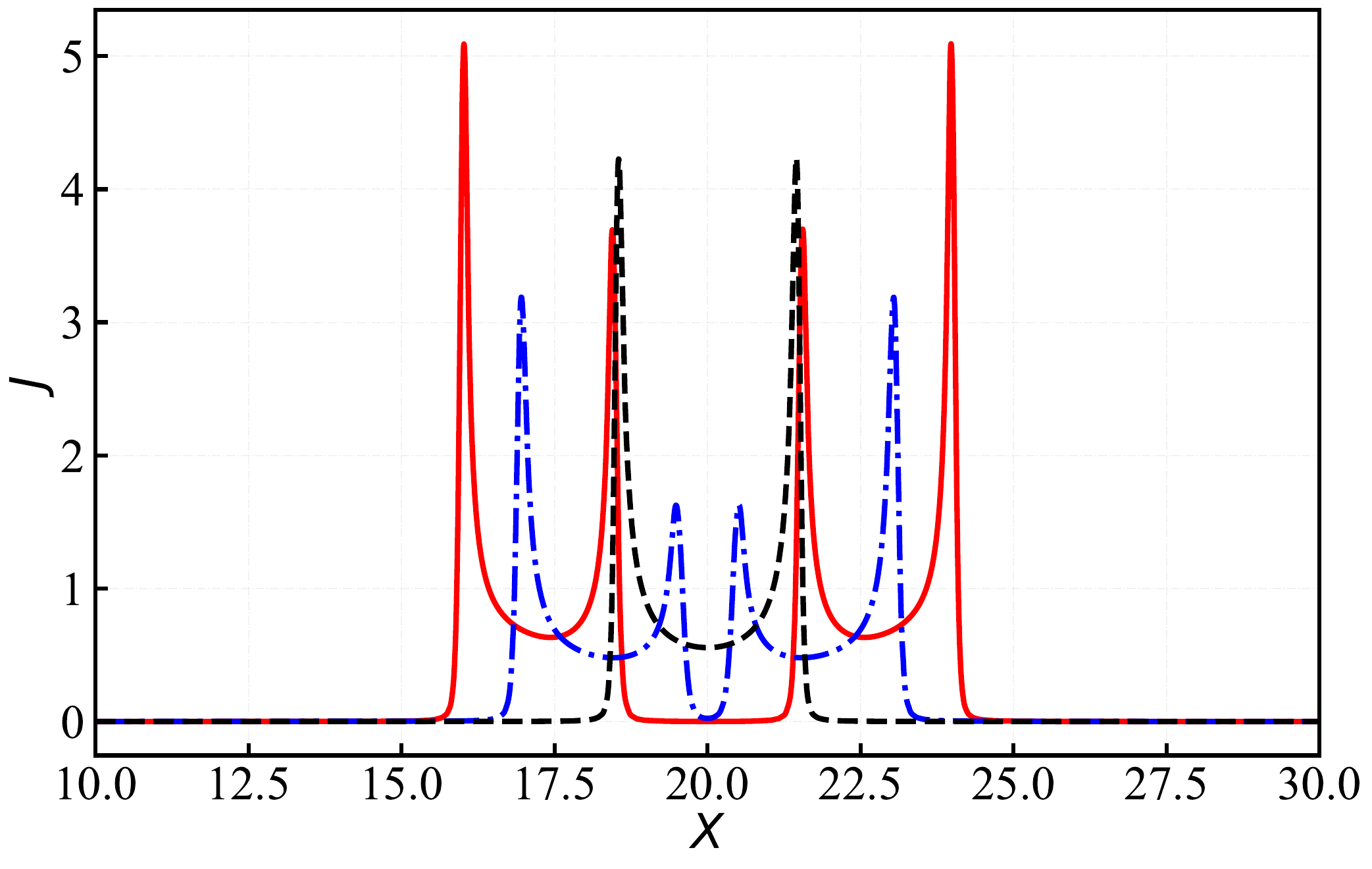}
        \subcaption{}
        \label{fig7b}
    \end{subfigure}
     \begin{subfigure}[t]{0.5\textwidth}
        \centering
        \includegraphics[width=6cm]{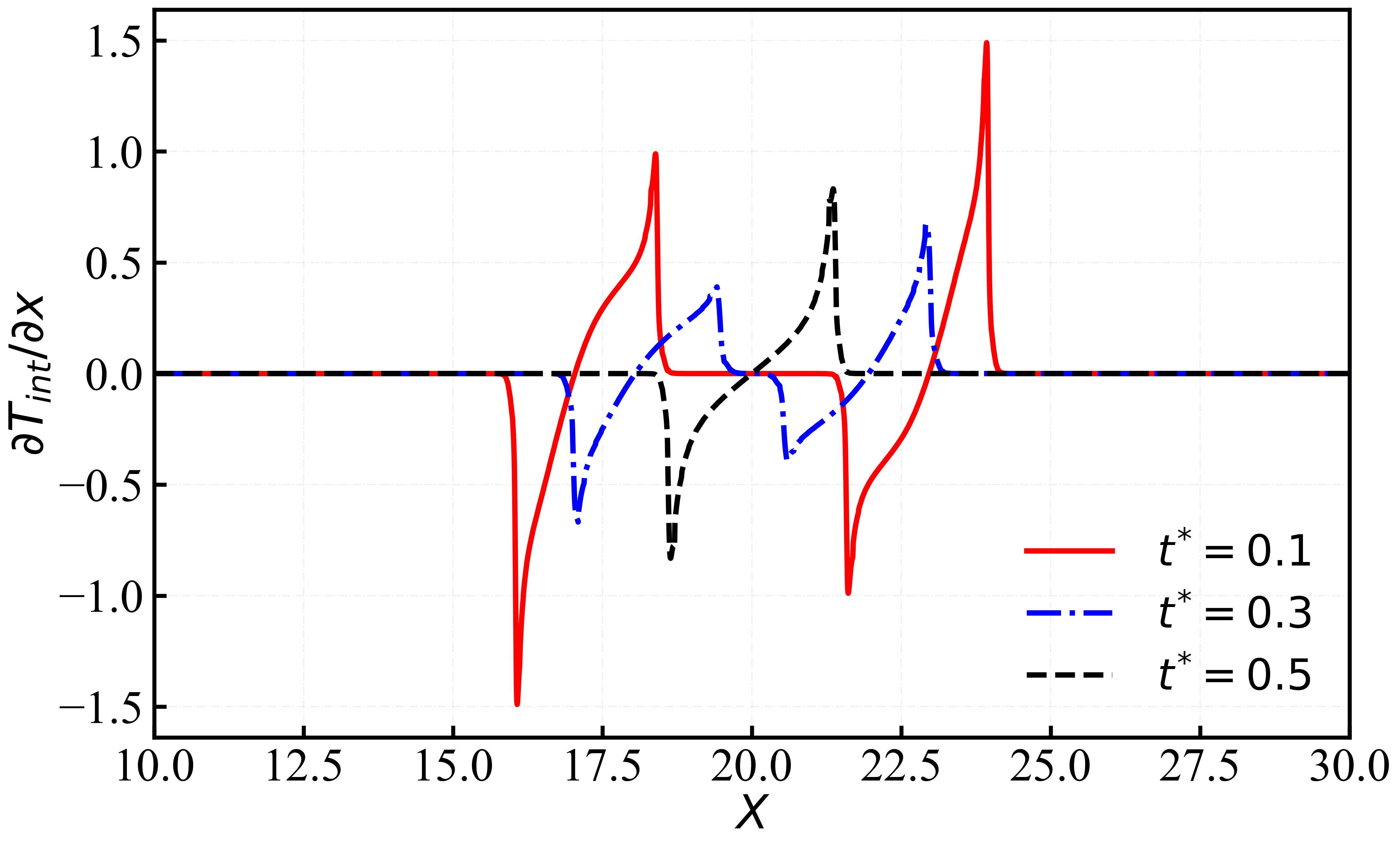}
        \subcaption{}
        \label{fig7c}
    \end{subfigure}
    \begin{subfigure}[t]{0.5\textwidth}
        \centering
        \includegraphics[width=6cm]{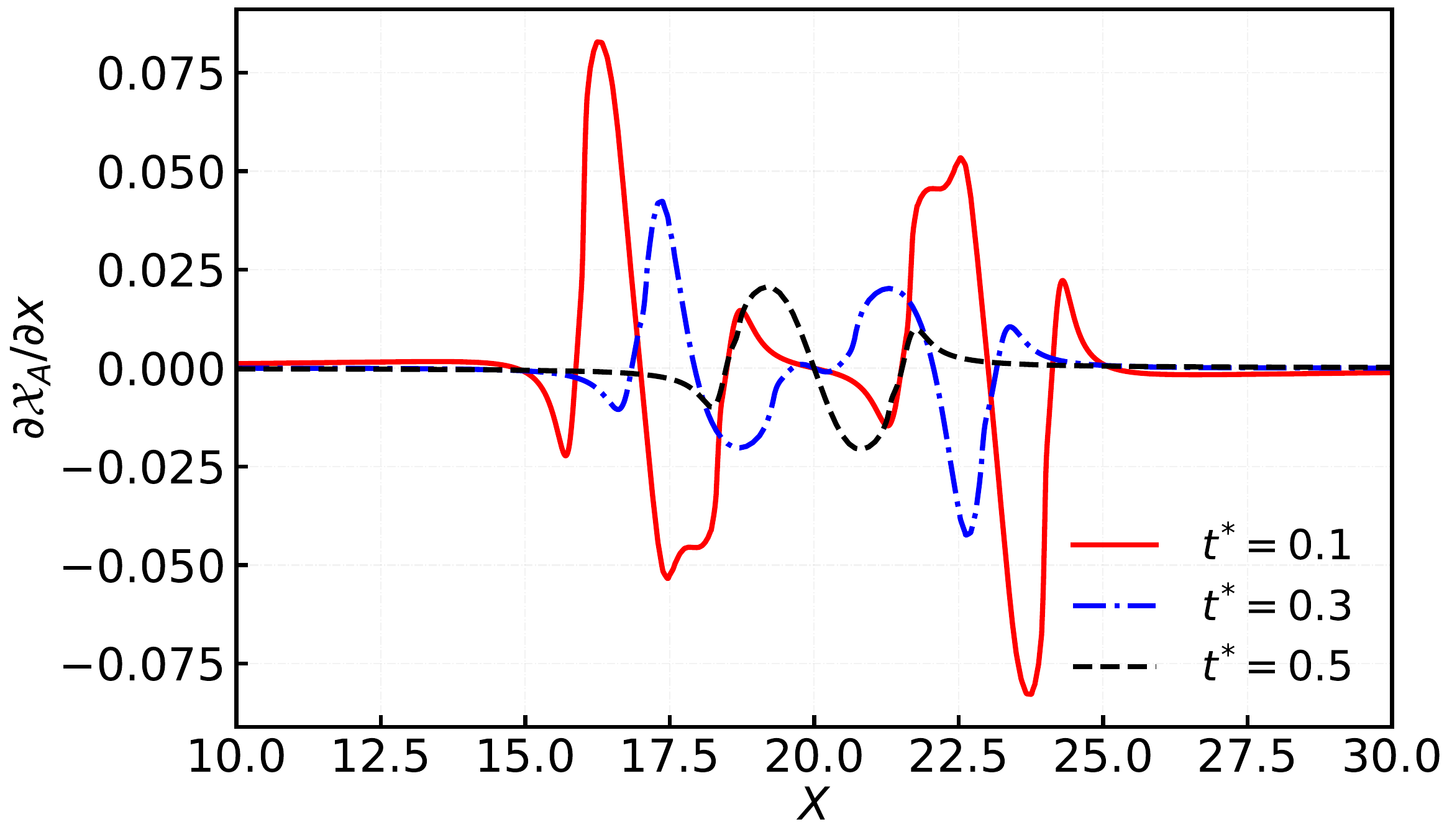}
        \subcaption{}
        \label{fig7d}
    \end{subfigure}
    \caption{(a) Space-time plot of two $50\%$ water-morpholine drops translating and coalescing with each other; (b) Distribution of total evaporation flux (J) across the droplet interface; (c) Gradient of interface temperature ($T_{int}$) along the droplet interface; (d) The gradient of concentration of more volatile liquid A (water) across the droplet interface; The non-dimensional numbers for the case is taken as $Kn = 0.001, E= 1 \times 10^{-4}, \delta = 10^{-5}, \psi =0.1, \delta x = 6, Pe_{v,A} =0.03, Pe_{v,B} =0.1, Pe =5 , Ma = 0.01, \mathcal{H} = 0.3$. }
    \label{fig7}
\end{figure}

\begin{figure}
    \begin{minipage}{\linewidth}
    \centering
    \includegraphics[width=\textwidth]{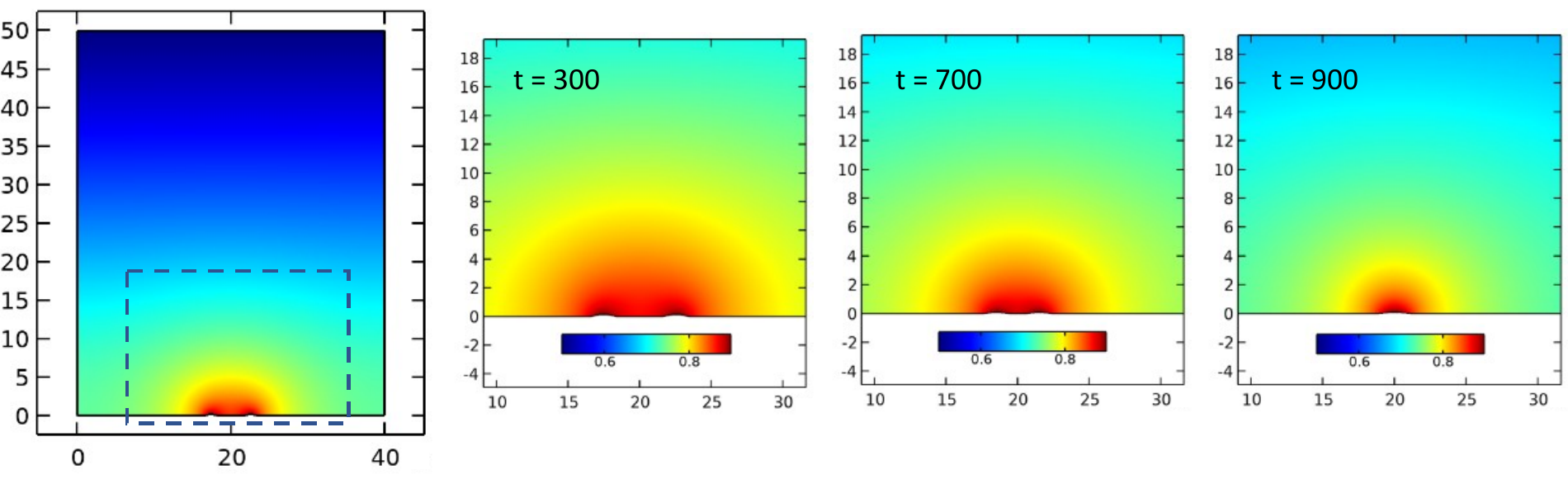}
    \end{minipage}
    \caption{Variation concentration of water vapour during the attraction of the droplets. }
    \label{fig:attarction_contour}
\end{figure}

Having investigated the evaporation dynamics of multiple pure droplets and isolated binary droplets, we now investigate the dynamics of a pair of binary mixture droplets (water-morpholine). It should be highlighted that while the attraction and repulsion behaviors in binary mixture droplets are akin to those seen in pure droplets, the underlying physical process is distinct because of the solutal Marangoni effect present in binary droplets. Here, we consider a case where each droplet initially consists of a $50\%$ mixture of water and morpholine. The property ratios of water (A) and morpholine (B) are shown in Table \ref{tab:4}. 
Figure \ref{fig7}a shows the space-time plot for two droplets composed of water-morpholine binary mixtures placed on a substrate at $25^{\circ}C$ ($Ma =0.01$). 
We consider a case with a high concentration of water ($50\%$) and observe that as the droplets evaporate, they approach each other and coalesce (at $t = 722$). The total evaporation flux ($J=J_A+J_B$) distribution is illustrated in figure \ref{fig7}b. Here, the evaporation flux is suppressed near the contact line of the droplets facing each other (proximal side). The evaporation flux profile shows very similar behaviour to that of the pure droplet already discussed in figure~\ref{F2}d, due to the accumulation of water vapour in the area between the drops; figure \ref{fig:attarction_contour} depicts this water vapour accumulation over the droplets and the area between the droplets. 
Due to suppressed evaporation, the concentration of the water (the more volatile component) in the liquid phase remains high at the proximal sides from the droplet apex. The opposite is true on the distal side of each droplet, where, due to enhanced evaporation of water, the water concentration inside the droplet decreases considerably, 
resulting in an asymmetric gradient of water concentration across the droplet, as depicted in figure~\ref{fig7d}. As shown there, the gradient of water concentration ($\partial \mathcal{X}_A/\partial x$) acquires a higher value on the far side of the droplets compared to the proximal side of the droplet. This asymmetry in concentration generates an asymmetric surface tension gradient and thereby solutal Marangoni stresses, which not only try to contract the droplet but also push the droplets to move toward each other. The effect of thermal Marangoni stresses is shown in figure~\ref{fig7c} where we depict the variation of the interfacial temperature gradient $\left(\frac{\partial T_{int}} {\partial x}\right)$. Because of the lower evaporation rate, the proximal side has a higher temperature than the distal contact points of the droplets. Therefore, it results in a higher magnitude of the temperature gradient from the apex to the distal side of the droplets. This asymmetry in the temperature gradient results in a variation of the surface tension gradient, prompting a repulsive force between the drops. 
To gain a deeper understanding of the mechanisms influencing the interaction of binary mixture droplets, we derive the average velocity in the z-direction $ \left( U = \frac{1}{h} \int_0^h u dz \right)$ and express it as the sum of its primary constituents, $U=U_c+U_t+U_\mathcal{X}$, which represent the flow induced by capillary pressure ($U_c$), solutal Marangoni stresses ($U_\mathcal{X}$), and thermo-capillarity ($U_t$), as demonstrated below. 
\begin{align}
    U_c = -\left( \frac{h^2}{2 \mu}\right) \frac{\partial p}{\partial x}, \\
    U_t = \frac{h}{\mu} \frac{\partial T_{int}}{\partial x} (\mathcal{X}_A + \gamma_R - \mathcal{X}_A \gamma_R),
\end{align}
\begin{equation}
    U_x = \left(\frac{h}{\mu Ma}\right) \frac{\partial \mathcal{X}_A}{\partial x} (1 - \sigma_R +Ma \gamma_R T_{int} -Ma T_{int}).
\end{equation}

\begin{figure}
    \centering 
    \begin{minipage}{0.7\textwidth}
    \centering
        \includegraphics[width=\linewidth]{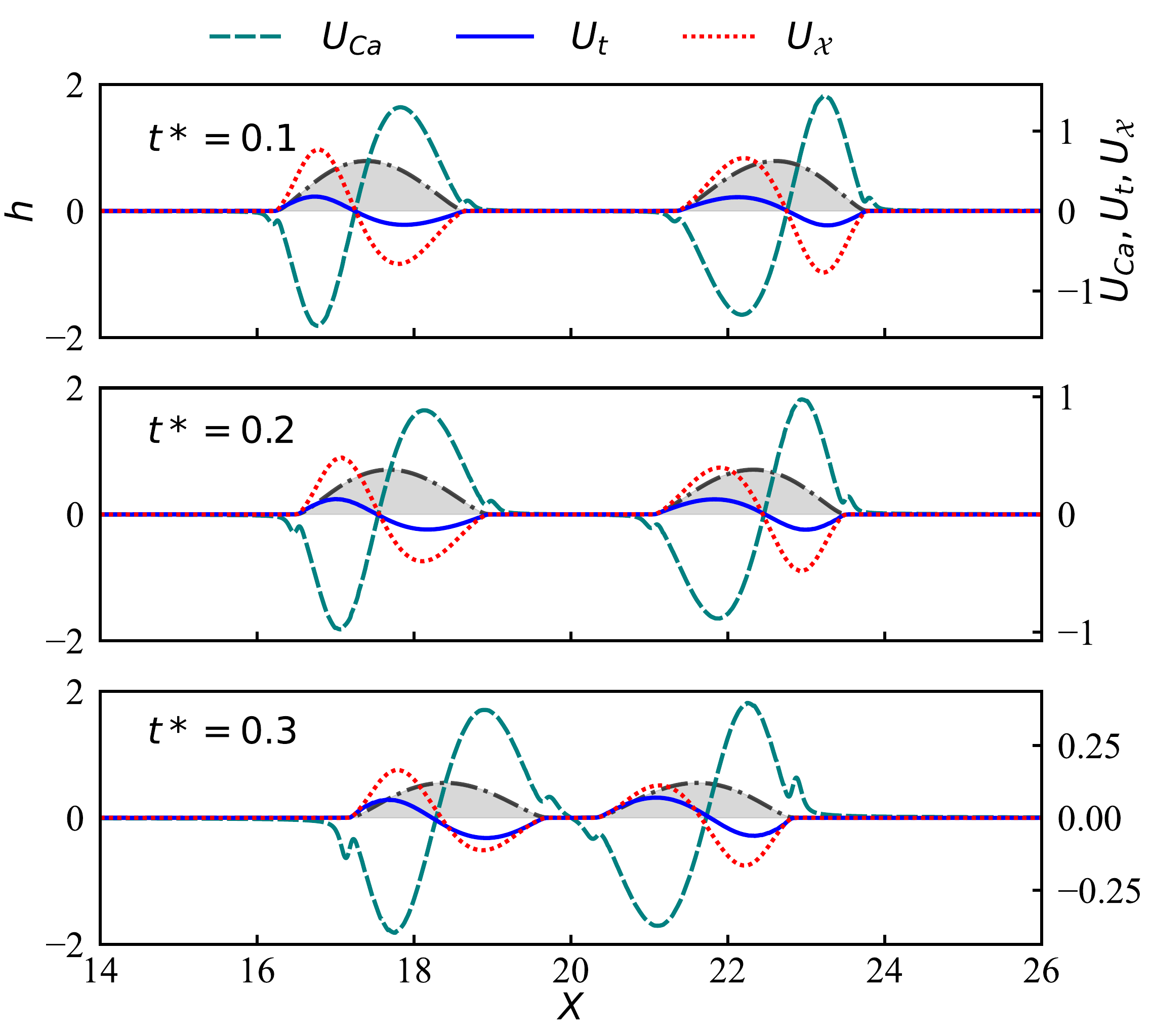}\\
        \footnotesize(a)
    \end{minipage}\hfill
    \begin{minipage}{0.5\textwidth}
    \centering
        \includegraphics[width=\linewidth]{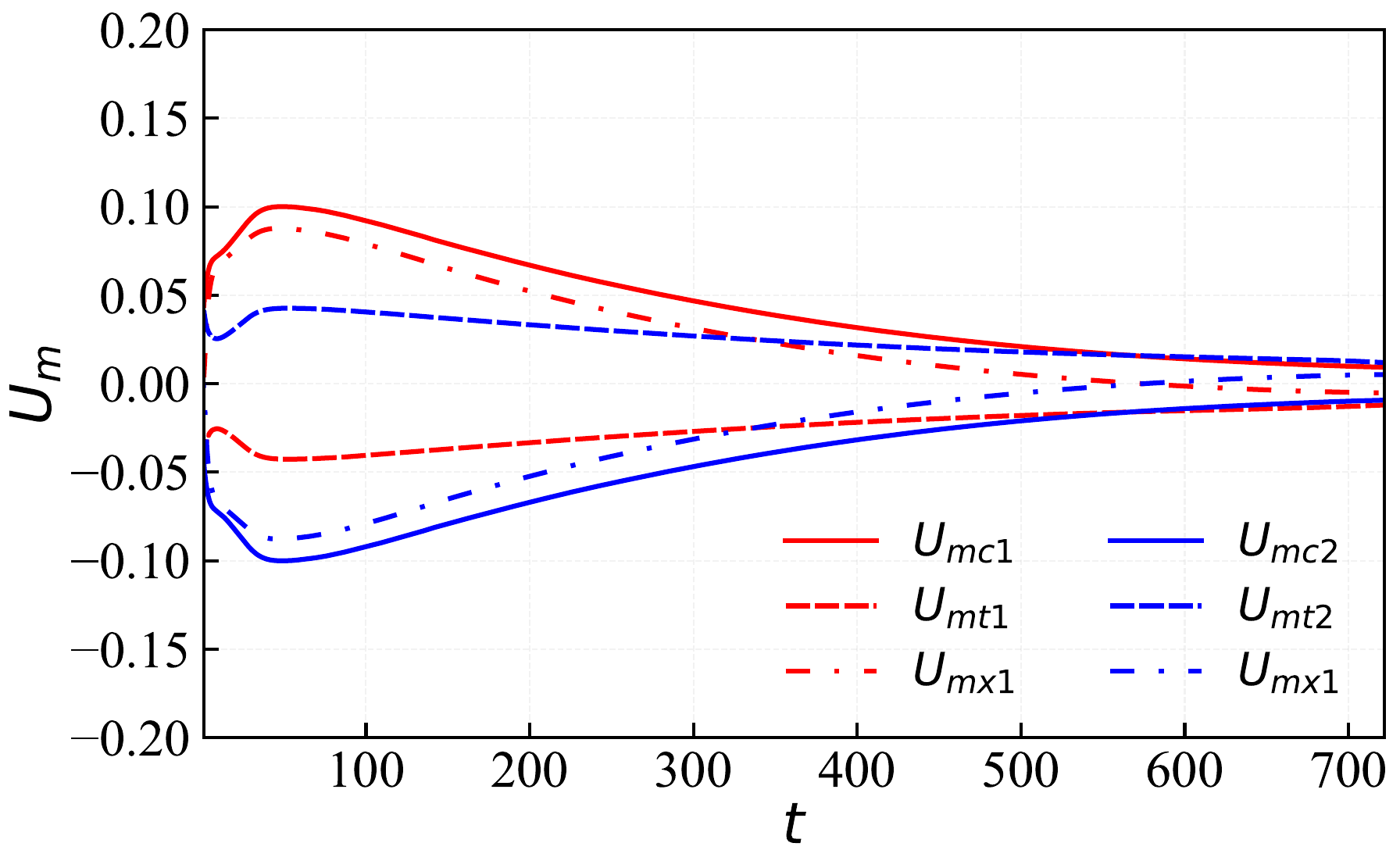}\\ \footnotesize(b)
    \end{minipage}
    \caption{(a) Axial distribution of $U_c$, $U_t$ and $U_\mathcal{X}$ at the interface of the droplets at different times, (b) time evolution of mean velocity components for both the droplets $(U_{mc},  U_{mt}, \rm{and} U_{mx})$.}
    \label{9d}
    \end{figure}

Figure \ref{9d}a represents the distribution of the velocity components ($U_c, U_t$ and $U_{\mathcal{X}}$) of the droplets
at different times as they approach one another. We observe that the capillary velocity is large and induces spreading for both droplets (for the droplet on the left, $U_c$ is positive to the right of the apex and negative to the left of the apex, and vice versa for the droplet on the right). The solutal Marangoni velocity ($U_\mathcal{X}$) acts opposite to the capillary velocity, inducing a contraction of the droplet footprints. The thermal Marangoni velocity ($U_t$) also tends to contract the droplets' footprints. In the present case, $U_t$ has a lower significance due to lower $Ma$. 
\medskip
In figure~\ref{9d}, as the droplets evaporate and come towards each other, we observe that the velocity profiles for all three components become asymmetric. This implies that each component has different values on both sides of the droplet's apex, generating a resultant velocity that moves the droplets. To analyse the resultant effect of the velocities, we calculate the mean ($U_{mc}, U_{mx}, \rm{and}$ $U_{mt}$) of each velocity component in the x-direction for both droplets:
\begin{eqnarray}
    U_{mc1} = \frac{\int_{x_1}^{x_2} h U_c} {\int_{x_1}^{x_2} h dx}, 
    U_{mt1} = \frac{\int_{x_1}^{x_2} h U_t} {\int_{x_1}^{x_2} h dx},
    U_{mx1} = \frac{\int_{x_1}^{x_2} h U_x} {\int_{x_1}^{x_2} h dx},\\
    U_{mc2} = \frac{\int_{x_3}^{x_4} h U_c} {\int_{x_3}^{x_4} h dx}, 
    U_{mt2} = \frac{\int_{x_3}^{x_4} h U_t} {\int_{x_3}^{x_4} h dx},
    U_{mx2} = \frac{\int_{x_3}^{x_4} h U_x} {\int_{x_3}^{x_4} h dx}, \\
    U_m = U_{mc} + U_{mt} +U_{mx},
\end{eqnarray} 
here, subscript $1$ indicates velocity components for the left droplet ($x_1$ and $x_2$ being their contact points) and subscript $2$ is for the right droplet ($x_3$ and $x_4$ are its contact points). 
Figure \ref{9d}b shows the time evolution of the mean velocity components, where $U_{mc1}, U_{mt1}, U_{mx1}, U_{mc2}, U_{mt2}, U_{mx2}$ are the mean capillary, thermal Marangoni, and solutal Marangoni velocities for the droplets. It is evident that for the left droplet, the values of $U_{mc1}$ and $U_{mx1}$ remain positive, whereas the values of $U_{mt1}$ are negative. In contrast, for the right droplet (droplet 2), this pattern is inverted. This suggests that capillary and solutal Marangoni velocities drive the droplets toward each other, whereas the thermal Marangoni velocity causes them to repel. Given the current scenario where thermal Marangoni effects are minimal ($Ma =0.01$), the thermal Marangoni velocity is significantly lower than the other two velocities, facilitating the attraction between the droplets.
\begin{figure}
    \begin{subfigure}[t]{0.5\textwidth}
        \centering
        \includegraphics[width=6cm]{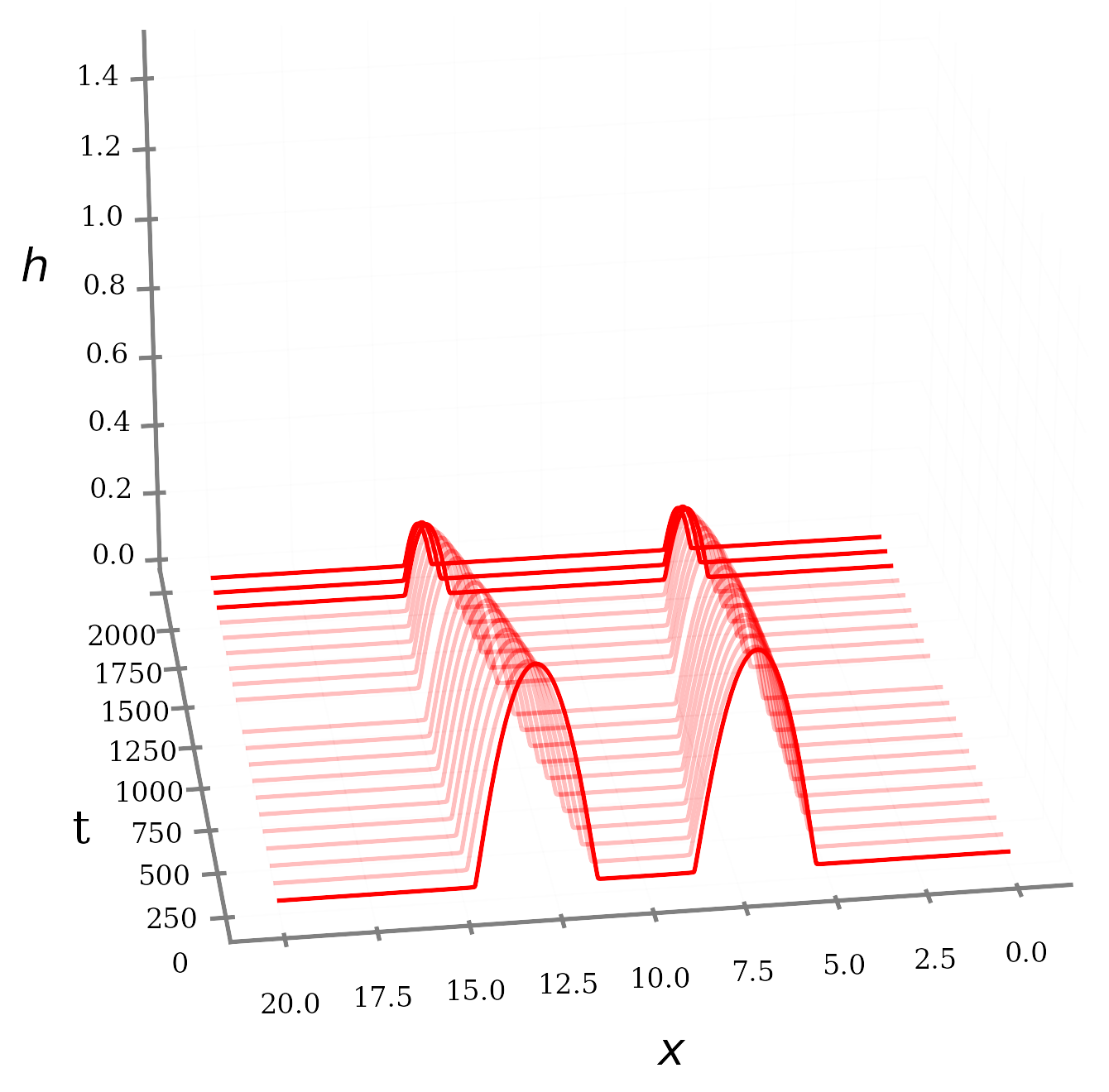}
        \subcaption{}
    \end{subfigure}
    \begin{subfigure}[t]{0.5\textwidth}
        \centering
        \includegraphics[width=6cm]{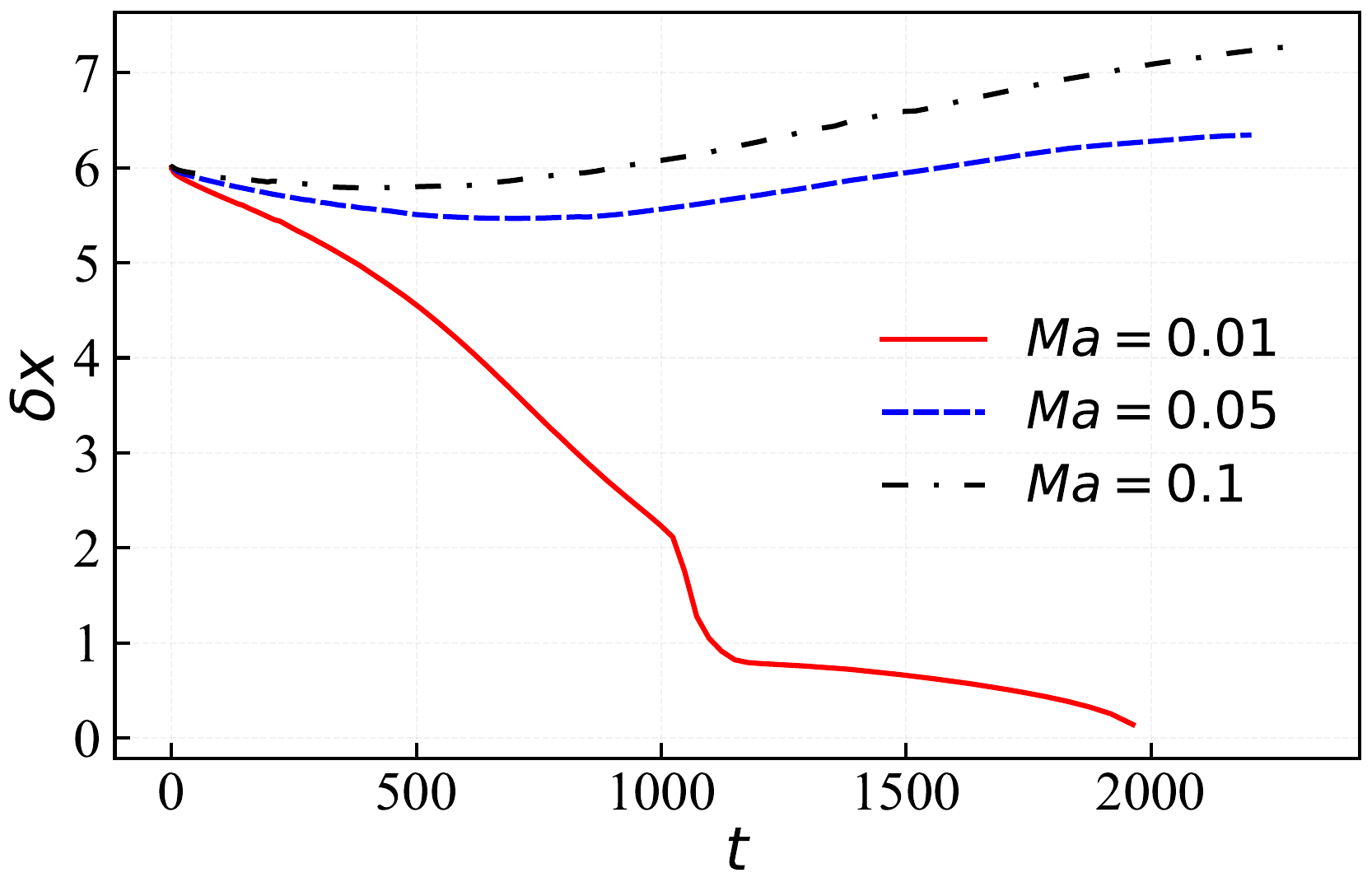}  
        \subcaption{}
    \end{subfigure}
    \begin{subfigure}[t]{0.5\textwidth}
        \centering
        \includegraphics[width=6cm]{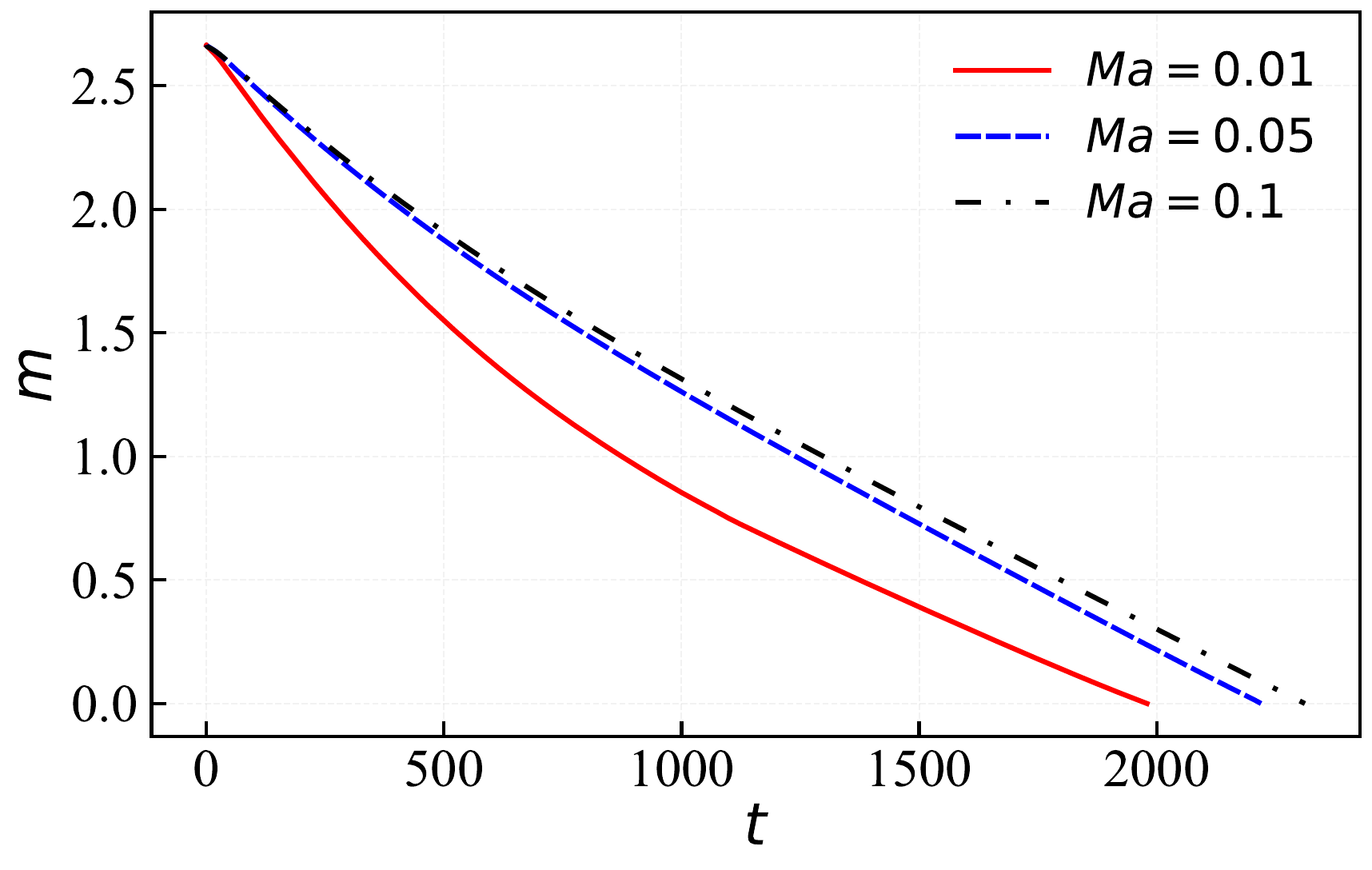}  
        \subcaption{}
    \end{subfigure}
    \begin{subfigure}[t]{0.5\textwidth}
        \centering
        \includegraphics[width=6cm]{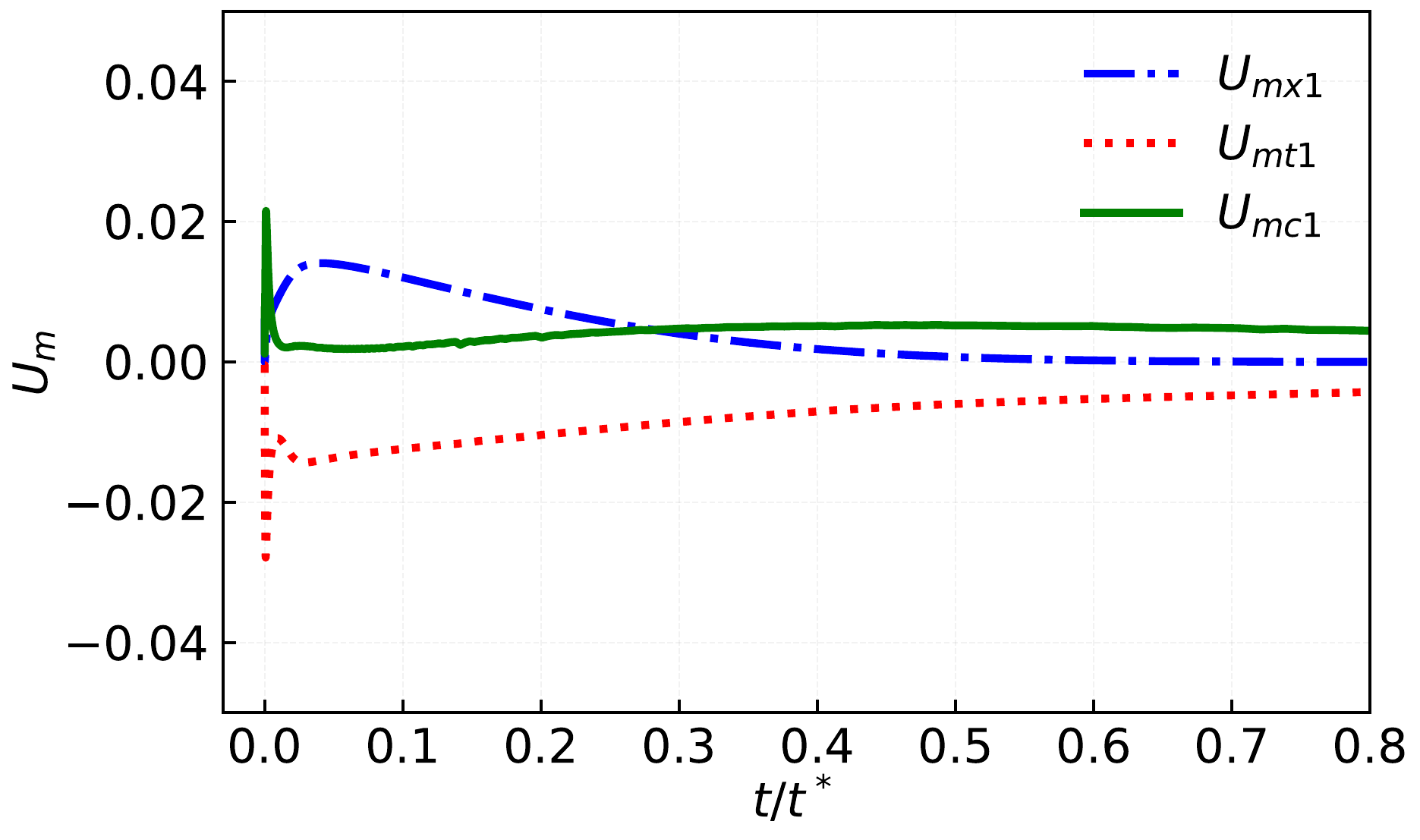} 
        \subcaption{}
    \end{subfigure}
    \caption{ (a) Space-time plot for the droplets for an enhanced thermal Marangoni effect ($ Ma= 0.1$), (b) Temporal variation of the distance between the droplet ($\delta x$) for varying Marangoni numbers, (c) Total mass ($m$) evolution of the droplets for varying Marangoni numbers, (d) Mean velocity components of $25\%$ water morpholine drops for $Ma =0.05$. Other properties are mentioned in Table \ref{tab:4}. 
    }
    \label{f10}
\end{figure}

\subsection{Effect of thermal and solutal Marangoni stresses}
 Having discussed the attraction of two binary droplets, we now check how the thermal Marangoni stresses affect their dynamics. To do this, we vary the Marangoni number ($Ma$) while keeping all other properties the same. Figure~\ref{f10}a depicts the space-time diagram of the droplets for two $25\%$ water morpholine droplets at $Ma =0.1$ (substrate temperature $ T_s = 65^{\circ} C$) (other parameter values are mentioned in Table \ref{tab:4}). 
 In this scenario, droplets drift away from one another. The temporal evolution of the distance between the centres of the droplets ($\delta x$) is presented in figure \ref{f10}b with varying thermal Marangoni numbers ($Ma$). It is noted that as $Ma$ increases from $0.01$ to $0.1$, 
 the droplets cease moving closer; instead, they exhibit repulsive behaviour. This repulsive behaviour can be characterised by mean velocity components for the left droplet (Droplet 1) as illustrated in figure \ref{f10}d for $Ma =0.05$. In this scenario, it is observed that at first, $U_{mx}$ and $U_{c}$ promote the droplets to approach each other. However, as the MVC (water) evaporates, $U_{mx}$ diminishes to zero, and $U_{mt}$ along with $U_{mc}$ become more influential. During this period, $U_{mc}$ and $U_{mt}$ neutralize one another, causing the droplets to cease moving toward each other. If the thermal Marangoni number ($Ma > 0.05$) increases further, the greater dominance of $U_{mt}$ over $U_{mc}$ after $U_{mx}$ becomes zero, causing the droplets to repel. 

As shown in figure~\ref{f10}c, the droplets with lower $Ma$ have smaller lifetimes. The compressive stress generated at $Ma=0.1$ is an order of magnitude higher than that at $ Ma=0.01$. 
As a result, droplets exhibit a flatter shape for $Ma = 0.01$ than for $Ma = 0.1$, resulting in an increased evaporation rate and a reduced droplet lifespan.
\begin{figure}
    \centering
    \begin{subfigure}[t]{0.48 \textwidth}
            \includegraphics[width=6cm]{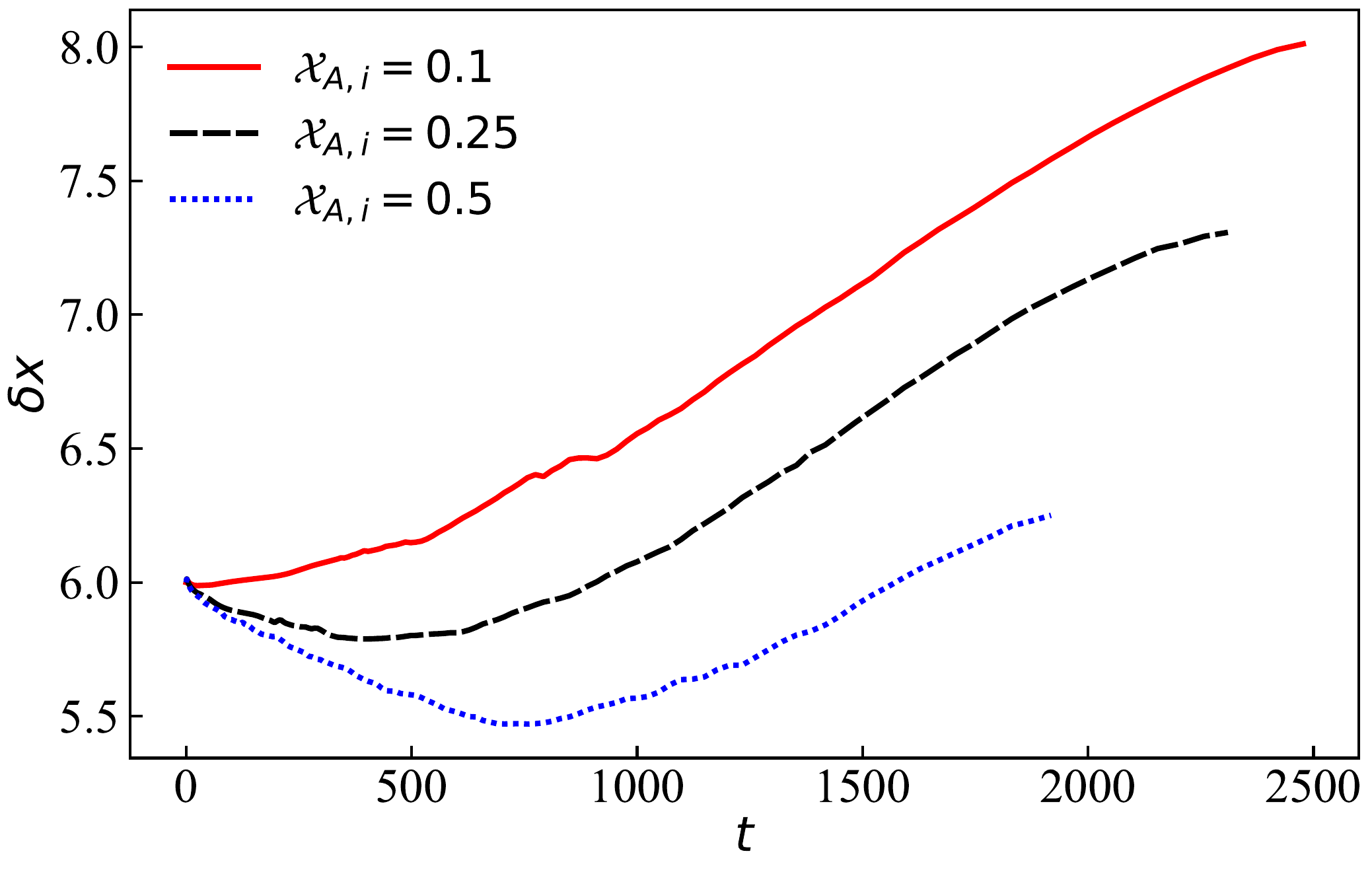}
            \subcaption{}
    \end{subfigure}
    \begin{subfigure}[t]{0.48 \textwidth}
            \includegraphics[width= 6cm]{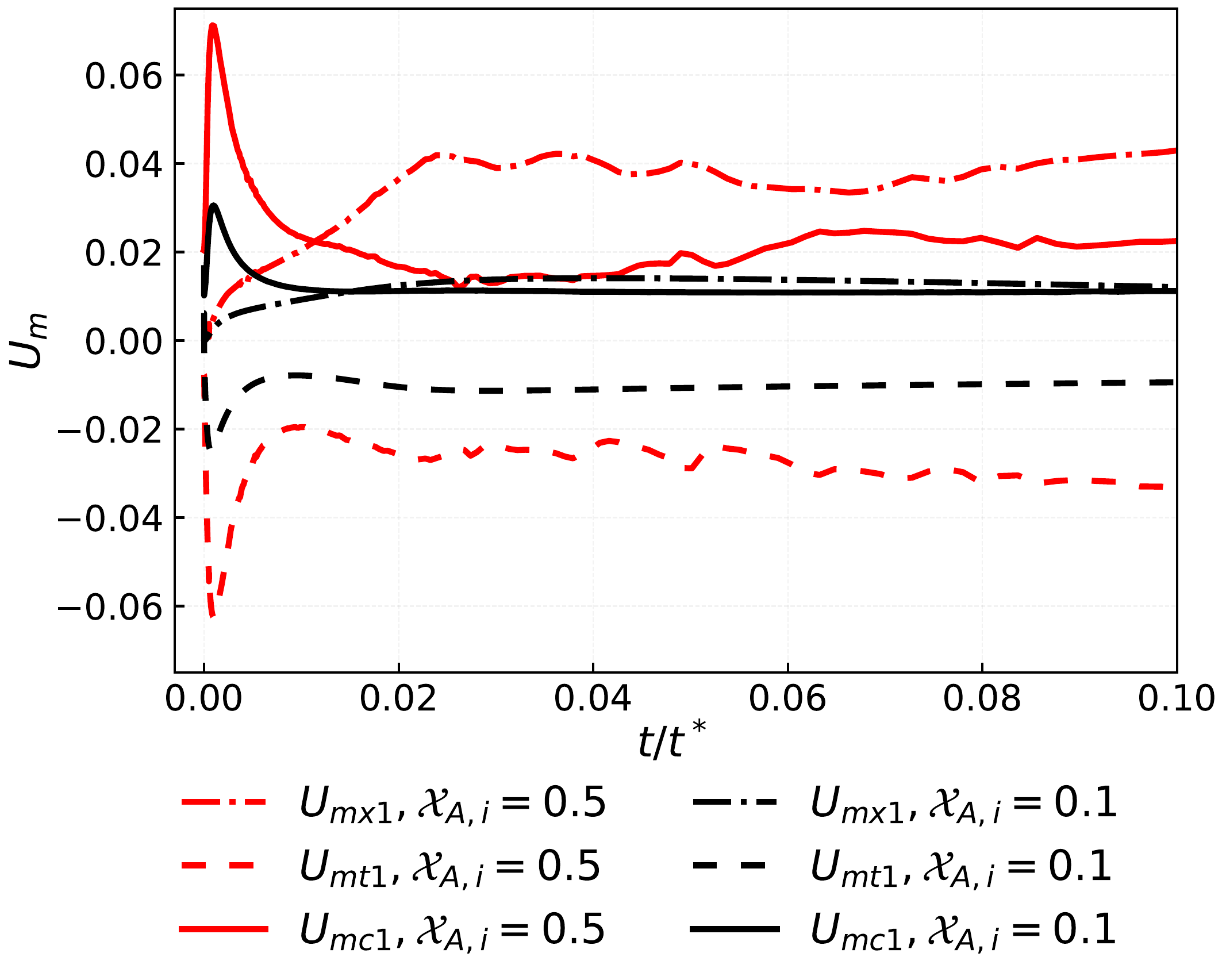}
            \subcaption{}
    \end{subfigure}
    \caption{Temporal evolution of the separation between droplet centers for two water–morpholine droplets with different initial compositions at $Ma = 0.06$, with all other parameters identical to those listed in Table \ref{tab:4}. (b) Corresponding variation of the mean velocity components for $10\%$ and $50\%$ water–morpholine droplets.}
    \label{fig:12}
\end{figure}
To investigate the impact of solutal Marangoni stresses, we analyze scenarios with varying concentrations of the most volatile component ($\mathcal{X}_{A,i}$), spanning from 0.1 to 0.5. In figure~\ref{fig:12}a, we depict the evolution of the separation between the centroids of the droplets ($\delta x$) throughout the evaporation process. As concentration increases, droplets have a greater tendency to move toward each other due to the strong solutal Marangoni effect. As the most volatile element (water) evaporates from the system, the solutal Marangoni effect diminishes, allowing thermal Marangoni stresses to take over, ultimately causing the droplets to move away from each other.

\begin{figure}
    \centering
    \begin{subfigure}[t]{0.8\textwidth}
    \centering
    \includegraphics[width=8cm]{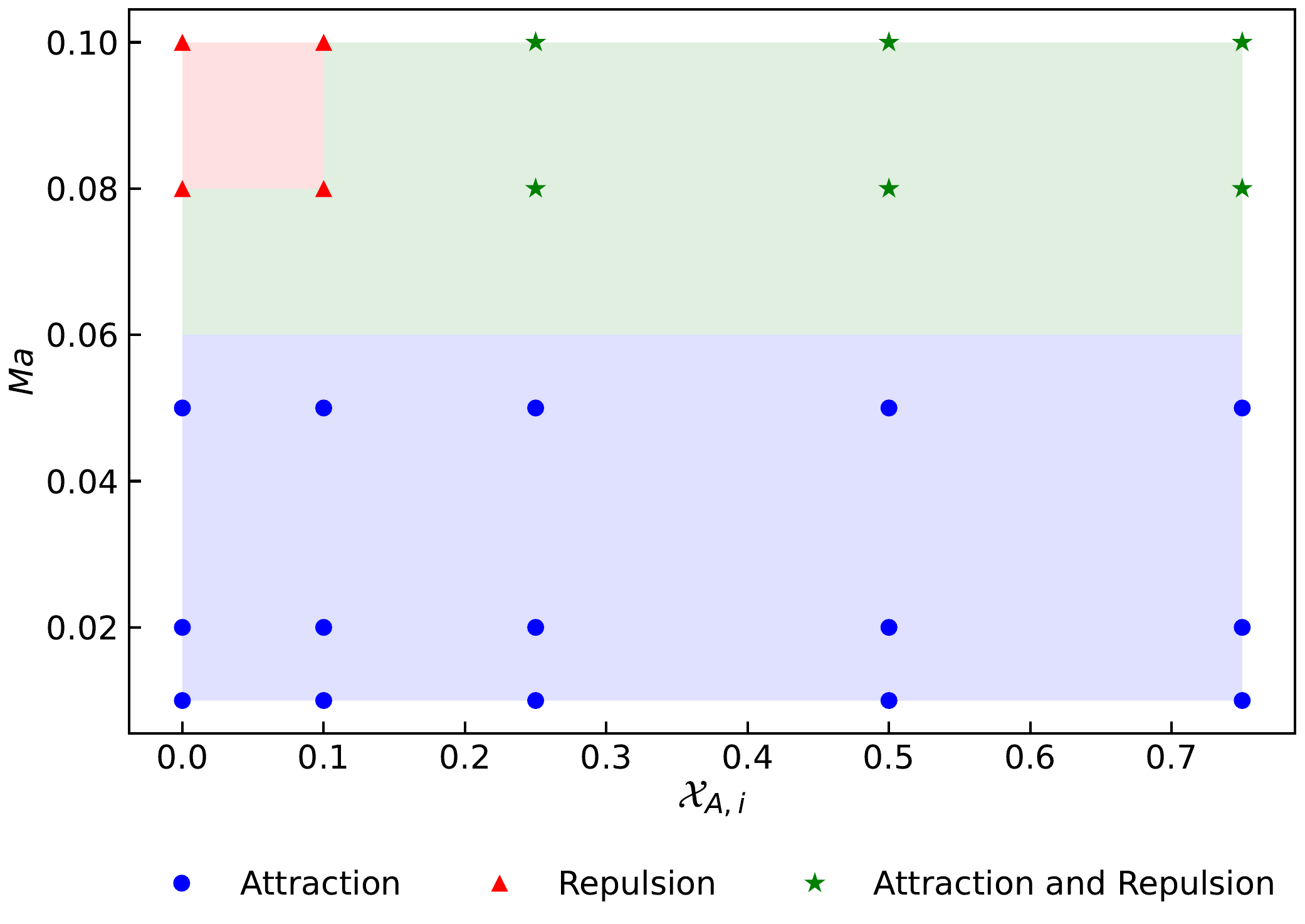}
    \subcaption{}
    \end{subfigure}
    \caption{Regime map of droplet attraction and repulsion based on $\mathcal{X}_{A,i}$ and $Ma$ for water morpholine droplets. }
    \label{fig12}
\end{figure}

Figure~\ref{fig:12}b illustrates the temporal evolution of the mean velocity components ($U_{mc}, U_{mx}$, and $U_{mt}$) for the left droplet in two distinct scenarios where $\mathcal{X}_{A} = 0.1$ and $\mathcal{X}_{A} = 0.5$ for $Ma =0.05$. For higher $\mathcal{X}_{A} = 0.5$, $U_{mx}$ has higher values compared $U_{mt}$. The positive values of $U_{mc}$, in conjunction with $U_{mx}$, help the droplets to attract one another. However, for a lower value of $\mathcal{X}_{A} = 0.1$, $U_{mx}$ is much less than that of $U_{mt}$, which means that weaker solutal Marangoni effects give way to stronger thermal Marangoni effects, prompting the droplets to repel.
This physical interplay between the solutal and thermal Marangoni forces and the capillary forces, resulting in different droplet translation movements, depends mainly on the initial composition ($\mathcal{X}_{A,i}$) of the droplets and the thermal Marangoni stresses ($Ma$). Figure~\ref{fig12} depicts the regime map of interactions between droplets, highlighting their attractive, mixed attraction-repulsion, and purely repulsive behaviors based on $\mathcal{X}_{A,i}$ and $Ma$. Below $Ma = 0.06$, droplets exhibit attraction irrespective of their initial composition. Conversely, for $Ma$ above $0.06$, droplets with $\mathcal{X}_{A,i} \ge 0.25$ initially move towards each other before repelling as the solutal Marangoni effect diminishes with the depletion of $\mathcal{X}_{A}$. In cases where $\mathcal{X}_{A,i} \leq 0.1$, droplets shows a repulsive behavior. 
\begin{figure}
    \centering
    \includegraphics[width=14cm]{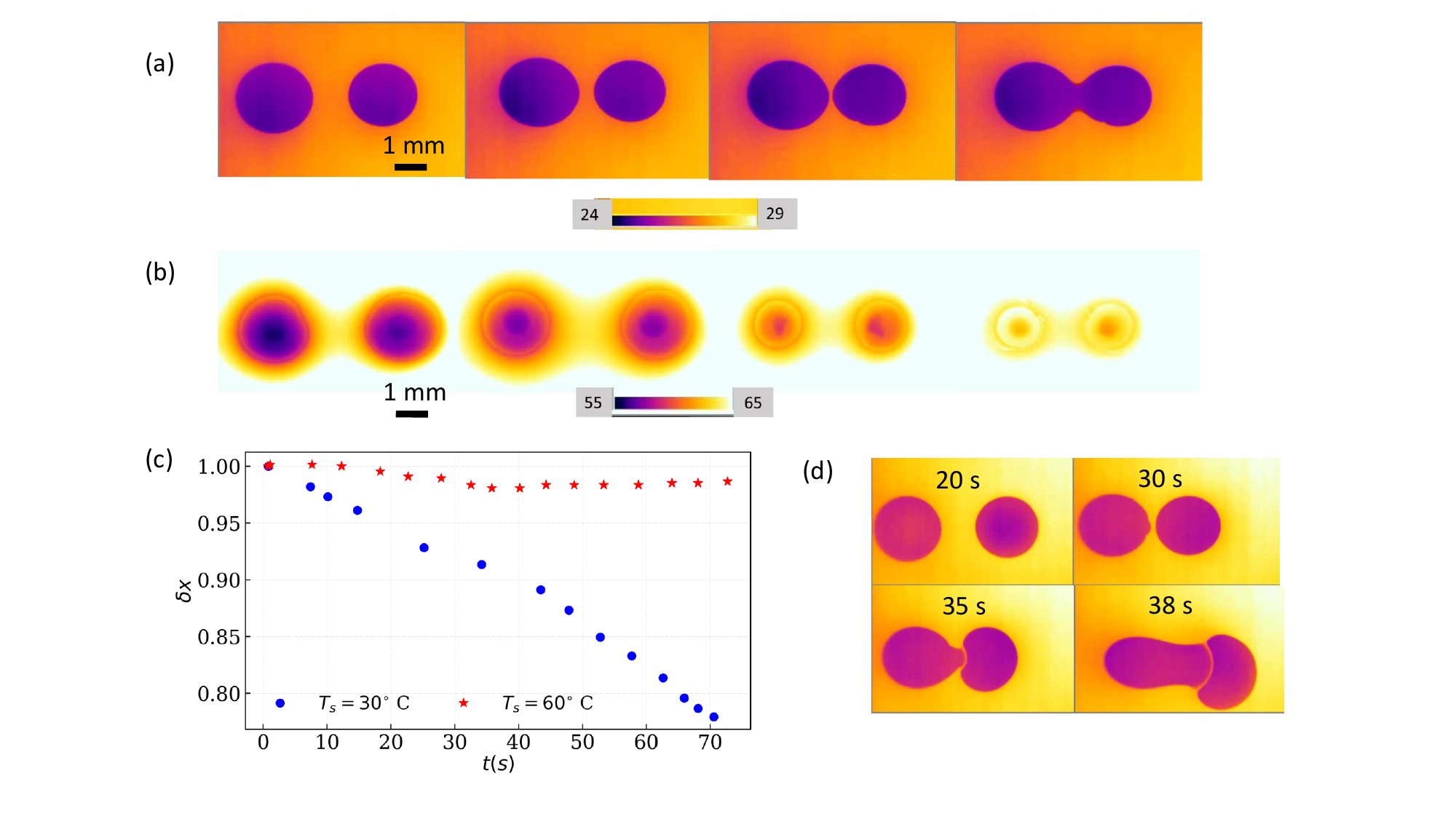}
    \caption{Experimental snapshots for two $25\%$ water morpholine droplets at (a) substrate temperature $30^\circ$C, (b) substrate temperature $60^{\circ}$C, (c) time evolution of the centroid distance ($\delta x$, which is non-dimensionalised with initial separating distance.) for the droplets, and (d) Time evolution images for two water-morpholine droplets having $10\%$ (left droplet) and $50\%$ (right droplet) water concentrations, respectively. }
    \label{fig:exp}
\end{figure}
\subsection{Experimental comparison}
We conducted experiments (see Appendix \ref{s5}) to evaluate the predictions generated by our model qualitatively. Figure~\ref{fig:exp}a depicts the time series infrared images of two water-morpholine droplets placed on a substrate heated to $30^{\circ}$C. It is clear that the droplets are converging and eventually coalescing. On the other hand, when the substrate temperature is increased to $60^{\circ}$ C, they almost remain in their original position, inhibiting a repulsive action due to a higher thermal Marangoni (figure~\ref{fig:exp}b). This is quantified in figure~\ref{fig:exp}c, where the evolution of the droplet centroid ($\delta x$) calculated using ImageJ is depicted. These two occurrences can be qualitatively compared to our model predictions for $Ma =0.016$ and $Ma =0.08$. The predictions also illustrate attraction and coalescence (figure~\ref{fig7}) and repulsion (figure~\ref{f10}) behaviours, respectively, in water-morpholine droplets. These observations can be perfectly placed in the regime map depicted in figure~\ref{fig12}.

Figure \ref{fig:exp}d presents infrared images of two water morpholine droplets at a substrate temperature of $30^{\circ}$C, with initial concentrations of $10\%$ water (left) and $50\%$ water (right). The droplet with lower water content approaches the other, but instead of merging, it chases it. This chasing behaviour of the droplets with different initial concentrations of water is illustrated in figure~\ref{f14}. We stress that the theoretical framework employed here is intrinsically two-dimensional. Specifically, it describes the evolution of liquid filaments that are uniform along the out-of-plane direction, rather than fully three-dimensional sessile droplets. In contrast, the experiments are genuinely three-dimensional. This distinction is crucial, particularly because vapour-mediated interactions are sensitive to the geometry and dimensionality of the diffusion field. Therefore, the comparison between simulations and experiments is meant to be qualitative rather than quantitative: the model serves to elucidate the fundamental interaction mechanisms and overall tendencies, but it is not intended to match the experimental dynamics in a strictly quantitative manner.
\begin{figure}
    \begin{subfigure}[t]{1\textwidth}
    \centering
    \includegraphics[width=9cm]{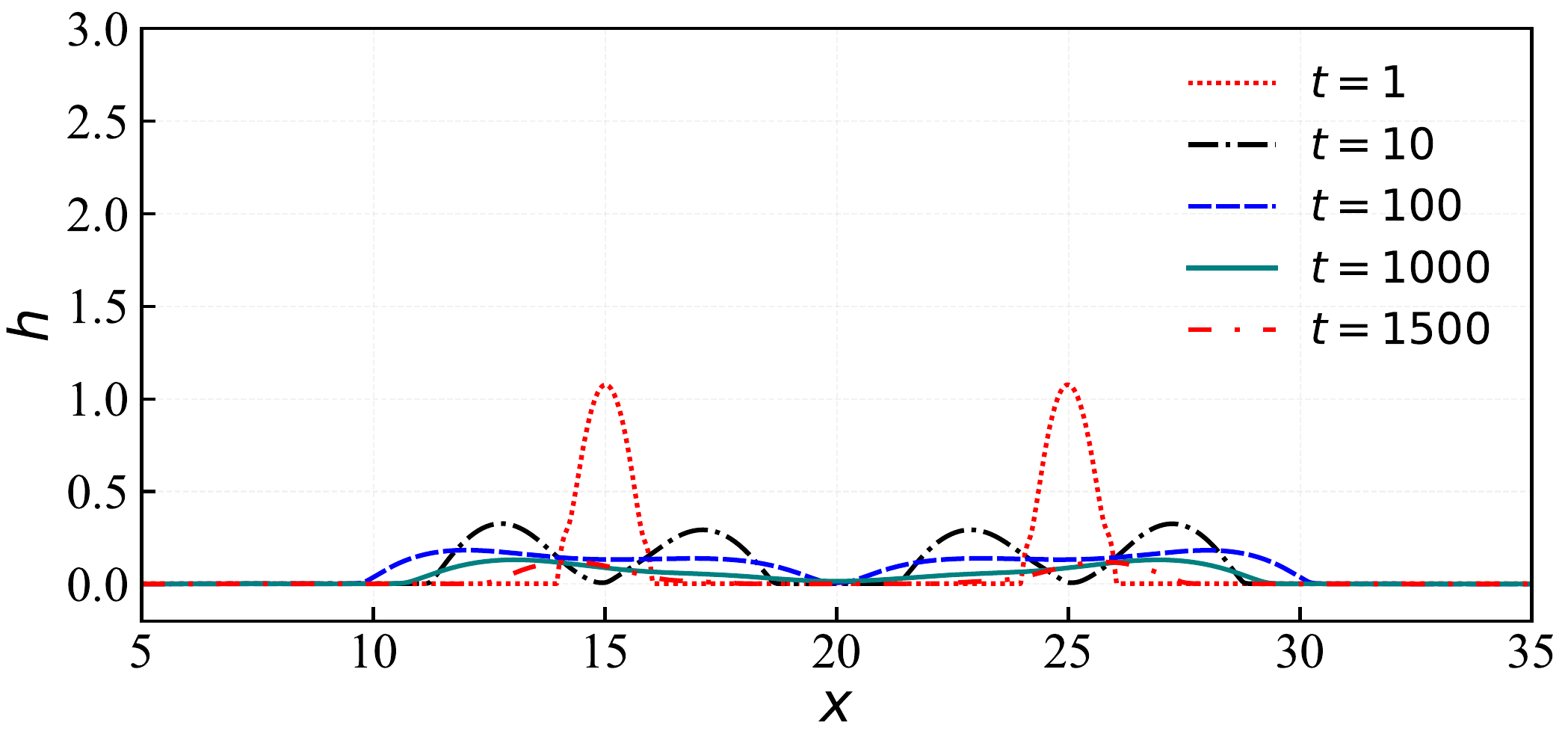}
    \subcaption{}
    \end{subfigure}
    \begin{subfigure}[t]{1\textwidth}
    \centering
    \includegraphics[width=9cm]{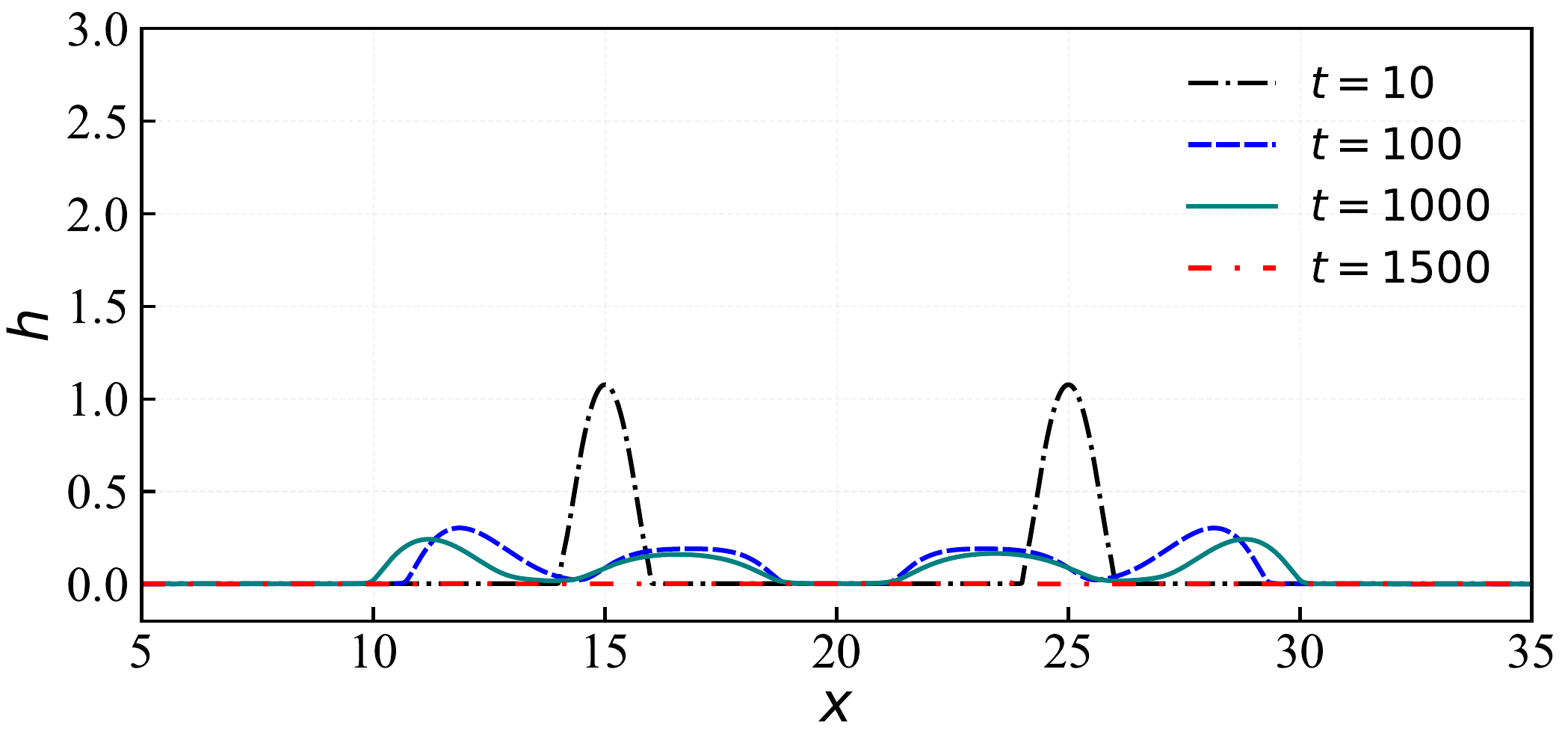}
    \subcaption{}
    \end{subfigure}
    \caption{Evolution of the drop shape for two $50\%$ ethanol-water droplets ($\mathcal{X}_{A,i} =50 \%$) for (a) $Ma = 0.01$ and (b) $Ma =0.08$.}
    \label{f12}
\end{figure}
\begin{figure}
   \begin{subfigure}[t]{0.5\textwidth}
    \centering
    \includegraphics[width=6cm]{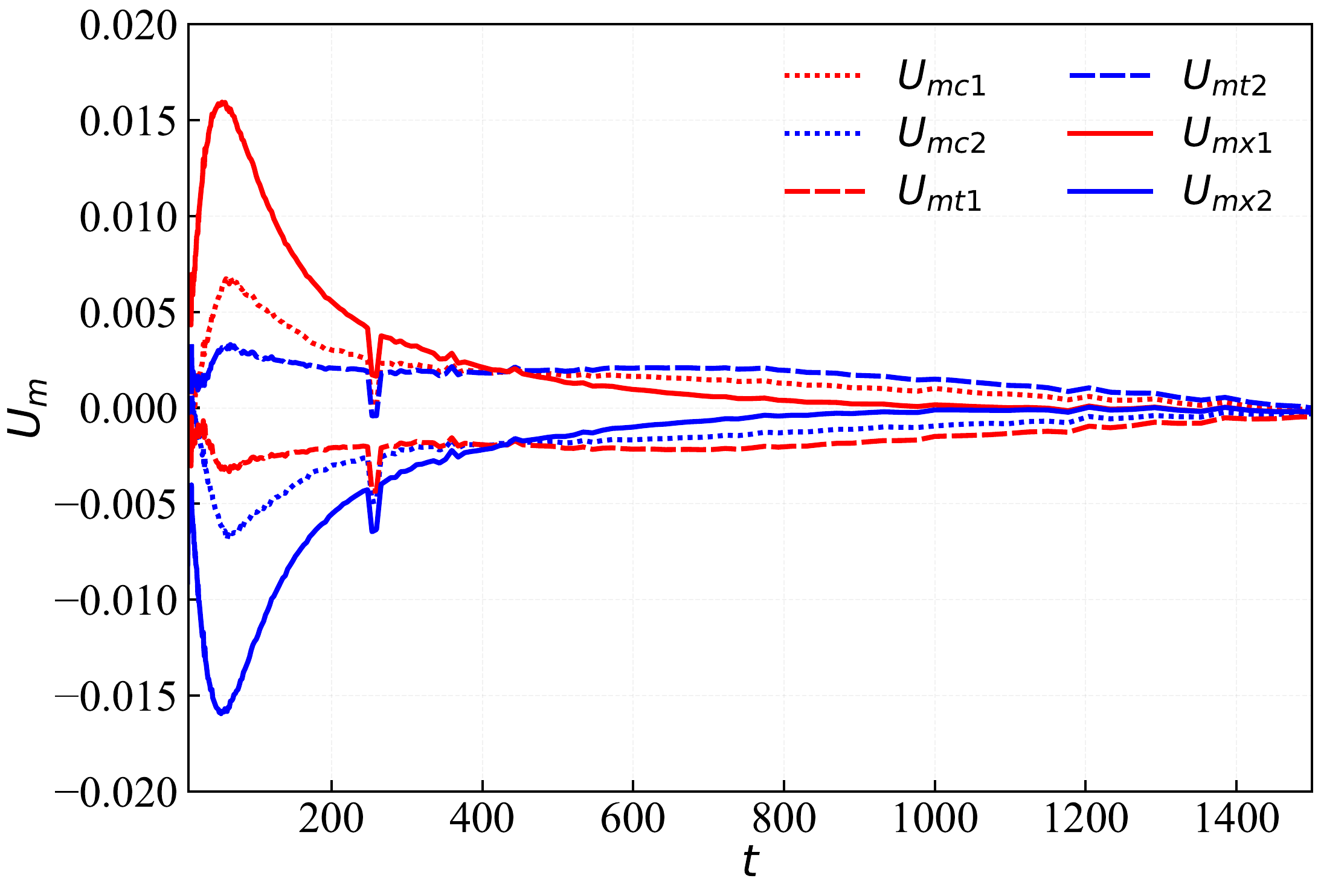}
    \subcaption{}
    \end{subfigure}
    \begin{subfigure}[t]{0.5\textwidth}
    \centering
    \includegraphics[width=7cm]{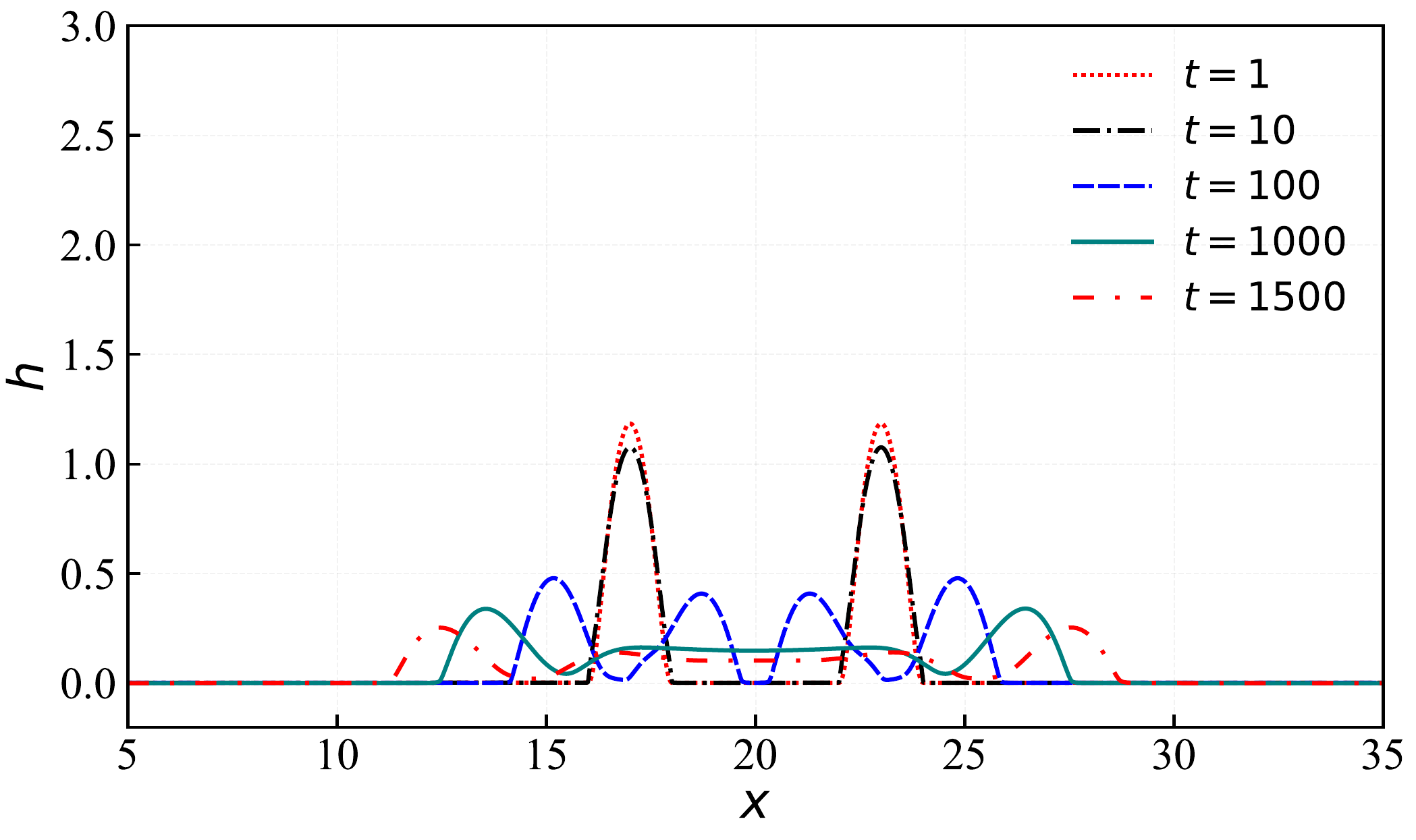}
    \subcaption{}
    \end{subfigure}
    \caption{Mean velocity components for two $50\%$ ethanol-water droplets at $Ma =0.08$, (b) Drop shape evolution of two $50\%$ ethanol-water droplets at $Ma =0.08$, with initial separating
    distance $\delta x =6$}
    \label{fig:15}
\end{figure}
\subsection{Effect of surface tension ratio}
 So far, we have explored the behaviour of binary mixture droplets with a surface tension ratio $\sigma_R < 1$. Given the previously observed variations in the behaviour of isolated binary droplets with differing surface tension ratios (figure~\ref{fig6}), we now examine the dynamics of droplets consisting of a binary mixture with a ratio greater than 1. Similarly to the cases shown in figure~\ref{fig6}, we consider two $50 \%$ ethanol-water droplets placed at an initial distance of $10R_0$. Since the droplets spread to a very high extent, short-range interactions end up in the coalescence of the droplets under any physical conditions. Figure~\ref{f12}(a) shows the dynamics of the droplets at lower $Ma =0.005$. We can see that both droplets spread to a high extent and become extremely thin. Approximately, at $t =1000$, the already thin droplets simply coalesce and evaporate very quickly. At this point, the precise translation of the droplets is unobservable, as they have already become extremely thin because of spreading.
 \begin{figure}
    \centering
    \includegraphics[width=14cm]{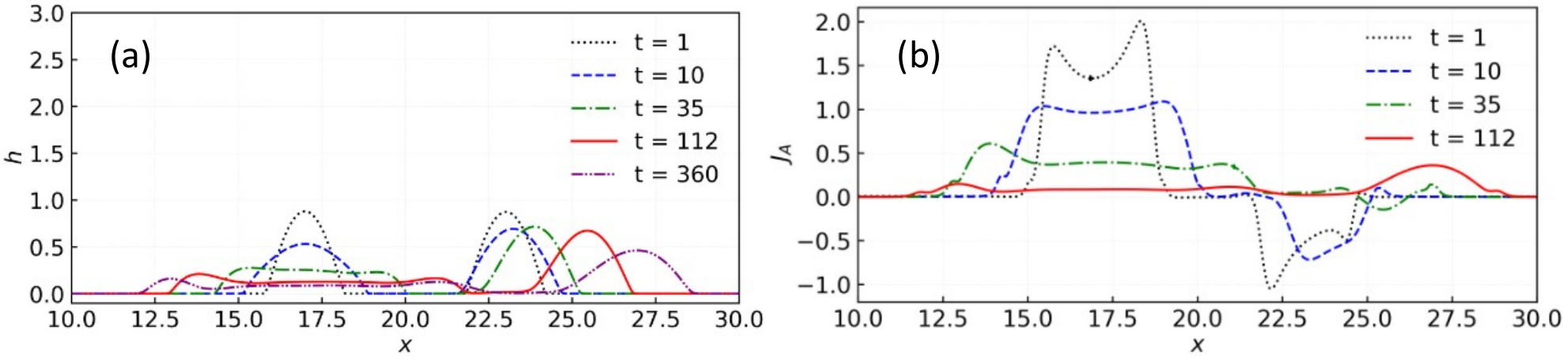}
    \caption{(a) Drop shape evolution of two ethanol-water droplets having $25\%$ ethanol (left droplet) and $10\%$ ethanol (right droplet), (b) Evaporation flux of ethanol ($J_A$) from the droplets. The non-dimensional parameter values for ethanol-water mixture are given in Table \ref{tab:5}. }
    \label{fig16}
\end{figure}
 On the other hand, for a higher thermal Marangoni condition ($Ma = 0.05$), the thermal stress reduces the footprint of the droplet, limiting the spreading. Moreover, the asymmetry of the droplet shapes ($t =1000, 1500$) is clearly observed as a result of the vapour shielding effect. In this case, we note that the droplets exhibit repulsive interactions and do not coalesce, in line with our previous discussions regarding the effect of higher thermal Marangoni on water-morpholine drops. We further analyse these behaviours of ethanol-water drops with the help of mean velocity components, as shown in figure~\ref{fig:15}a. The velocities are plotted for $50\%$ ethanol-water droplets at $Ma = 0.08$. Here, we can observe that the droplets have very high $U_{mx}$ at the beginning while they spread rapidly. As they spread and the ethanol depletes, $U_{mx}$ almost goes to zero, whereas $U_{mt}$ has higher values than $U_{mx}$ at this point onwards (for $t>400$). This thermal Marangoni velocity does not allow the droplet to coalesce at higher $Ma$ as observed in figure~\ref{f12}a.

Unlike the Marangoni contracted drops, the ethanol-water drops spread rapidly (see Appendix \ref{appC}) to form a very thin film. 
The thin films do not show this translation toward one another, although their evaporation is suppressed due to the vapour shielding effect. 
Also, due to their rapid spreading, if the initial distance between the droplets is less than their maximum spreading diameter, the droplets coalesce into a single droplet at any mixture composition and Marangoni number ($Ma$) as shown in figure~\ref{fig:15}b. 
\begin{figure}
    \centering
    \includegraphics[width = 12cm]{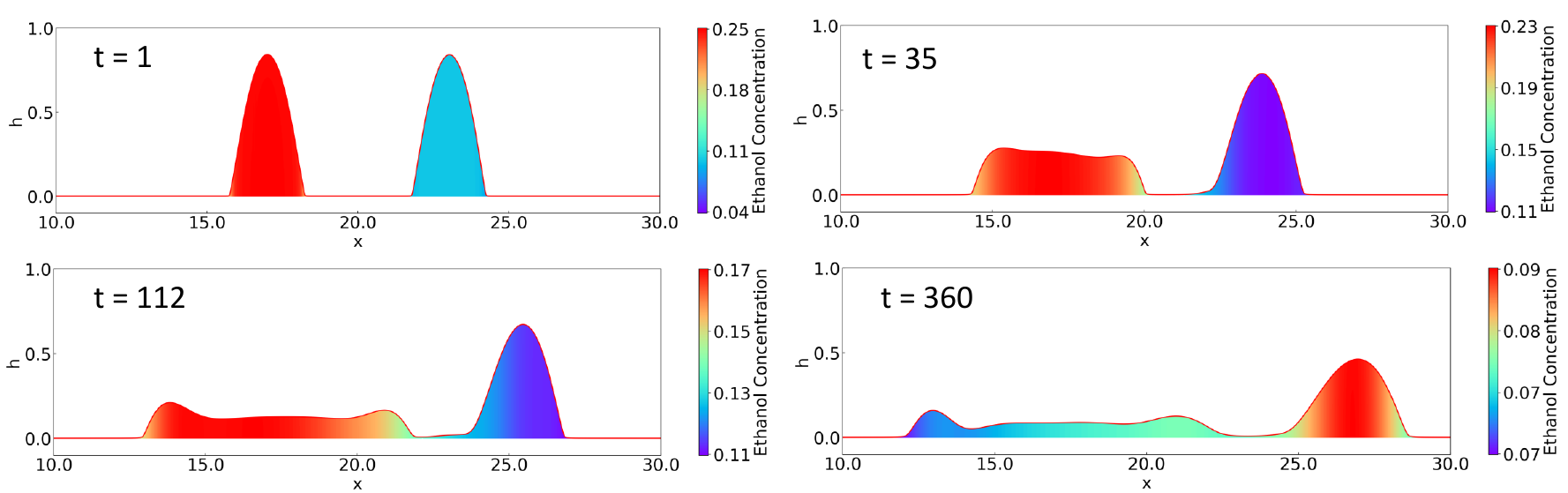}
    \caption{Ethanol concentration of  the droplets for two ethanol-water droplets having $\mathcal{X}_{A,i} = 0.25$ (left) and $\mathcal{X}_{A,i} = 0.10$ (right). }
    \label{fig17}
\end{figure}

\begin{figure}
    \begin{subfigure}[t]{0.5\textwidth}
    \centering
      \includegraphics[width=6cm]{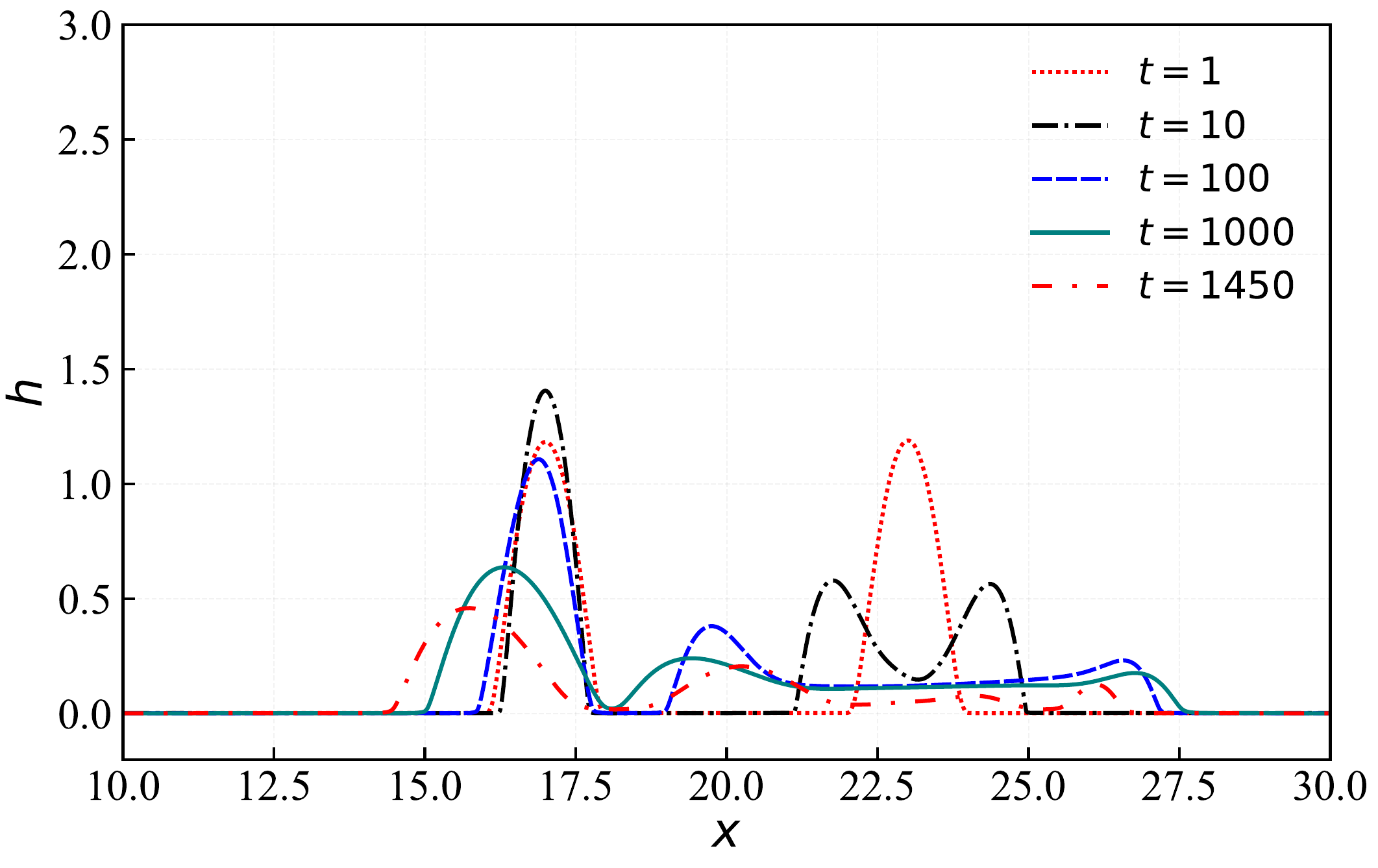}  
      \subcaption{}
    \end{subfigure}
    \begin{subfigure}[t]{0.5\textwidth}
    \centering
      \includegraphics[width=6cm]{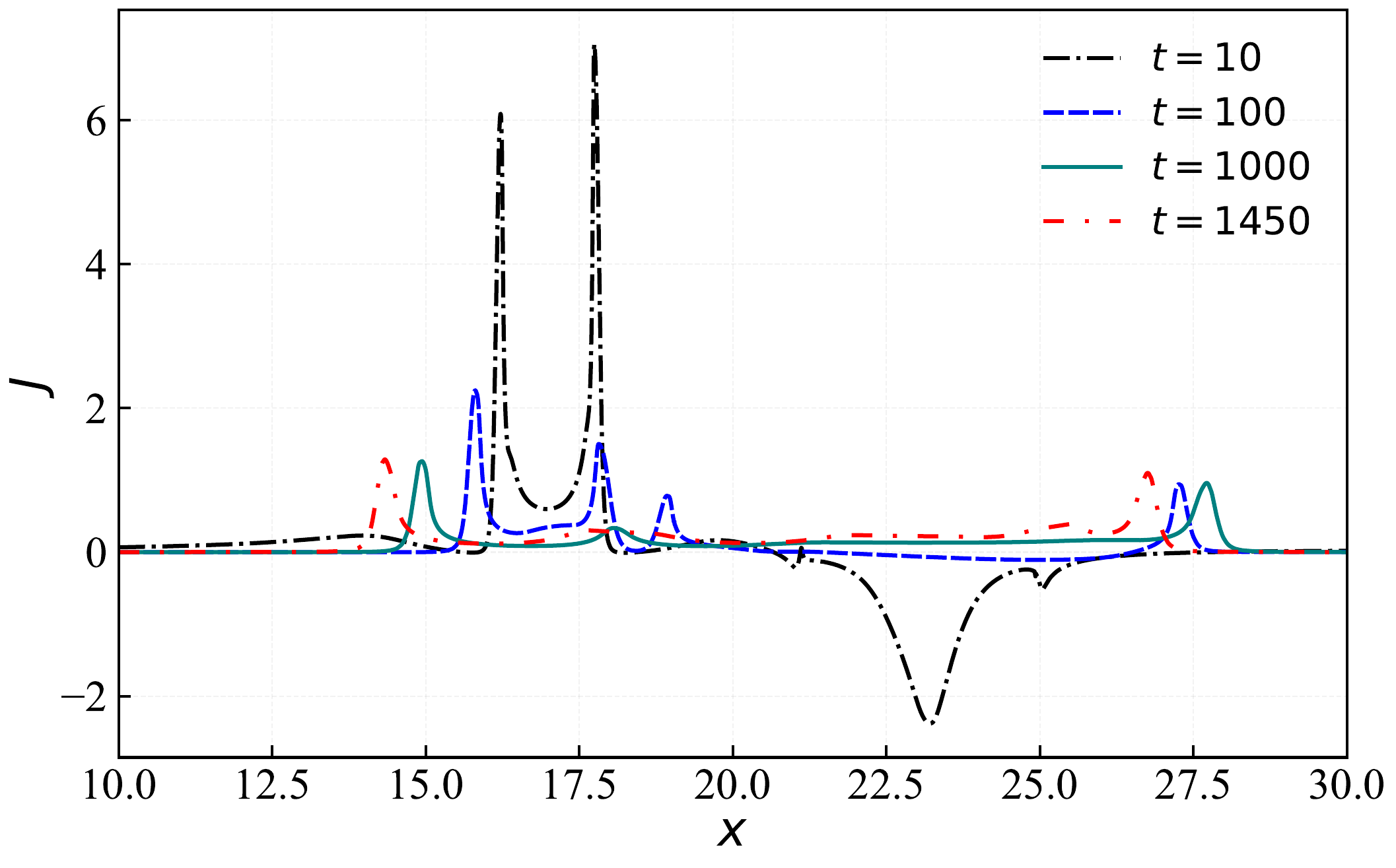}  
      \subcaption{}
    \end{subfigure}
    \begin{subfigure}[t]{1\textwidth}
    \centering
      \includegraphics[width=6cm]{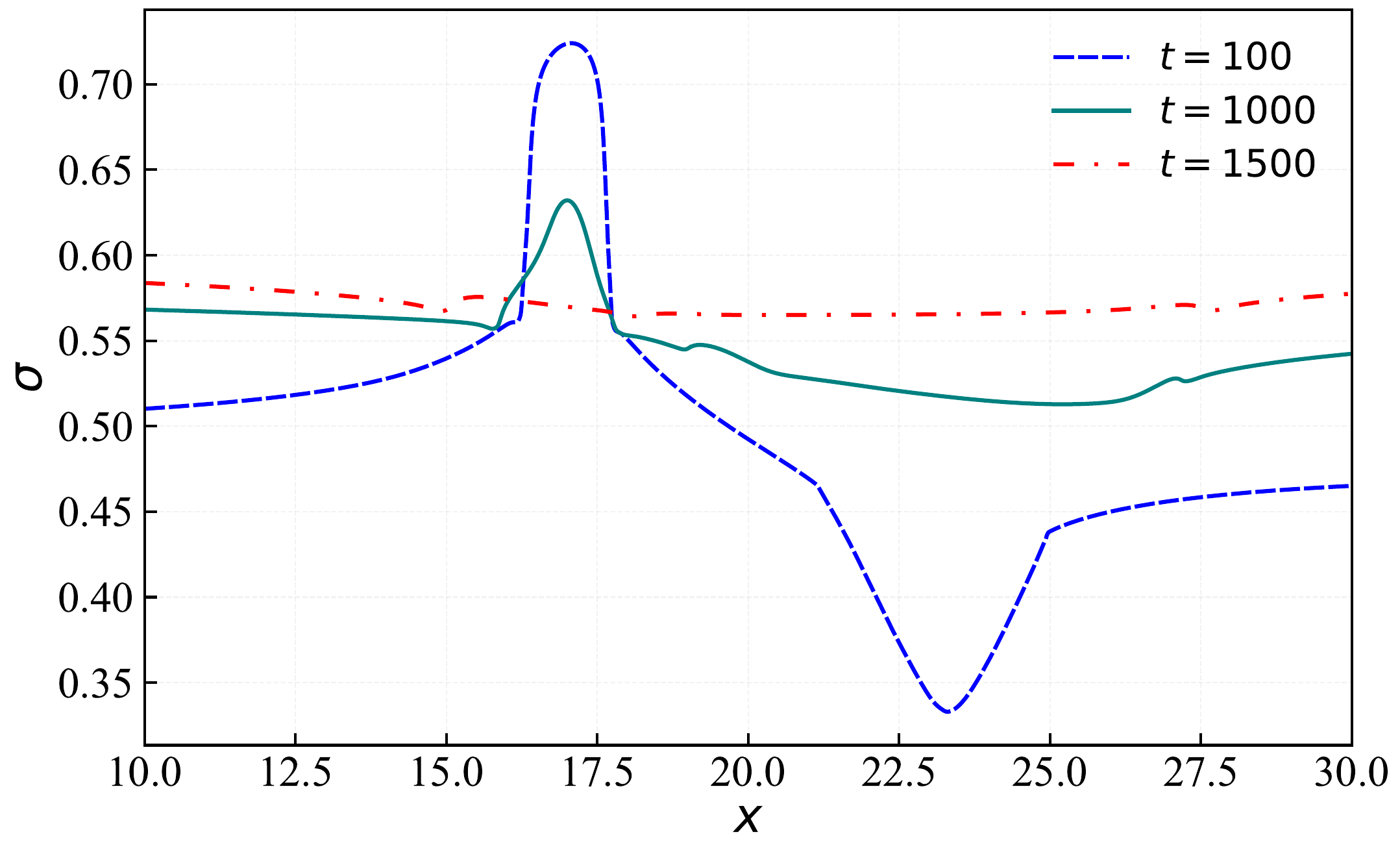}  
      \subcaption{}
    \end{subfigure}
    \begin{subfigure}[t]{1\textwidth}
        \centering
      \includegraphics[width=12cm]{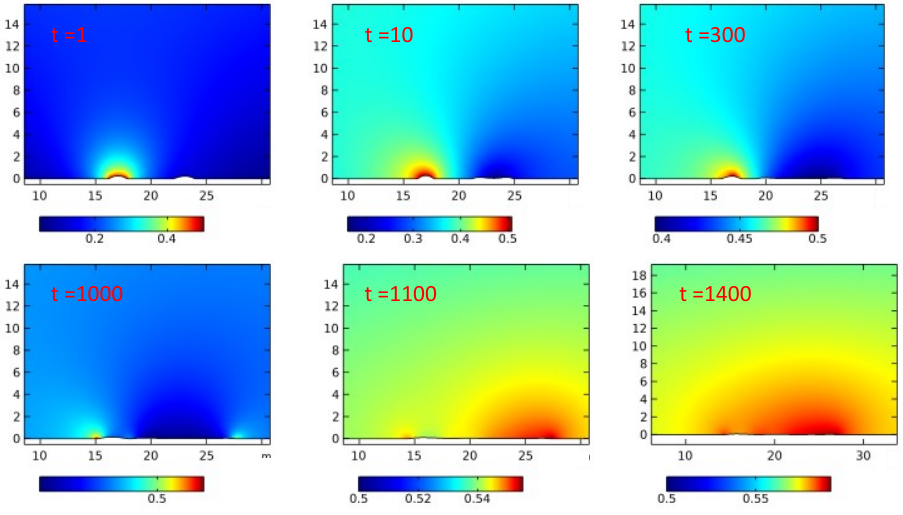}  
      \subcaption{}
    \end{subfigure}
    
    \caption{(a) Drop shape evolution, (b) Total evaporation flux distribution, (c) Surface tension across the drop interface, (d) Water vapour concentration at different times for two water-morpholine drops having $\mathcal{X}_{A,i} =0.5$ for the droplet on the left and $\mathcal{X}_{A,i} =0.1$ for droplet on the right. The parameter space for this cases are: $Kn = 0.001, E=2 \times 10^{-4}, \delta = 10^{-5}, \psi =0.1, \delta x = 6, Pe_{v,A} =0.03, Pe_{v,B} =0.1, Pe =5 , Ma = 0.01, \mathcal{H} = 0.5$.}
    \label{f14}
    \end{figure}
\subsection{Droplets with different initial concentrations}

Thus far, we have examined the cases in which both droplets have the same initial composition. We now move on to investigate the dynamics of the binary droplets that have different initial mixture compositions. 

\subsubsection{Ethanol-water drops with different initial concentration}
We have analysed a pair of ethanol-water droplets, each composed of distinct initial compositions. Figure~\ref{fig16}a illustrates how the shapes of these droplets change with time, where the left droplet begins with $\mathcal{X}_{A,i} = 0.25$ and the right droplet with $\mathcal{X}_{A,i} = 0.10$. Here, $\mathcal{X}_{A,i}$ is the initial concentration of ethanol. The droplet with a higher ethanol content (left droplet) begins to spread at $t = 10$, while the droplet on the right, containing less ethanol, retains its shape. Simultaneously (at $t = 10$), the droplet on the left undergoes rapid ethanol evaporation, resulting in a significant accumulation of ethanol vapour surrounding the droplets (generating a vapour gradient). This vapour subsequently condenses in the right droplet with a lower ethanol concentration, as illustrated in figure~\ref{fig16}b, where $J_A$ denotes the evaporation flux of ethanol. As time progresses, the left droplet continues to spread, pushing the other droplet away ($ t = 35, 112,$  and $360$). This behaviour is attributed to the surface tension gradient generated from the condensation of ethanol in the right droplet. As shown in figure~\ref{fig17}, at $t = 35$ and $112$, the ethanol concentration of the right droplet increased on the proximal side (side facing the left droplet) of the droplet due to higher condensation. Because of the surface tension gradient arising from the concentration gradient of ethanol and water, the left droplet pushes the right droplet away.

\subsubsection{Water-Morpholine drops with different initial concentrations}
Figure~\ref{f14} illustrates the behaviour of two drops comprised of water and morpholine, with the left drop containing $50\%$ water, whereas the right drop contains $10\%$ water. In figure~\ref{f14}a, we observe that the droplet on the left maintains its shape; however, the droplet on the right starts to spread and deform ($t =1$ to $t = 100$). Later, at $t =1000$ when the droplets are very close, the droplet with a lower water concentration pushes the other droplet rather than coalescing. This kind of chasing behaviour is also observed by \cite{Cira2015} in their experiments with water-PG droplets of different concentrations. We also observed the same chasing behaviour in our experiments shown in figure~\ref{fig:exp}d. This kind of behaviour of the droplets can be understood by looking into the evaporation flux profile for both droplets presented in figure~\ref{f14}b. Here we observe that as the droplet on the left evaporates, there is significant condensation on the droplet on the right ($t =10, 100$). As a result, the differential concentrations generated create a differential surface tension between the drops, as shown in figure~\ref{f14}c. 
Furthermore, we plot the evolution of the water vapour concentration in figure~\ref{f14}d. It can noted that due to faster evaporation from the left droplet, the concentration of water vapour is rapidly increasing (from $t =10$ to $t =1000$). As a result of the increase in the water vapour concentration, condensation was observed on the right droplet. It should be noted that morpholine is also evaporated, although at a much slower rate, contributing to the total evaporation flux and hence to the surface tension gradient.

When comparing ethanol-water and water-morpholine drops, we find that in ethanol-water mixtures, droplets with more ethanol (less water) push those with less ethanol (more water). Conversely, in water-morpholine mixtures, droplets with less water push those with more water. This is due to ethanol having a lower surface tension than water ($\sigma_R > 1$), while water's surface tension is greater than morpholine's ($\sigma_R < 1$). Thus, in both scenarios, droplets with initially lower surface tension push those with initially higher surface tension, aligning with experimental results.
\section{Conclusions \label{s7}}
Evaporating droplets situated near another evaporating droplet mutually affect each other’s rates of evaporation. This interaction also impacts the evaporative cooling (thermal Marangoni) and shape of the droplets, resulting in their motion. Additionally, when these droplets comprise a binary mixture, the solutal Marangoni effect becomes significant as well. We have developed a model-based lubrication theory for flat pure and binary droplets evaporating in the presence of each other. We employed a precursor film for the free-moving contact line of the droplets to observe their translation under different physical conditions. In addition, we performed some experiments (infrared imaging) of two water-morpholine binary mixture drops placed on a heated hydrophilic surface.

From a theoretical standpoint, our model demonstrated that multiple droplets can inhibit each other's evaporation rates compared to when they evaporate in isolation. We could also show the interplay between the capillary and thermal Marangoni, making the droplets attract or repel each other in the limit of pure fluid droplets. Moreover, by incorporating binary mixtures into the droplets, we demonstrated the behavior of these mixtures during evaporation under isolated conditions, revealing patterns of either spreading (e.g., ethanol-water) or contraction (e.g., water-morpholine). Our model precisely identified the fundamental mechanisms, such as the interaction among capillary, solutal, and thermal Marangoni effects, illustrating the attractive and repulsive behaviours of these binary droplets when situated in proximity to one another. We explicitly note that the solutal Marangoni effects encourage the droplets to move towards one another, whereas the thermal Marangoni forces act as a repulsive force. By considering the composition of the mixtures and the thermal influences, we can develop a regime map that illustrates the conditions under which the droplets either attract or repel each other. Moreover, our model allows us to anticipate the chasing phenomenon of the droplets when there are different initial mixture compositions.
Experimentally, we investigated the water-morpholine droplets at different surface temperatures. We found that at lower surface temperature ($30^{\circ}$C) the droplets always attract each other. However, their propensity for attraction diminishes at an elevated substrate temperature ($60^{\circ}$C). Our model also forecasts this identical behaviour. Further, we performed experiments by placing two water-morpholine drops in proximity but with different initial water concentrations, which showed that the droplet with a lower water concentration chases the droplet with the higher water concentration. Our model also predicts this significant scenario. 
We emphasize that the present model is two-dimensional, rather than a fully three-dimensional representation of sessile droplets. Because vapor-mediated interactions are highly sensitive to the spatial geometry and the dimensionality of the diffusion field, the comparison with experiments should be regarded as qualitative. Accordingly, we did not carry out a quantitative comparison between our model and experimental measurements; however, the model’s predictions reproduce the same phenomena observed in experiments within the corresponding parameter regime. This model lays the foundation for a more complex full-scale three-dimensional model, which can be qualitatively compared with the observations from experiments.

\backsection[Declaration of interests]{The authors report no conflict of interest.}
\backsection[Funding] {The authors gratefully acknowledge the support received from the ThermaSMART project of European Commission (grant no. EC-H2020-RISE-ThermaSMART-778104).}
\newpage
\appendix
\makeatletter
\@addtoreset{figure}{section}
\@addtoreset{table}{section}
\makeatother
\renewcommand{\thefigure}{\thesection\arabic{figure}}
\renewcommand{\thetable}{\thesection\arabic{table}}

\section{Weak forms}\label{appA}
The weak forms of the evolution equations (Eqs. \ref{2.40}, \ref{2.54}, \ref{2.55}, \ref{4.3}, \ref{4.4}) are derived as:
\begin{align}
R = \int_0^L \left\{(p+\Pi)\zeta - \frac{\epsilon^2}{Ma} 
    \frac{\partial h}{\partial x}\left( \frac{\partial \sigma}{\partial x} \zeta + \frac{\partial \zeta}{\partial x} \sigma \right) \right\} dx \\
R = \int_0^L \{Kn J_A - (\mathcal{X}_A \rho^{ve,A} -                
   \rho^{v,A}|_{int})\} \zeta dx  \\
   R = \int_0^L \left\{Kn J_B - \left(\frac{\rho^v_R }{\sqrt{M_R}}\right) \left[ (1-\mathcal{X}_A)\rho^{ve,B} - \rho^{v,B}|_{int}\right]\right\} \zeta dx \\ 
R = \int_0^L \left\{\frac{\partial h}{\partial t}+ \frac{\partial q} 
    {\partial x} + EJ\right\} \zeta dx 
\end{align}
\begin{align}
        R = \int_0^L  \left\{\frac{\partial\mathcal{X}_{A0}}{\partial t} + \frac{q}{h}\frac{\partial \mathcal{X}_{A0}}{\partial x} +q^*\frac{\partial \mathcal{X}_{A1}}{\partial x} + \frac{ \partial q^*}{\partial x}\mathcal{X}_{A1} +\frac{2}{3}EJh^2\mathcal{X}_{A1} \right\}\zeta dx \nonumber \\ 
    - \int_0^L \left\{\frac{J_A - J \mathcal{X}_{A0}}{ Pe' h^2 \left(\frac{J}{3} - \frac{1}{Pe'Eh}\right)}\right\} \zeta dx 
\end{align}
\section{Effect of domain size}
\label{appB}
Here, we present the effect of domain size on the dynamics of droplet translation dynamics. The ideal far-field boundary condition is to set a constant concentration value at $z \to \infty$. To reduce the numerical redundancy, here, we consider a finite height of the far-field boundary and set a Dirichlet condition for the concentrations. In order to demonstrate the impact of this simplified approach, we adjust the domain size in the $z$-axis, illustrating its influence on both droplet motion and mass reduction (figure~\ref{fig20}).

\medskip
\begin{center}
\centering
\begin{minipage}{0.48\textwidth}
  \centering    \includegraphics[width=\linewidth]{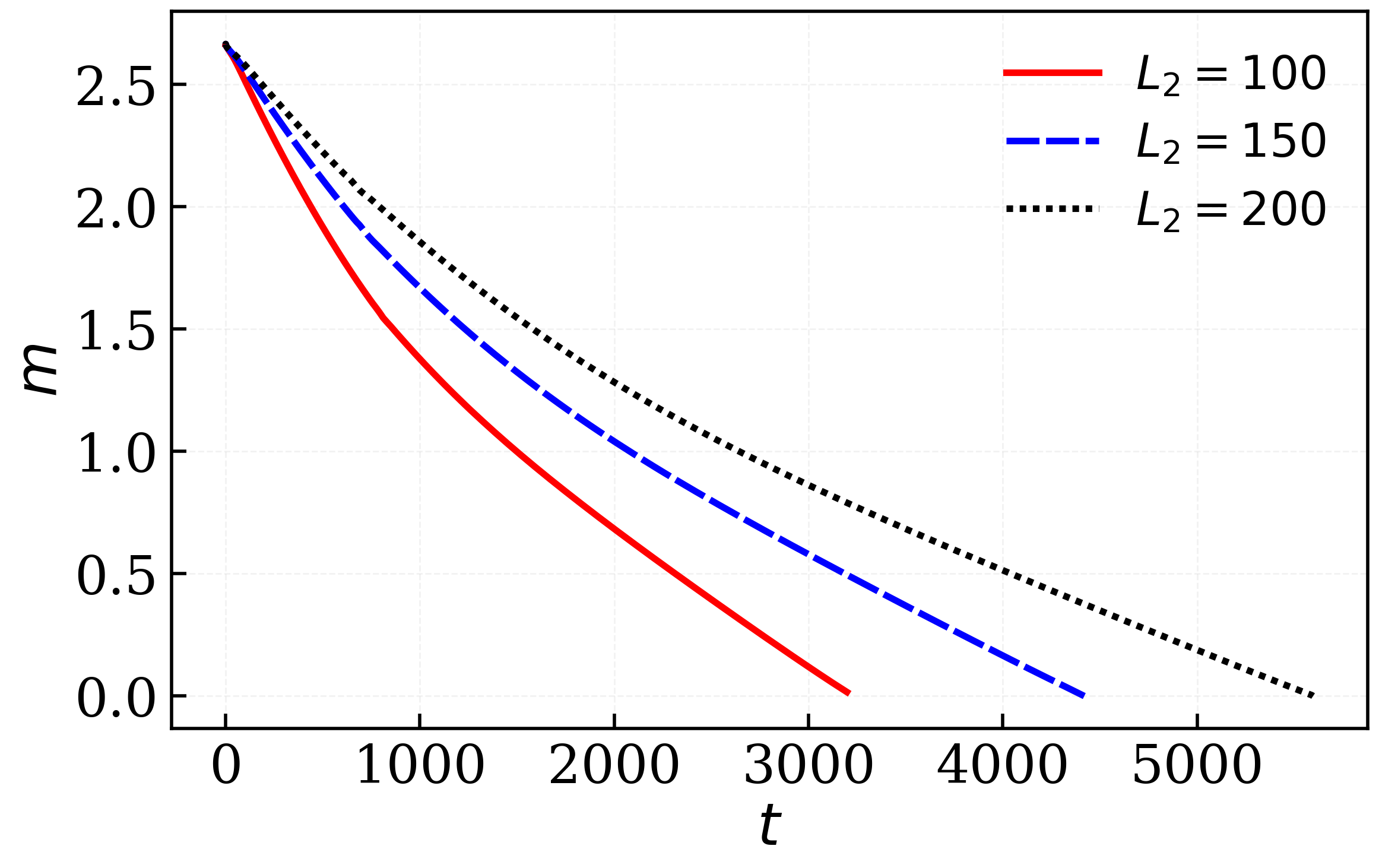}\\
  \footnotesize{(a)} 
\end{minipage}\hfill
\begin{minipage}{0.48\textwidth}
  \centering    \includegraphics[width=\linewidth]{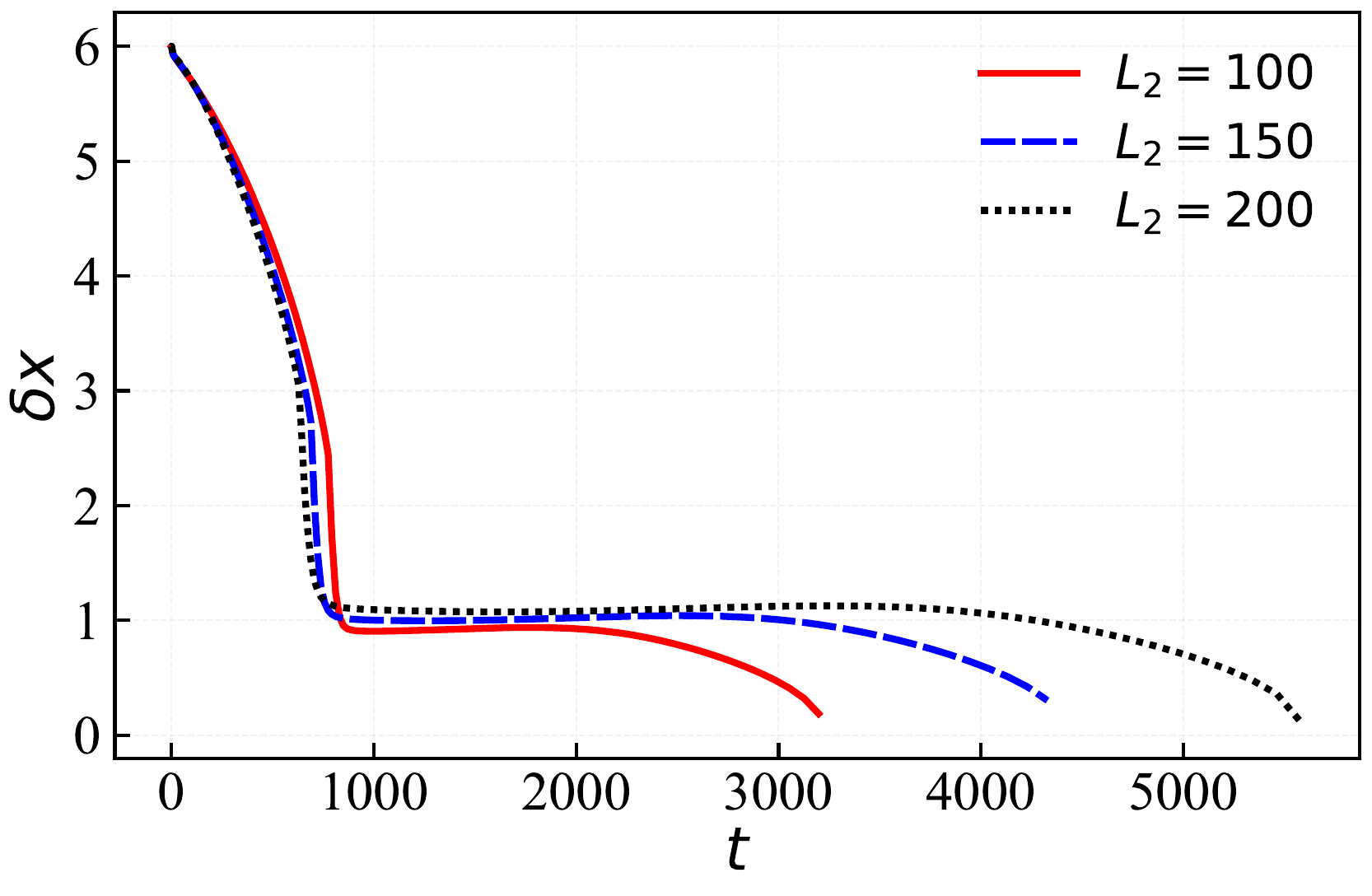}\\
  \footnotesize(b)
\end{minipage}
\captionof{figure}{Time evolution of (a) the total mass of the system, and (b) separating distance between the drops ($\delta x$) for different domain heights ($L_2$), in the case of two $50\%$ water-morpholine droplets at $Ma =0.01$, and other parameters are the same as figure \ref{fig7}.}
\label{fig20}
\end{center}


\medskip
As shown in figure~\ref{fig20}a, the mass depletion becomes faster as the length of the far-field boundary ($L_2$) is increased from $100$ up to $200$. We observe a rather slow convergence for the evaporation rate of the droplets. To investigate the effect of $L_2$ on the dynamics of the drops, the time evolution of the separating distance of the droplets ($\delta x$) is depicted in figure \ref{fig20}b. We can clearly observe that the translation dynamics (in this case attraction) do not change with varying size of the far-field boundary.

\section{Spreading in the absence of Marangoni transport}
\label{app:no_marangoni}
\begin{figure}
    \centering
    \begin{minipage}{1.1\textwidth}
        \centering
        \includegraphics[width=\linewidth]{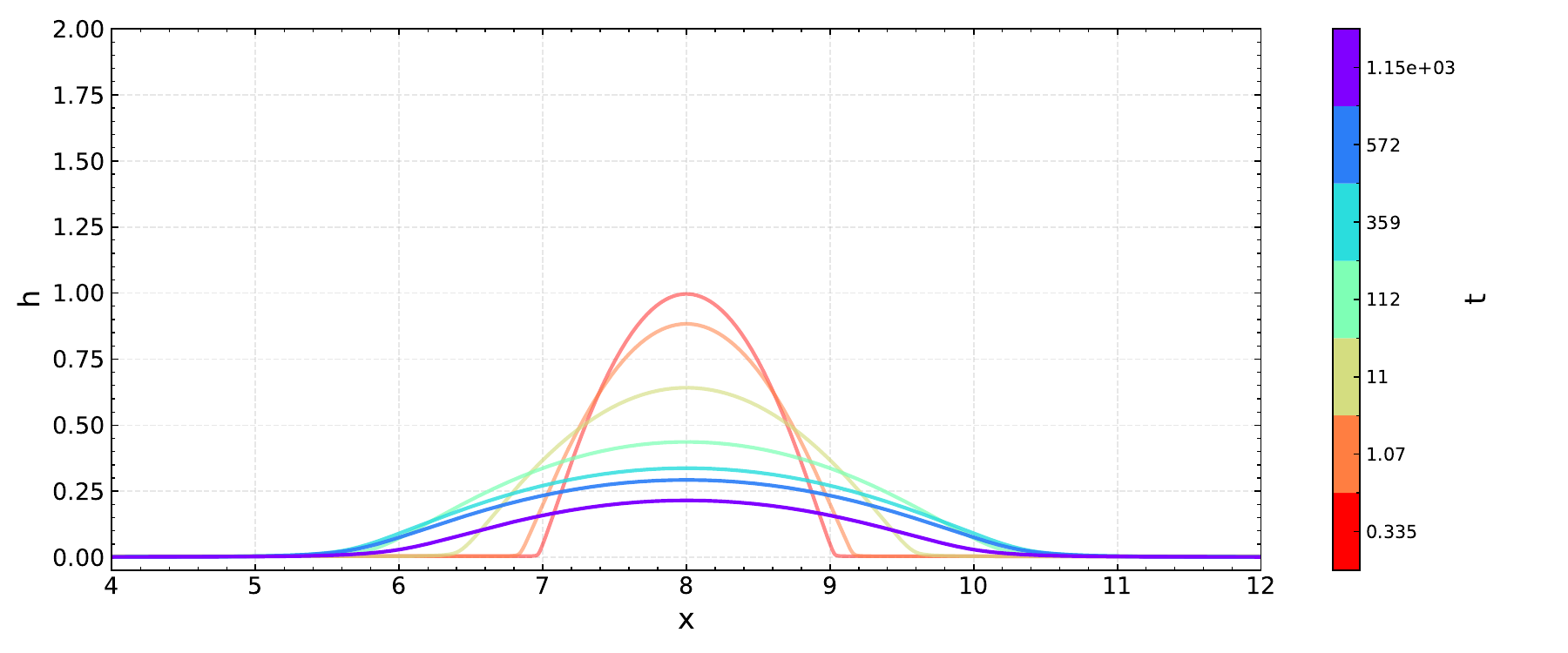} \par (a)
    \end{minipage}
    \hfill
    \begin{minipage}{0.6\textwidth}
        \centering
        \includegraphics[width=\linewidth]
        {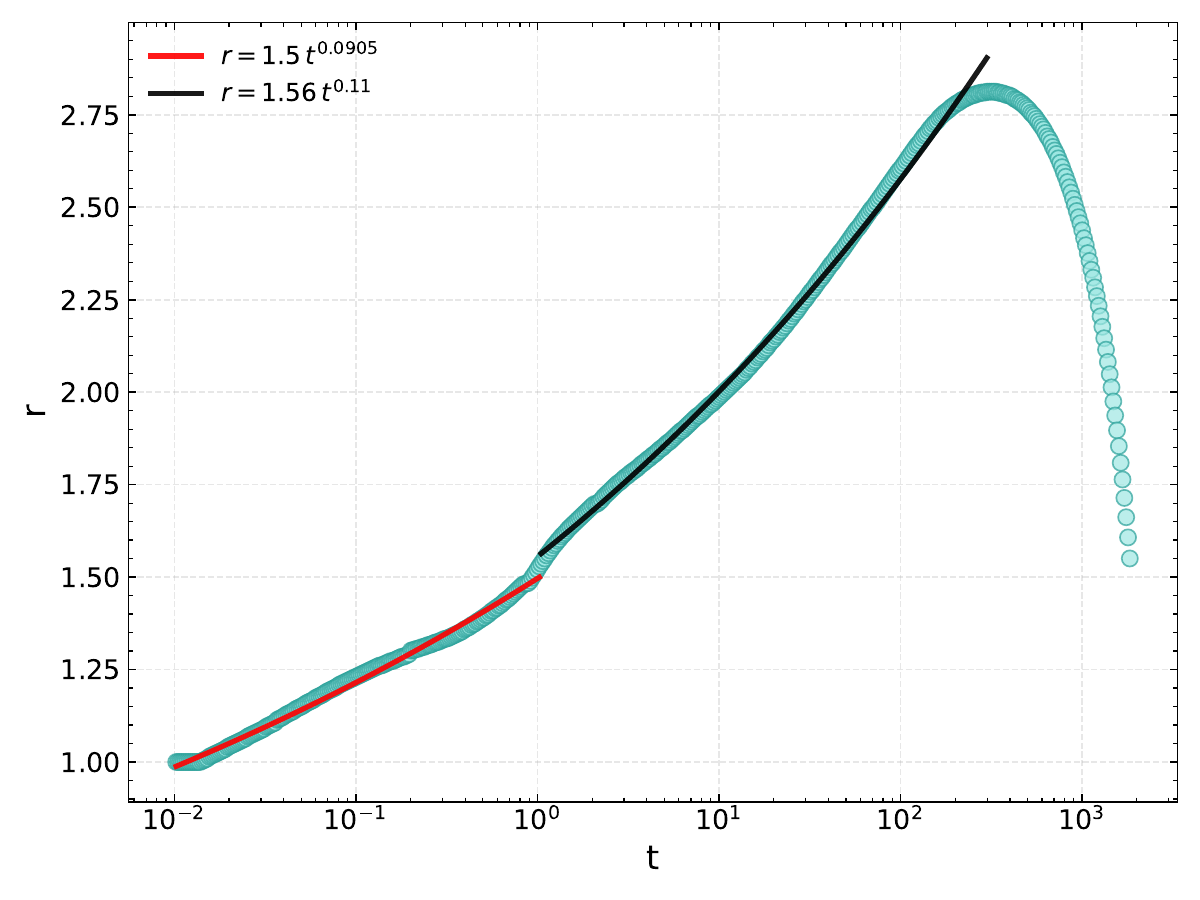} \par (b)
    \end{minipage}
    \caption{Spreading of an isolated water--morpholine
    drop with the Marangoni contribution to the liquid
    flux set to zero.
    (a) Evolution of the drop profile; colours indicate
    time as specified by the colour bar.
    (b) Normalised contact radius and empirical power-law
    fits. All other model parameters and initial
    conditions are retained from the corresponding
    full-model simulation.}
    \label{fig:no_marangoni}
\end{figure}
To evaluate how Marangoni transport contributes to limiting spreading, we reran the complete-wetting simulation with the Marangoni term in the liquid flux removed, while keeping all other conditions the same as in the case shown in figure~\ref{fig6}a. The
pressure-driven flux, including capillary and
disjoining-pressure contributions, was retained.
The modified flow rate becomes
\begin{equation}
    q= \left[ -\frac{h^3}{3 \mu} \frac{\partial p}{\partial x} \right]. \tag{C1}
\end{equation}
Figure~\ref{fig:no_marangoni}(a) shows that the drop
broadens and its height decreases when Marangoni
transport is suppressed. In contrast, the full model
exhibits an extended interval of approximately constant
contact radius before retraction. This comparison
demonstrates that Marangoni transport is essential to
the suppression of spreading under the simulated
conditions.

The contact-radius evolution without Marangoni
transport, shown in figure~\ref{fig:no_marangoni}(b),
is approximately described by
\begin{equation}
\frac{r}{R_0} \simeq
\begin{cases}
1.5(t^*)^{0.090}, & t^* \lesssim 1,\\
1.56(t^*)^{0.11}, & t^* \gtrsim 1,
\end{cases} \tag{C2}
\label{eq:empirical_spreading}
\end{equation}
where $t^*=\hat{t}/\hat{T}_0$ and $\hat{T}_0= \hat{R}_0/\hat{U}^* = 8.42$~ms.
These expressions are empirical fits to the numerical
results and are not interpreted as a verification of
Tanner's law for the present two-dimensional geometry.

For a reproducible measure of substantial spreading,
we define $\hat{t}_{sp}^{sim}$ as the elapsed time
required for the contact radius to increase by 50\%
from its initial value. Using the early-time fit gives
\begin{equation}
t_{sp}^{*}
\simeq
\left(\frac{1.5}{1.5}\right)^{1/0.090}
= 1,
\qquad
\hat{t}_{sp}^{sim}
\simeq 1.0T_0
\simeq 8.42~\mathrm{ms}.
\tag{C3}
\label{eq:simulation_spreading_time}
\end{equation} 
For comparison with the experimental geometry,
Tanner's law provides a classical reference for
viscous spreading of a non-volatile, completely
wetting, axisymmetric drop in the absence of
Marangoni stresses \citep{tanner1979spreading}.
Writing
$\hat{R}=C_T
[\hat{\sigma}\hat{V}^{3}(\hat{t}+\hat{t}_v)/
\hat{\mu}]^{1/10}$,
where $C_T$ is the spreading prefactor and
$\hat{t}_v$ accounts for the finite initial radius,
the corresponding time for a 50\% increase is
\begin{equation}
\hat{t}_{sp}^{T}
\simeq
\frac{\hat{\mu}_A\hat{R}_0^{10}}
     {C_T^{10}\hat{\sigma}_A\hat{V}^{3}}
\left(1.5^{10}-1\right).
\label{eq:tanner_spreading_time}
\tag{C4}
\end{equation}
Using $\hat{R}_0=2.1$~mm, $\hat{V}=3~\mu$l,
$\hat{\mu}_A=10^{-3}$~Pa\,s and
$\hat{\sigma}_A=72\times10^{-3}$~N\,m$^{-1}$ gives
\begin{equation}
\frac{\hat{\mu}_A\hat{R}_0^{10}}
     {\hat{\sigma}_A\hat{V}^{3}}
=0.858~\mathrm{ms},
\qquad
\hat{t}_{sp}^{T}
\simeq 48.6\,C_T^{-10}~\mathrm{ms}.
\tag{C5}
\end{equation}
Here, $C_T$ is a dimensionless prefactor that depends
on the microscopic contact-line description.
We take $C_T=1$ solely to report a nominal,
prefactor-free scaling estimate, giving
$\hat{t}_{sp}^{T}\sim48.6$~ms; this is not an
experimentally determined value of $C_T$.
The dependence on $C_T$ is retained explicitly
because it enters the inferred time to the tenth
power. Accordingly, the Tanner estimate provides
a classical reference for the experimental
geometry, rather than a quantitative prediction
or proof of a separation of timescales.
Our direct evidence for the role of Marangoni
transport is the controlled numerical comparison:
with the other parameters unchanged, suppressing
the Marangoni contribution to the liquid flux
restores spreading.

The characteristic solutal Marangoni transport time,
based on the reference surface-tension contrast
$\Delta\hat{\sigma}_{sol}
=(1-\sigma_R)\hat{\sigma}_A \mathcal{X}_{A,i}$, is
\begin{equation}
\begin{aligned}
\hat{t}_{sol}
&=
\frac{\hat{\mu}_A\hat{R}_0}
{\epsilon(1-\sigma_R)\hat{\sigma}_A \mathcal{X}_{A,i}}
\\
&=
\frac{10^{-3}(2.1\times10^{-3})}
{0.2(0.5)(72\times10^{-3})(0.5)}
\simeq 0.61~\mathrm{ms}.
\end{aligned}
\label{eq:solutal_transport_time}
\tag{C6}
\end{equation}
Consequently,
\begin{equation}
\frac{\hat{t}_{sol}}
     {\hat{t}_{sp}^{sim}}
\simeq 0.067,
\qquad
\frac{\hat{t}_{sol}}
     {\hat{t}_{sp}^{T}}
\simeq 0.0125\,C_T^{10}.
\tag{C7}
\end{equation}
For $E=10^{-4}$ (from Table~4), the nominal evaporation timescale
is $\hat{t}_{ev}=\hat{T}_0/E=84.2$~s. The numerical
comparison therefore gives
\begin{equation}
\hat{t}_{sol}\simeq0.61~\mathrm{ms}
<
\hat{t}_{sp}^{sim}\simeq8.4~\mathrm{ms}
\ll
\hat{t}_{ev}=84.2~\mathrm{s}.
\tag{C8}
\end{equation}
The solutal transport scale is approximately fifteen
times shorter than the time required for a 50\%
increase in radius without Marangoni transport.

Together, the timescale estimates and the controlled
flux-removal calculation support the interpretation
that inward Marangoni transport can oppose substantial
outward spreading. The direct numerical evidence is
that suppressing Marangoni transport restores
spreading. This behaviour is consistent with the
competition between capillary and Marangoni fluxes
described by \citet{karpitschka_marangoni_2017}.
The numerical spreading time remains specific to
the simulated geometry and precursor-film parameters
and is not transferred directly to the experiments.

\section{Experimental Method\label{s5}}

\begin{figure}
\centering
    \begin{minipage}{\linewidth}
    \centering
    \includegraphics[width=0.55\linewidth]{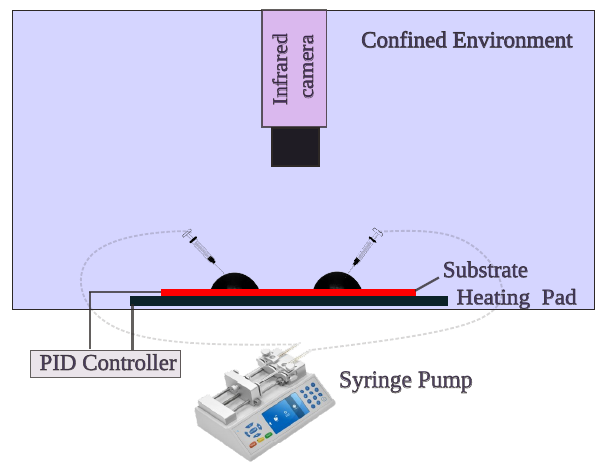}
    
    \end{minipage}
    \caption{Schematic representation of the experimental setup.}
    \label{fig2}
\end{figure}

As shown in figure~\ref{fig2}, experiments are performed in a closed chamber with Infrared imaging under constant ambient conditions ($22^{\circ} C$ and $30\%$ relative humidity). An infrared camera is used (FLIR A320, 30 Hz) to take a top view of the droplets. The droplets are generated simultaneously using a double-barrel syringe pump. Two $3 \pm 0.2 \mu l$ drops are generated simultaneously and deposited on the substrate at a centre-to-centre distance of $10$ mm. Glass slides (75 mm $ \times $ 25 mm and 1 mm thick) are used as the substrate. They were treated with oxide plasma for 30 seconds to ensure very high wettability and to remove organic impurities from the surface. The glass slides were placed over a heating pad. The temperature of the substrate is controlled using a PID controller. The temperature of the surface is measured using a thermocouple attached to the substrate and maintained at the desired set point by the controller. The experiments are performed for three different compositions of water and morpholine mixtures, with water concentrations of $10\%, 25\%$ and $50\%$. For all the compositions, experiments are performed for substrate temperatures of $30^{\circ} \pm 3^{\circ}C) C$ and $60^{\circ} \pm 3^{\circ}C) $. The estimated error, considering the mixture percentages, substrate temperature, and calculation of the centroid of the droplets, is measured to be within $10\%$.

\section{Modes of evaporation} 
\label{appE}

\begin{figure}
    \centering
    \begin{minipage}{0.45\textwidth}
    \centering
          \includegraphics[width=\linewidth]{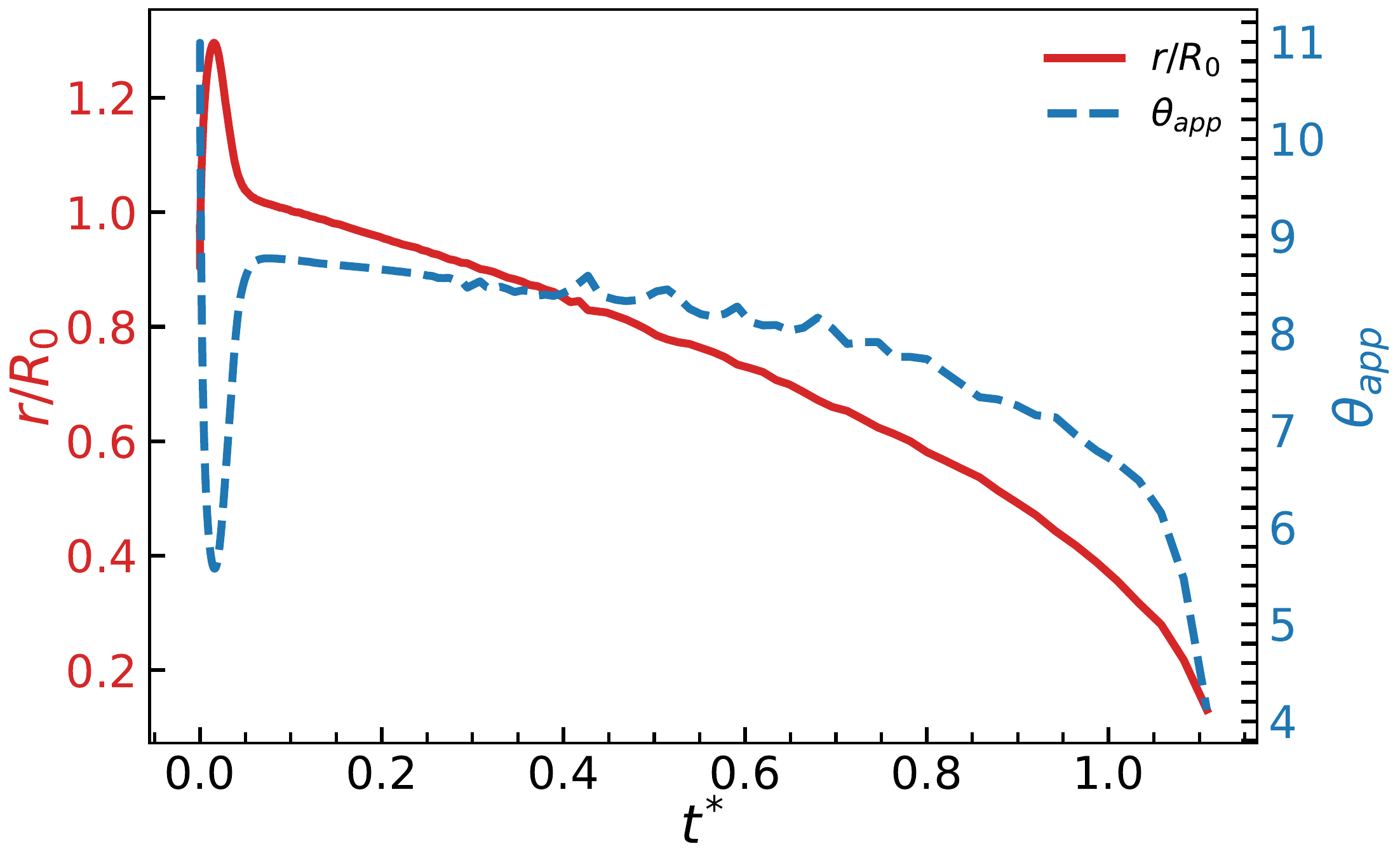}\\ 
          \footnotesize(a)
         
    \end{minipage}
    \begin{minipage}{0.45\textwidth}
    \centering
          \includegraphics[width=\linewidth]{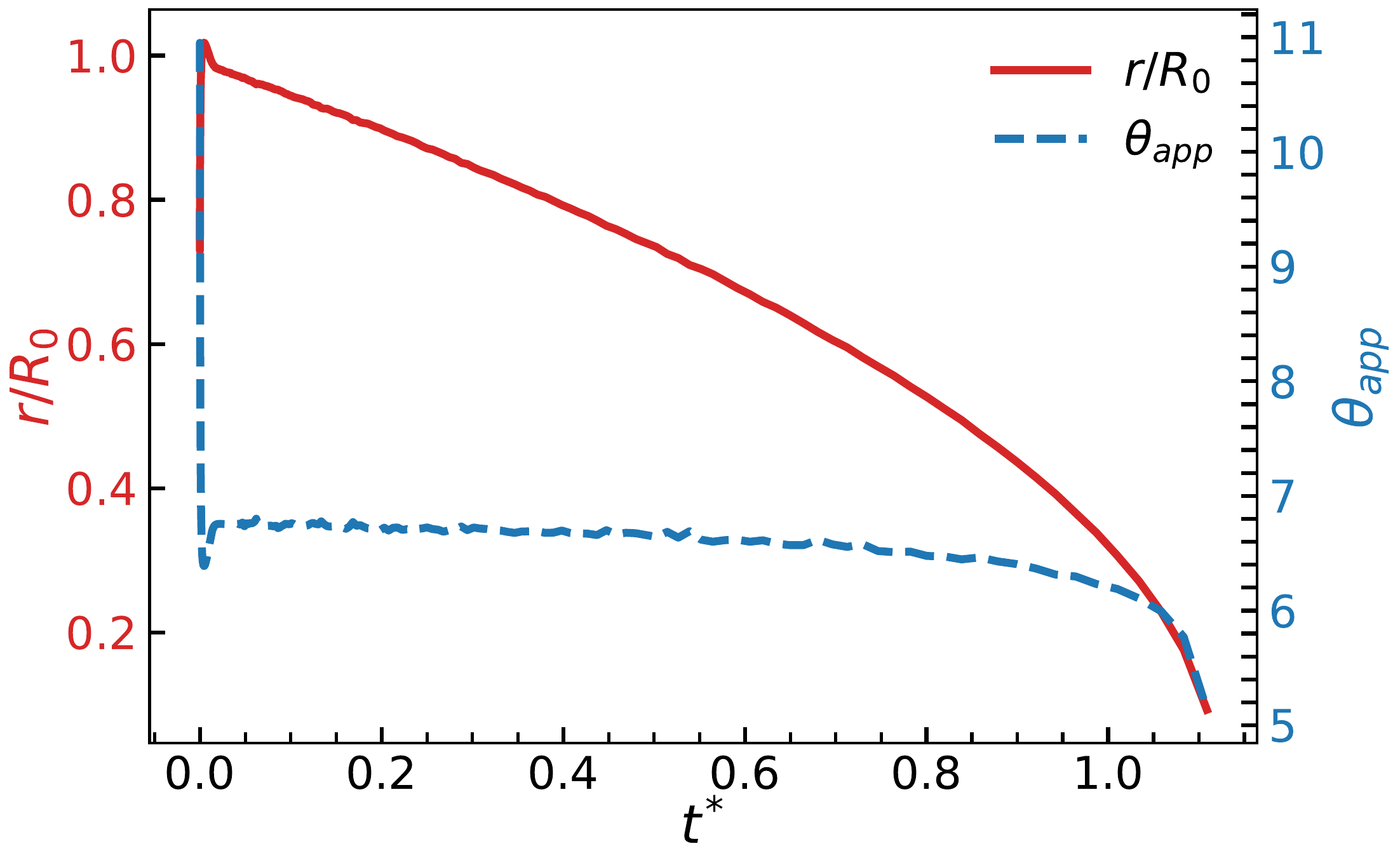} \\
          \footnotesize(b)
    \end{minipage}    
    \caption{Spreading radius $(r/R_0)$ and apparent contact angle variation with time $t^{*}$ for systems of (a) pure isolated droplet and (b) two pure droplets evaporating simultaneously, with the base case parameter values $Kn =  1 \times 10^{-3}$, $E = 2 \times 10^{-4}$, $\delta = 10^{-4}$, $\psi = 0.1$, $Ma =0.005$, $Pe_{v} =0.1, \mathcal{A} = 100$. The time $t^* = \frac{t}{t_e}$, $t_e$ is the time when $90\%$ of the drop is evaporated.}
    \label{f.E1}
\end{figure}

\begin{figure}
    \centering
    \begin{minipage}{0.48\textwidth}
    \centering
          \includegraphics[width=6cm]{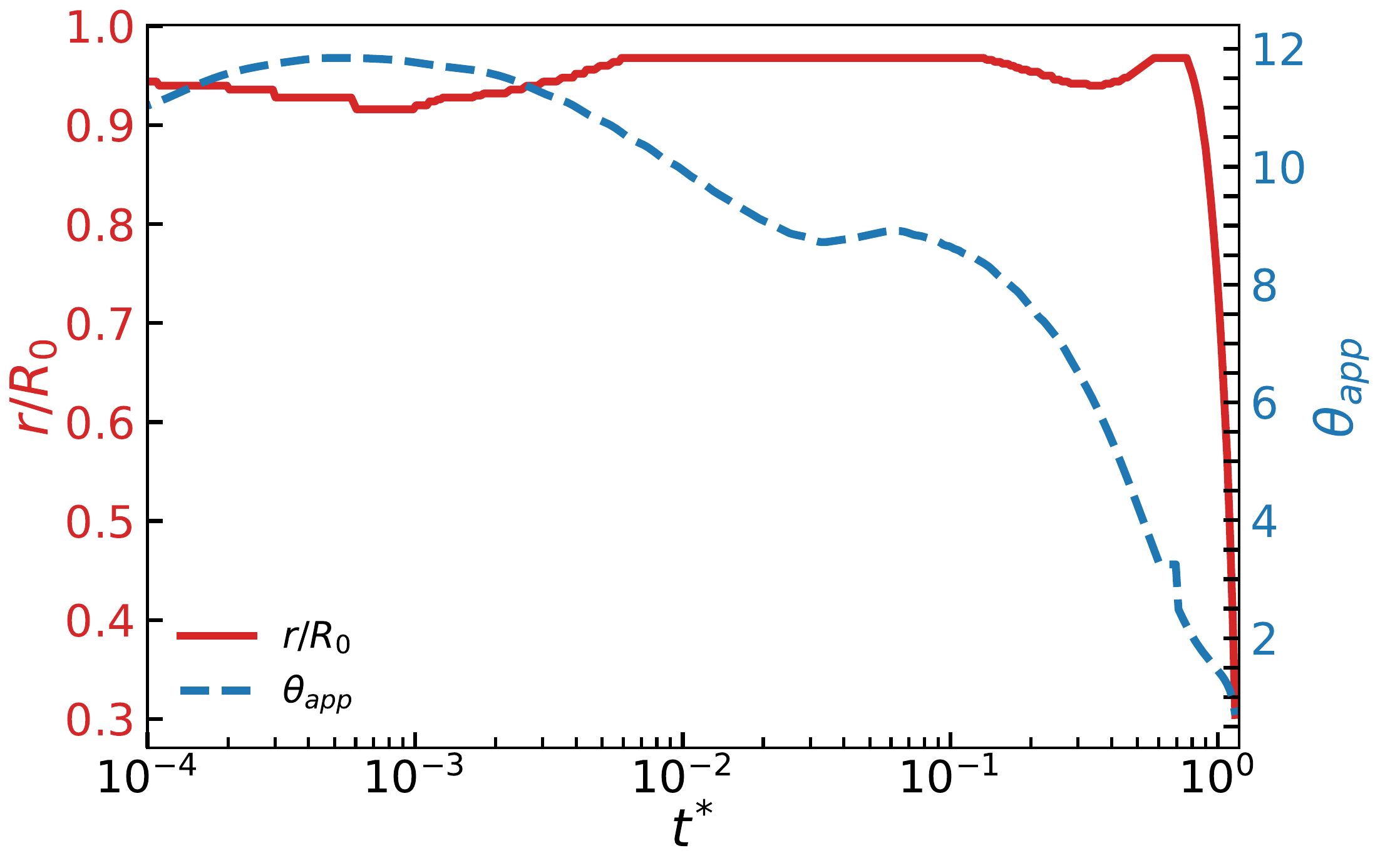} 
          \\ \footnotesize(a)
    \end{minipage}
    \begin{minipage}{0.48\textwidth}
    \centering
          \includegraphics[width=6cm]{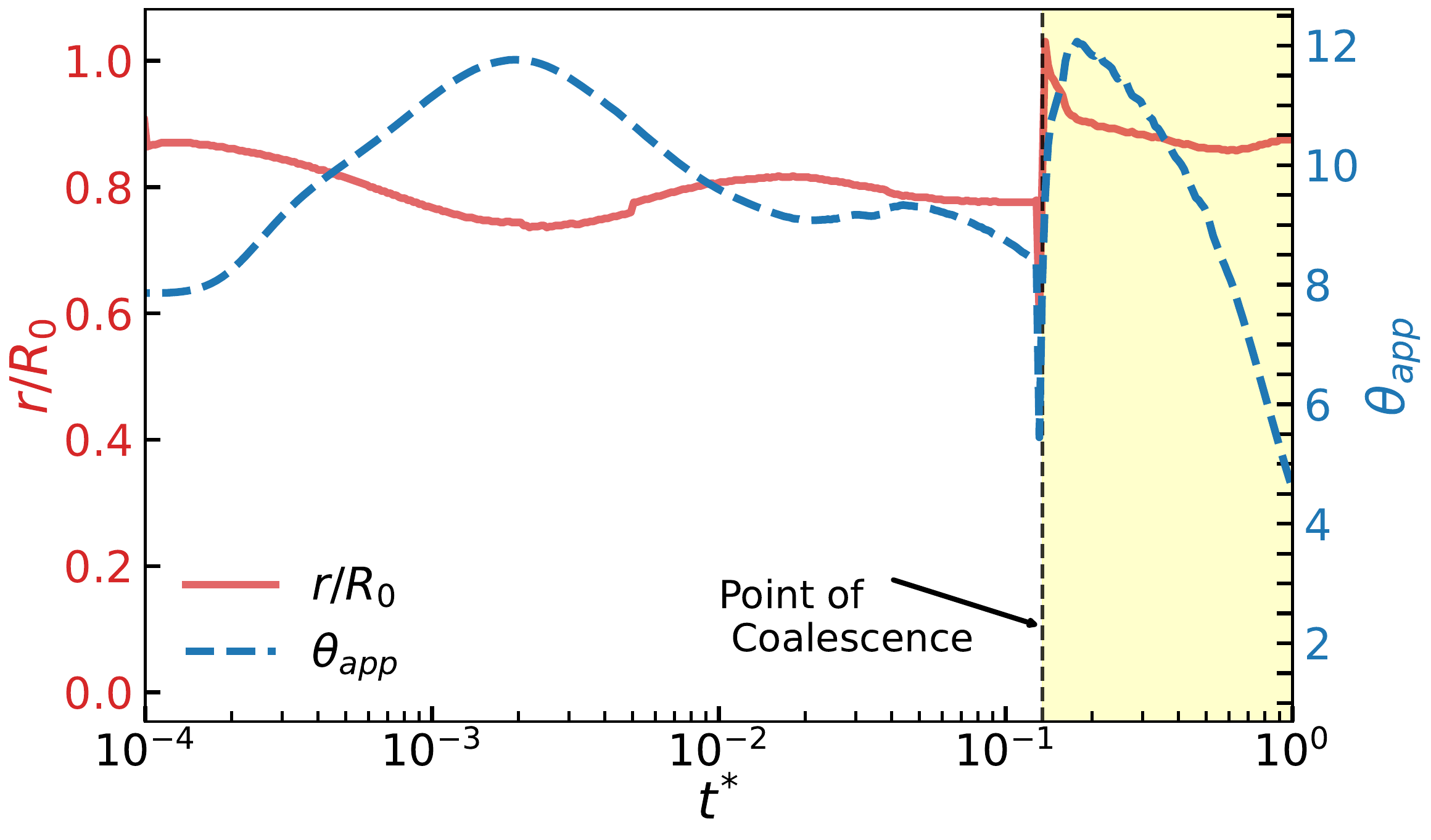} 
          \\\footnotesize(b)
    \end{minipage}
    \begin{minipage}{0.48\textwidth}
    \centering
          \includegraphics[width=6cm]{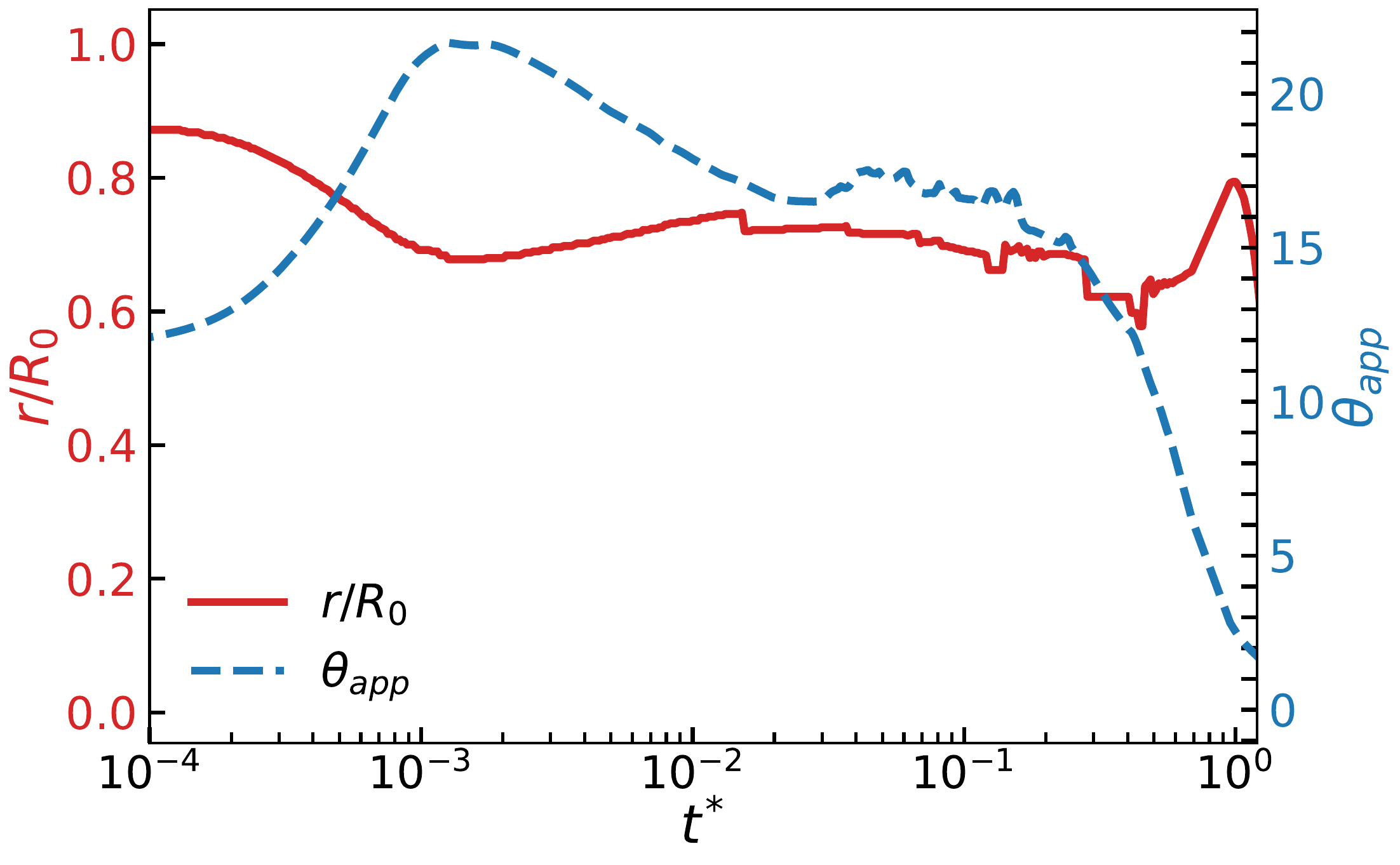} 
          \\\footnotesize(c)
    \end{minipage}
    \caption{Spreading radius $(r/R_0)$ and apparent contact angle ($\theta_{app}$) variation with time $t^{*}$ for $50\%$ water-morpholine droplets: (a) in isolation, (b) as a pair for $Ma = 0.01$ (attraction and coalescence) and (c) as a pair for $Ma = 0.08$ (no coalescence). }
    \label{F.E2}
\end{figure}

The modes of evaporation for multiple droplets evaporating simultaneously are investigated compared to when they evaporate in isolation. Figure \ref{f.E1}a represents the contact radius ($r/R_0$) and apparent contact angle evolution ($\theta_{app}$) with time for a pure droplet evaporating in isolation. We observe from figure~\ref{f.E1} that after initial rapid spreading, the droplet evaporates in constant contact angle mode (CCA) until $t^* = 0.7$. Afterwards, both the contact angle and the contact radius change until the drop is fully evaporated. For two pure droplets placed at a distance of $6R_0$ from each other, the time evolution of contact radius and contact angle of one of the droplets is reported in figure~\ref{f.E1}b. We note that the droplets exhibit similar behaviour, spreading initially and then evaporating in a constant contact angle (CCA) mode for the majority of the droplet lifetime.

In the case of binary mixture droplets, a water-morpholine droplet in isolation maintains a constant contact radius (CCR) mode of evaporation up to about $t^* = 0.7$, after the initial retraction of the contact line at early times (figure~\ref{F.E2}a). The same water-morpholine droplets evaporating in pairs show attraction and coalescence, as already discussed in the paper. In figure~\ref{F.E2}b, we observe that up to the coalescence time of the drop at $t^* =0.2$, each droplet follows a constant contact radius (CCR) evaporation mode. Post-coalescence, the behaviour remains similar until the droplet's lifetime ends in isolation. We also investigated the evaporation mode for the droplet pair evaporation at a higher $Ma$ of $0.07$, where droplets do not coalesce (figure~\ref{F.E2}c). Here, we observe a similar CCR evaporation mode for the droplets as well. At higher $Ma$, the main change is that the droplet footprints retract more, increasing the initial $\theta_{\rm{app}}$. Nonetheless, the evaporation mode remains CCR following the initial contraction until the water in the mixture is fully evaporated.

In the case of binary mixture droplets, a water-morpholine droplet in isolation maintains a constant contact radius (CCR) mode of evaporation up to about $t^* = 0.7$, after the initial retraction of the contact line at early times (figure~\ref{F.E2}a). The same water-morpholine droplets evaporating in pairs show attraction and coalescence, as already discussed in the paper. In figure~\ref{F.E2}b, we observe that up to the coalescence time of the drop at $t^* =0.2$, each droplet follows a constant contact radius (CCR) evaporation mode. Post-coalescence, the behaviour remains similar until the droplet's lifetime ends in isolation. We also investigated the evaporation mode for the droplet pair evaporation at a higher $Ma$ of $0.07$, where droplets do not coalesce (figure~\ref{F.E2}c). Here, we observe a similar CCR evaporation mode for the droplets as well. At higher $Ma$, the main change is that the droplet footprints retract more, increasing the initial $\theta_{\rm{app}}$. Nonetheless, the evaporation mode remains CCR following the initial contraction until the water in the mixture is fully evaporated.

\section{Rapid spreading of ethanol-water drops} 
\label{appC}

\begin{figure}
    \centering
    \begin{subfigure}{0.45\textwidth}
        \centering
        \includegraphics[width=\linewidth]{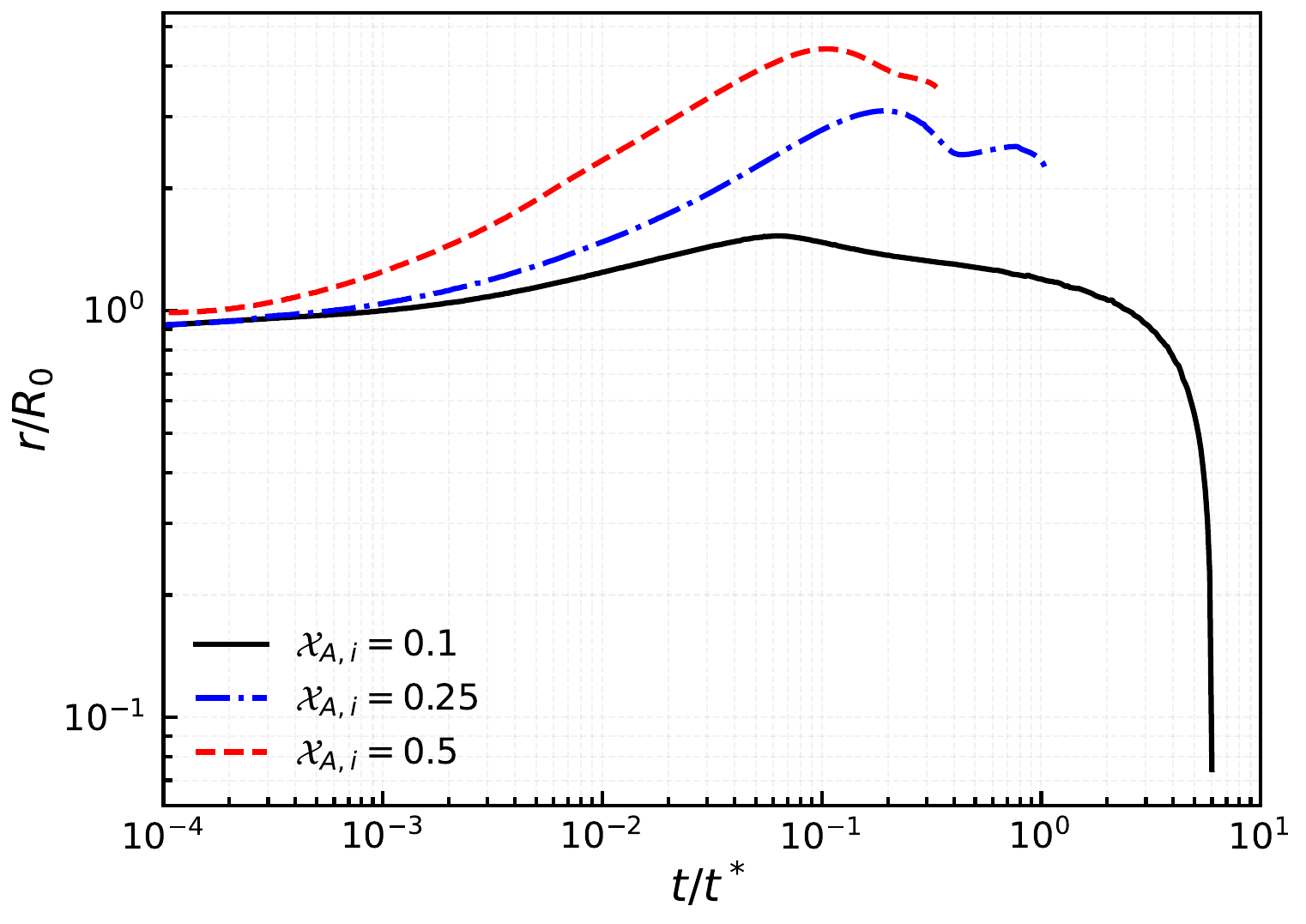}\\
        \footnotesize(a)
    \end{subfigure}\hfill
    \begin{subfigure}{0.45\textwidth}
        \centering
        \includegraphics[width=\linewidth]{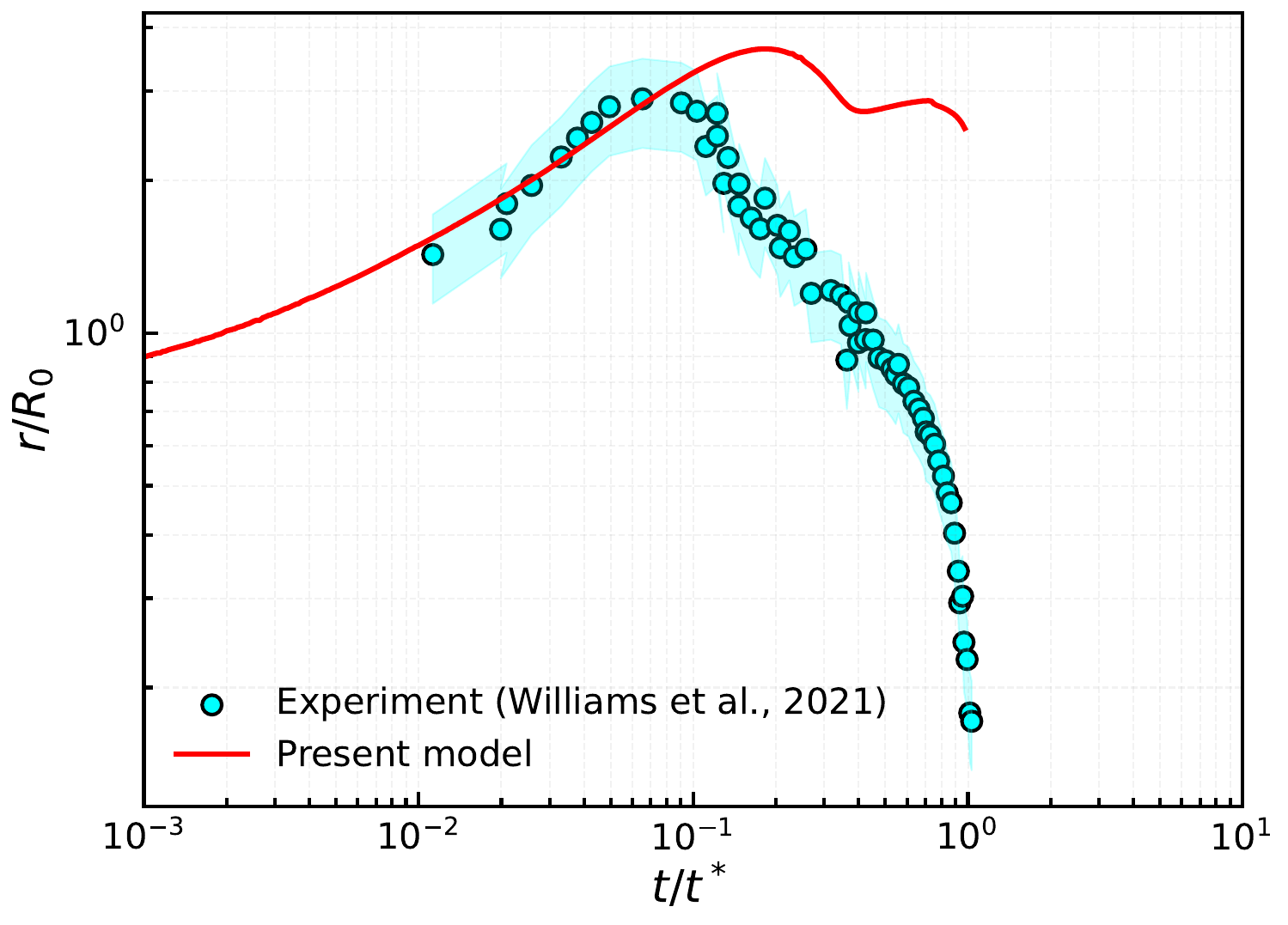}\\
        \footnotesize(b)
    \end{subfigure}
    \caption{(a) Contact radius evolution with time ($t^{*}$ is the time of evaporation for the fastest evaporating droplet) of an isolated ethanol-water droplet for varying initial ethanol concentration, (b) Comparison of contact radius evolution with the experiments of \cite{Williams2021} for a $25 \%$ ethanol-water droplet deposited on a $70^{\circ} C$ glass substrate. The non-dimensional parameters for this physical scenario are $Kn = 8.84 \times 10^{-4} , E= 2.66 \times 10^{-4}, \delta = 10^{-5}, \psi =0.1, Pe_{v,A} =0.01, Pe_{v,B} =0.1, Pe =20 , Ma = 0.16, \mathcal{H} = 0.3$.}
    \label{fig:20}
\end{figure}

\begin{table}
    \centering
    \begin{tabular}{ccccccc}
    &\multicolumn{2}{c}{$\mathcal{X}_{A,i} =0.11$} &
    \multicolumn{2}{c}{$\mathcal{X}_{A,i} =0.25$} &
    \multicolumn{2}{c}{$\mathcal{X}_{A,i} =0.5$} \\
    \hline
     & \makecell{Experiment} & Present model & \makecell{Experiment } & Present model &\makecell{Experiment} & Present model \\
    $n_1$ & $0.54$ &$0.5$&$1.15$& $0.9$ &$1.36$ & $1.6$ \\
    $n_2$ & $0.3$& $0.15$ & $0.45$ & $0.5$ & $0.59$ & $0.7$  \\
    \end{tabular}
    \caption{Comparison between experiments of \cite{Williams2021} and the present model for spreading exponent of an isolated ethanol-water droplet of various concentrations deposited on a $70^{\circ}$ substrate. Model parameters for the cases comapred are $Kn = 8.84 \times 10^{-4} , E= 2.66 \times 10^{-4}, \delta = 10^{-5}, \psi =0.1, Pe_{v,A} =0.01, Pe_{v,B} =0.1, Pe =20 , Ma = 0.16, \mathcal{H} = 0.3$}
    \label{tab6}
\end{table}

Ethanol-water droplets show a super-spreading behaviour when deposited on a heated surface, as depicted in figure~\ref{fig6}b. Here, we investigate the spreading behaviour of the droplet by calculating and comparing its spreading exponents with the experiments of \cite{Williams2021}. Figure \ref{fig:20}a illustrates the evolution of the contact radius for ethanol-water droplets with varying initial ethanol concentrations ($10, 25,$ and $50\%$). The droplets spread more rapidly as the initial concentration of ethanol increases, attributed to
the enhanced influence of solutal Marangoni. Figure \ref{fig:20}b shows the comparison of contact radius evolution between the present model and the experiment of \cite{Williams2021}. 
Here, the time is nondimensionalised with the evaporation time of the droplet in experiment. The present model found to be very accurate for the initial spreading of the droplets, however, it over predicts the spreading radius  as well as the evaporation time of the droplet. This prediction is much closer (in time scale) than the one-sided model presented in \cite{Williams2021}. Further, a comparison of the spreading exponents is shown in table \ref{tab6}. Here, $n_1$ and $n_2$ are two distinct spreading exponents $\left( \frac{r}{R_0} \sim \left( \frac{t}{t^*}\right)^{n_i (i =1,2)}\right)$ observed in the spreading phase of the droplets. We can observe very close agreement of these exponents with experiments that solidify the accuracy of our model.

\section{Contact line singularities for a binary evaporating droplet}
In this appendix, we perform a complete local 
asymptotic analysis of all singularities at 
the contact line of the binary evaporating 
droplet considered in the present work. The 
domain is $0\leq x\leq L$, the droplet is 
centered at $x_1 = L/2$ with contact radius 
$\ell_0$, and the interface is described by 
a fourth-order polynomial profile symmetric 
about $x = L/2$. We have performed this analysis for a single droplet. The composition field is 
represented by a parabolic-in-$z$ profile 
consistent with the lubrication approximation. 
The model applies a single Robin boundary 
condition uniformly across the entire 
interface -- including both droplet and 
precursor film regions -- and the near-zero 
evaporation in the precursor film emerges 
naturally from the large disjoining pressure 
in the thin precursor layer. The analysis 
follows \citet{Saxton2017} and 
\citet{Colinet2011}, adapted to this 
physically consistent formulation.

The domain is $0\leq x\leq L$. The droplet 
is centred at $x_1 = L/2$ and is symmetric 
about this point. The contact radius is 
$\ell_0$, so the contact lines are at:
\begin{equation}
    x_{cl}^L = \frac{L}{2} - \ell_0 
    \quad\text{(left)}, \qquad
    x_{cl}^R = \frac{L}{2} + \ell_0 
    \quad\text{(right)}.
    \label{eq:contact_lines}
\end{equation}
The precursor film occupies 
$0\leq x < L/2-\ell_0$ and 
$L/2+\ell_0 < x\leq L$, and the droplet 
occupies $L/2-\ell_0 < x < L/2+\ell_0$.

The interface height is taken as a 
fourth-order polynomial:
\begin{align}
    h(x) &= 
    \frac{15(V_0-\beta\ell_0)}{8\ell_0^5}
    x^4
    - \frac{15(L/2)(V_0-\beta\ell_0)}
    {2\ell_0^5}x^3 
    + \frac{(45(L/2)^2-15\ell_0^2)
    (V_0-\beta\ell_0)}{4\ell_0^5}x^2
    \notag\\
    &\quad
    + \frac{(-15(L/2)^3+15(L/2)\ell_0^2)
    (V_0-\beta\ell_0)}{2\ell_0^5}x
    + \frac{15((L/2)^2-\ell_0^2)^2
    (V_0-\beta\ell_0)}{8\ell_0^5} + \beta,
    \label{eq:quartic}
\end{align}
where $\beta\geq 0$ is the precursor film 
thickness and $V_0 = 2/3$. With 
$\mathscr{A} = 15(V_0-\beta\ell_0)/\ell_0^5$, 
the factored form is:
\begin{equation}
    h = \mathscr{A}(\ell_0^2-\xi^2)^2,
\end{equation}
where $\xi$ denotes the distance from the droplet centre, has the following leading-order behaviour near the contact line at $\xi=\ell_0$. Introducing $s=\ell_0-\xi\geq 0$ as the distance from the contact line into the droplet gives
\begin{equation}
    h \sim 4\mathscr{A}\ell_0^2 s^2
    \equiv Cs^2,
    \qquad
    C = 4\mathscr{A}\ell_0^2 .
    \label{eq:h_quartic}
\end{equation}
Thus, the quartic profile meets the substrate tangentially, with zero local slope. In what follows, the same symbol $\xi$ is used for the inner coordinate measured from the contact line. The corresponding curvature at the contact line is
\begin{equation}
    \kappa_{CL}
    =
    -\left.
    \frac{\partial^2 h}{\partial \xi^2}
    \right|_{\xi=0}
    =
    -8\mathscr{A}\ell_0^2 < 0 .
    \label{eq:kappa_CL}
\end{equation}

The liquid pressure is given by
\begin{equation}
    p =
    -\frac{\varepsilon^2}{Ma}\,2\sigma\kappa
    - \Pi(h),
    \qquad
    \Pi(h)
    =
    \mathcal{A}
    \left[
    \left(\frac{\mathcal{B}}{h}\right)^m
    -
    \left(\frac{\mathcal{B}}{h}\right)^n
    \right],
    \label{eq:pressure}
\end{equation}
with $m>n>0$. The equilibrium vapour densities are
\begin{equation}
    \rho^{ve,A}
    =
    1+\delta p+\psi T_{int},
    \qquad
    \rho^{ve,B}
    =
    1+M_R\delta p+\mathcal{L}M_R\psi T_{int},
    \label{eq:rhove}
\end{equation}
where $p$ is the non-dimensional liquid pressure. The evaporative fluxes satisfy the kinetic Robin condition
\begin{equation}
    J_i = \Lambda_i G^i,
    \qquad
    G^i =
    \mathcal{X}_A^{int}\rho^{ve,i}
    -
    \rho^{v,i}, \qquad \Lambda_i  = \frac{Pe_{v,i}}{Kn},
\end{equation}
The parabolic composition assumption gives the exact implicit relation
\begin{equation}
    \mathcal{X}_{A1}
    =
    \frac{
    \tfrac{1}{2}\mathcal{N} h\,\Delta C
    }{
    1-\tfrac{1}{3}\mathcal{N} Jh
    },
    \qquad
    \Delta C =
    \mathcal{X}_{A0}J-J_A .
    \label{eq:XA1}
\end{equation}
In the present model, $\mathcal{N} Jh\sim 10^{-1}$--$10^{-4}$, so that the thin-film approximation
\begin{equation}
    \mathcal{X}_{A1}
    \approx
    \tfrac{1}{2} \mathcal{N} h\,\Delta C
\end{equation}
is uniformly valid.

For the model used in this work, the precursor film has thickness $\beta>0$ and the disjoining pressure is retained. In the flat precursor region, $h=\beta$ and $\kappa=0$, so that the pressure reduces to
\begin{equation}
     p_0
    =
    -\Pi(\beta)
    =
    -\mathcal{A}
    \left[
    \left(\frac{\mathcal{B}}{\beta}\right)^m
    -
    \left(\frac{\mathcal{B}}{\beta}\right)^n
    \right].
    \label{eq:dp_precursor}
\end{equation}
The capillary contribution vanishes because the precursor film is locally flat. The dominant disjoining-pressure contribution,
$-\mathcal{A}(\mathcal{B}/\beta)^m<0$, lowers the equilibrium vapour density below unity and thereby naturally suppresses evaporation in the thin precursor film, without imposing an additional cut-off. For fixed $\beta>0$, all terms in \eqref{eq:dp_precursor} are finite constants. Consequently,
\begin{align}
    \rho^{ve,A}_0
    &=
    1
    -
    \mathcal{A}
    \left(\frac{\mathcal{B}}{\beta}\right)^m
    +
    \mathcal{A}
    \left(\frac{\mathcal{B}}{\beta}\right)^n
    +
    \psi T_{int,0},
    \label{eq:rhoA0}
    \\
    \rho^{ve,B}_0
    &=
    1
    -
    M_R\mathcal{A}
    \left(\frac{\mathcal{B}}{\beta}\right)^m
    +
    M_R\mathcal{A}
    \left(\frac{\mathcal{B}}{\beta}\right)^n
    +
    \mathcal{L}M_R\psi T_{int,0}.
    \label{eq:rhoB0}
\end{align}
It follows that
\begin{equation}
    G^i_0 =
    \mathcal{X}_{A,0}^{int}\rho^{ve,i}_0
    -
    \rho^{v,i}_0
\end{equation}
is finite. The kinetic Robin condition therefore gives
\begin{equation}
    J_i \to J_i^0
    =
    \Lambda_i G^i_0
    =
    \text{finite}.
    \label{eq:J_finite}
\end{equation}
This is consistent with Section~3.2, eqs.~(22)--(24) of \citet{Saxton2017}, who showed that a Robin boundary condition produces a finite evaporation rate at the contact line, in contrast to the $\xi^{-1/2}$ singularity obtained in the lens model with a Dirichlet condition. The near-contact-line correction satisfies
\begin{equation}
    J_i(\xi)-J_i^0
    =
    O(\xi\log\xi),
    \qquad
    \frac{\mathrm{d}J_i}{\mathrm{d}\xi}
    =
    O(\log\xi),
    \label{eq:dJ_log}
\end{equation}
so that the derivative is only logarithmically singular and remains integrable.

The viscous stress is regularised by the finite precursor thickness. Since $h\to\beta>0$,
\begin{equation}
    \tau_{visc}
    =
    \frac{3\mu q}{h^2}
    \to
    \frac{3\mu q}{\beta^2}
    =
    \text{finite}.
\end{equation}
Thus, the precursor film removes the viscous contact-line singularity by preventing the local film height from vanishing.

For $Bi=0$, the interfacial temperature is
\begin{equation}
    T_{int}
    =
    -\frac{(J_A+\mathcal{L}J_B)h}{k}
    + T_s .
\end{equation}
Using $h\to\beta$ and $J_i\to J_i^0$ gives
\begin{equation}
    T_{int}
    \to
    T_{int}^{CL}
    =
    -\frac{(J_A^0+\mathcal{L}J_B^0)\beta}{k}
    + T_s
    =
    \text{finite}.
\end{equation}
The gradient $\partial T_{int}/\partial \xi$ receives contributions from both $\partial J_i/\partial \xi=O(\log\xi)$ and $\partial h/\partial \xi=O(\xi)$. The logarithmic contribution from the flux derivative is dominant, giving
\begin{equation}
    \tau_{th}
    =
    \gamma_T
    \frac{\partial T_{int}}{\partial \xi}
    =
    O\left(
    \frac{\beta}{k}\log\xi
    \right).
    \label{eq:tau_th_precursor}
\end{equation}
This singularity is integrable, since
\begin{equation}
    \int_0^\varepsilon |\log\xi|\,\mathrm{d}\xi
    =
    \varepsilon(1-\log\varepsilon)
    \to 0
    \qquad
    \text{as }
    \varepsilon\to 0 .
\end{equation}

The depth-averaged composition remains regular at the contact line. With $h\to\beta$ and $J_i\to J_i^0$, the depth-integrated composition equation reduces locally to a balance between advection and the evaporation source,
\begin{equation}
    q
    \frac{\partial \mathcal{X}_{A0}}{\partial \xi}
    \approx
    -\Delta C_0,
\end{equation}
where $\Delta C_0$ is finite. The diffusive contribution,
$\partial_\xi(h\mathcal{D}_A\partial_\xi \mathcal{X}_{A0})$, is subdominant in this limit. Hence
\begin{equation}
    \mathcal{X}_{A0}
    =
    \mathcal{X}_{A,0}^{int}
    +
    a_1\xi
    +
    O(\xi^2\log\xi),
    \qquad
    a_1
    =
    -\frac{\Delta C_0}{Q}.
    \label{eq:XA0_Taylor}
\end{equation}
The leading composition gradient is therefore finite. The parabolic correction also remains bounded, since
\begin{equation}
    \mathcal{X}_{A1}
    \approx
    \tfrac{1}{2}\mathcal{N}\beta\Delta C_0
    =
    \text{finite},
    \qquad
    \frac{\partial \mathcal{X}_{A1}}{\partial \xi}
    =
    O(\mathcal{N}\log\xi),
\end{equation}
with this correction being $O(\varepsilon^2)$ smaller. The total solutal Marangoni stress is therefore
\begin{equation}
    \tau_{sol}
    =
    S_\sigma
    \frac{\partial}{\partial \xi}
    \left(
    \mathcal{X}_{A0}
    +
    \tfrac{2}{3}\mathcal{X}_{A1}
    \right)
    \to
    S_\sigma a_1
    =
    \text{finite}.
    \label{eq:tau_sol_precursor}
\end{equation}
For a single-component droplet, $\Delta C_0=0$ and hence $\tau_{sol}=0$. For a binary droplet with differential volatility (all the compositions considered in the present model), $\Delta C_0\neq 0$, so that the solutal Marangoni stress is generally nonzero. Importantly, however, the binary nature of the droplet produces a finite contact-line solutal stress, not a divergent singularity. The precursor film is therefore not required for the solutal Marangoni stress to remain finite; its essential role is the regularisation of the viscous stress.

We next consider the hypothetical case of a plain binary droplet without a precursor film and without disjoining pressure, corresponding to $\beta=0$ and $\Pi=0$. This case is not the model used in the present work, but it is useful for identifying whether the binary nature of the droplet introduces any additional contact-line singularity. With $\Pi=0$, the pressure reduces to
\begin{equation}
    p
    =
    -\frac{\varepsilon^2}{Ma}\,2\sigma\kappa .
\end{equation}
For the quartic profile, $h\sim C\xi^2$ and the curvature is approximately constant near the contact line, $\kappa\approx -2C$. Hence
\begin{equation}
     p
    \approx
    -\frac{\varepsilon^2}{Ma}\,2\sigma(-2C)
    =
    \frac{4\varepsilon^2 C\sigma}{Ma}
    =
    O(1).
    \label{eq:dp_noPi}
\end{equation}
Since $\varepsilon^2/Ma=O(1)$, this capillary pressure contribution is finite and non-negligible. The equilibrium vapour densities remain bounded, so that $G^i_0$ is finite and
\begin{equation}
    J_i
    \to
    \Lambda_i G^i_0
    =
    \text{finite}
    \qquad
    (\Pi=0,\ \beta=0).
    \label{eq:J_noPi}
\end{equation}
Thus, the Robin boundary condition regularises the evaporative flux independently of the precursor film, provided that the pressure, and hence the Kelvin correction $\delta p$, remains bounded. If the disjoining pressure were retained while the precursor film were removed, then $\Pi(h)$ would diverge as $h\to 0$, making the model ill-posed in this limit. In the present hypothetical case, however, $\Pi=0$ and the flux remains finite.

The viscous stress is not regularised in the absence of a precursor film. Since $h\sim C\xi^2$,
\begin{equation}
    \tau_{visc}
    =
    \frac{3\mu q}{h^2}
    \sim
    \frac{3\mu q}{C^2\xi^4}
    \to
    \infty .
    \label{eq:tau_visc_noPi}
\end{equation}
This $\xi^{-4}$ singularity is non-integrable. It is present for both single-component and binary droplets, because it arises solely from the vanishing film height and is independent of the number of volatile components.

The thermal Marangoni stress, by contrast, vanishes at the contact line in this case. With $J_i\to J_i^0$ finite and $h\sim C\xi^2$, the interfacial temperature satisfies
\begin{equation}
    T_{int}
    =
    -\frac{(J_A+\mathcal{L}J_B)h}{k}
    +
    T_s
    \sim
    -\frac{(J_A^0+\mathcal{L}J_B^0)C\xi^2}{k}
    +
    T_s
    \to
    T_s .
\end{equation}
Therefore,
\begin{equation}
    \tau_{th}
    =
    \gamma_T
    \frac{\partial T_{int}}{\partial \xi}
    \sim
    -\frac{
    2\gamma_T(J_A^0+\mathcal{L}J_B^0)C
    }{k}\,\xi
    \to
    0 .
    \label{eq:tau_th_noPi}
\end{equation}
This result is a direct consequence of the quartic contact-line geometry. Since $h\sim \xi^2$, the departure of $T_{int}$ from $T_s$ vanishes quadratically, while the flux remains finite. This differs from the precursor-film case, where $h\to\beta>0$ and the thermal Marangoni stress is logarithmically singular but integrable.

The solutal Marangoni stress also remains finite without the precursor film. With $J_i\to J_i^0$ finite and $h\sim C\xi^2$, the leading near-contact-line balance in the depth-integrated composition equation is
\begin{equation}
    Q
    \frac{\partial \mathcal{X}_{A0}}{\partial \xi}
    -
    \frac{\partial}{\partial \xi}
    \left(
    C\xi^2 \mathcal{D}_A
    \frac{\partial \mathcal{X}_{A0}}{\partial \xi}
    \right)
    =
    -\Delta C_0 .
\end{equation}
The diffusive term vanishes as $\xi\to 0^+$ because $h\sim C\xi^2\to 0$. The leading-order balance is therefore
\begin{equation}
    q
    \frac{\partial \mathcal{X}_{A0}}{\partial \xi}
    =
    -\Delta C_0,
    \qquad
    \frac{\partial \mathcal{X}_{A0}}{\partial \xi}
    \to
    -\frac{\Delta C_0}{Q}
    =
    A_1
    =
    \text{finite}.
    \label{eq:XA0_noPi}
\end{equation}
The parabolic correction satisfies
\begin{equation}
    \mathcal{X}_{A1}
    \approx
    \tfrac{1}{2}\mathcal{N} h\,\Delta C_0
    \sim
    \tfrac{1}{2}\mathcal{N} C\xi^2\Delta C_0
    \to 0,
\end{equation}
and its derivative is subdominant:
\begin{equation}
    \frac{2}{3}
    \frac{\partial \mathcal{X}_{A1}}{\partial \xi}
    =
    O(\mathcal{N} \xi)
    \to 0 .
\end{equation}
Thus,
\begin{equation}
    \tau_{sol}
    \to
    S_\sigma A_1
    =
    -\frac{S_\sigma\Delta C_0}{Q}
    =
    \text{finite}.
    \label{eq:tau_sol_noPi}
\end{equation}
For a single-component droplet, $\Delta C_0=0$ and the solutal Marangoni stress vanishes. For a binary droplet, $\Delta C_0\neq 0$ in general, and the solutal Marangoni stress is finite and nonzero. The multicomponent nature of the droplet therefore introduces a finite solutal Marangoni stress at the contact line, but it does not generate an additional divergent or non-integrable singularity.

In summary, the kinetic Robin condition regularises the evaporative flux in both cases considered here, provided that the pressure remains bounded. In the plain binary droplet without disjoining pressure, the only non-integrable singularity is the viscous stress,
$\tau_{visc}\sim \xi^{-4}$, which is also present for a single-component droplet and is caused by the vanishing film height. The precursor film regularises this singularity by maintaining $h=\beta>0$ at the contact line. The thermal Marangoni stress vanishes in the plain-droplet case because the quartic geometry gives $h\sim \xi^2$, whereas it is logarithmically singular but integrable in the precursor-film case. The solutal Marangoni stress is finite in both cases: it is zero for a single-component droplet and generally nonzero for a binary droplet, but it does not constitute an additional contact-line singularity. It should also be emphasized that in a purely diffusion-controlled scenario (lens model) with no kinetic contributions, corresponding to $Pe_v \to \infty$, the evaporation flux still exhibits a singularity \citep{Colinet2011} when no precursor film is present for solutal Marangoni effects in binary mixture droplets ($\tau_{sol} \sim \xi^{-1/2}$).


\bibliographystyle{jfm}

\bibliography{JFM_initials_corrected}

@Article{azzam_modeling_2024,
	author  = {Azzam, A. and Kempers, R. and Amirfazli, A.},
	title   = {Modeling of the evaporation process of a pair of sessile droplets using a point source model (PSM)},
	journal = {International Communications in Heat and Mass Transfer},
	year    = {2024},
	volume  = {157},
	pages   = {107733},
	doi     = {10.1016/j.icheatmasstransfer.2024.107733},
}

@article{Saxton2017,
  author  = {Saxton, M. A. and Vella, D. and Whiteley, J. P. and Oliver, J. M.},
  title   = {Kinetic effects regularize the 
             mass-flux singularity at the contact 
             line of a thin evaporating drop},
  journal = {Journal of Engineering Mathematics},
  volume  = {106},
  number  = {1},
  pages   = {47--73},
  year    = {2017},
  doi     = {10.1007/s10665-016-9892-4}
}

@article{Colinet2011,
  author  = {Colinet, P. and Rednikov, A.},
  title   = {On integrable singularities and 
             apparent contact angles within a 
             classical paradigm},
  journal = {European Physical Journal 
             Special Topics},
  volume  = {197},
  pages   = {89--113},
  year    = {2011},
  doi     = {10.1140/epjst/e2011-01443-x}
}

@article{haut2005surface,
  title={Surface-tension-driven instabilities of a pure liquid layer evaporating into an inert gas},
  author={Haut, B. and Colinet, P.},
  journal={Journal of colloid and interface science},
  volume={285},
  number={1},
  pages={296--305},
  year={2005},
  publisher={Elsevier}
}

@article{SULTAN2005,
	title = {Evaporation of a thin film: diffusion of the vapour and {Marangoni} instabilities},
	volume = {543},
	issn = {0022-1120},
	url = {http://www.journals.cambridge.org/abstract_S0022112005006348},
	doi = {10.1017/S0022112005006348},
	number = {-1},
	urldate = {2022-11-21},
	journal = {Journal of Fluid Mechanics},
	author = {Sultan, E. and Boudaoud, A. and Ben Amar, M.},
	month = nov,
	year = {2005},
	note = {Publisher: Cambridge University Press},
	pages = {183},
}

@article{oron1997long,
  title={Long-scale evolution of thin liquid films},
  author={Oron, A. and Davis, S. H. and Bankoff, S. G.},
  journal={Reviews of modern physics},
  volume={69},
  number={3},
  pages={931},
  year={1997},
  publisher={APS}
}

@Article{zhao_long-range_2025,
	author  = {Zhao, H. and Orejon, D. and Sefiane, K. and Shanahan, M. E. R.},
	title   = {Long-Range Vapor-Mediated Interactions between Adjacent Droplets},
	journal = {Langmuir},
	year    = {2025},
	volume  = {41},
	number  = {6},
	pages   = {3986--3994},
	doi     = {10.1021/acs.langmuir.4c04255},
    note = {Publisher: American Chemical Society},

}

@Article{malachtari_dynamics_2024,
	author  = {Malachtari, A. and Karapetsas, G.},
	title   = {Dynamics of the interaction of a pair of thin evaporating droplets on compliant substrates},
	journal = {Journal of Fluid Mechanics},
	year    = {2024},
	volume  = {978},
	pages   = {A8},
	doi     = {10.1017/jfm.2023.855},

}

@article{pham2017drying,
  title={Drying of droplets of colloidal suspensions on rough substrates},
  author={Pham, T. and Kumar, S.},
  journal={Langmuir},
  volume={33},
  number={38},
  pages={10061--10076},
  year={2017},
  publisher={ACS Publications}
}

@article{deGennes1985,
  author  = {de Gennes, P. G.},
  title   = {Wetting: statics and dynamics},
  journal = {Reviews of Modern Physics},
  volume  = {57},
  number  = {3},
  pages   = {827--863},
  year    = {1985},
  doi     = {10.1103/RevModPhys.57.827}
}

@article{davis1974motion,
  title={On the motion of a fluid-fluid interface along a solid surface},
  author={Davis, S. H. and others},
  journal={Journal of Fluid Mechanics},
  volume={65},
  number={1},
  pages={71--95},
  year={1974},
  publisher={Cambridge University Press}
}

@article{huh1971hydrodynamic,
  title={Hydrodynamic model of steady movement of a solid/liquid/fluid contact line},
  author={Huh, C. and Scriven, L. E.},
  journal={Journal of colloid and interface science},
  volume={35},
  number={1},
  pages={85--101},
  year={1971},
  publisher={Elsevier}
}

@book{israelachvili2011intermolecular,
  title={Intermolecular and surface forces (3rd ed)},
  author={Israelachvili, J. N.},
  year={2011},
  publisher={Academic press}
}

@Article{malachtari_evaporation_2025,
	author  = {Malachtari, A. and Tsakelidis, I. and Karapetsas, G.},
	title   = {Evaporation of a thin particle-laden sessile droplet on a soft viscoelastic substrate},
	journal = {Phys. Rev. Fluids},
	year    = {2025},
	volume  = {10},
	number  = {6},
	pages   = {063601},
	doi     = {10.1103/PhysRevFluids.10.063601},

}

@misc{kavuri_evaporation-driven_2024,
	title = {Evaporation-driven coalescence of two droplets undergoing freezing},
	url = {http://arxiv.org/abs/2408.09827},
	doi = {10.48550/arXiv.2408.09827},
	urldate = {2024-12-19},
	publisher = {arXiv},
	author = {Kavuri, S. and Karapetsas, G. and Sharma, C. S. and Sahu, K. C.},
	month = nov,
	year = {2024},
	note = {arXiv:2408.09827 [physics]},
}

@article{tanner1979spreading,
  title={The spreading of silicone oil drops on horizontal surfaces},
  author={Tanner, Lawrence H},
  journal={Journal of Physics D: Applied Physics},
  volume={12},
  number={9},
  pages={1473--1484},
  year={1979}
}

@article{wilson2023evaporation,
  title={Evaporation of sessile droplets},
  author={Wilson, Stephen K and D'Ambrosio, Hannah-May},
  journal={Annual Review of Fluid Mechanics},
  volume={55},
  number={1},
  pages={481--509},
  year={2023},
  publisher={Annual Reviews}
}

@article{karapetsas_evaporation_2016,
	title = {Evaporation of {Sessile} {Droplets} {Laden} with {Particles} and {Insoluble} {Surfactants}},
	volume = {32},
	issn = {0743-7463},
	url = {https://doi.org/10.1021/acs.langmuir.6b01042},
	doi = {10.1021/acs.langmuir.6b01042},
	number = {27},
	urldate = {2024-11-20},
	journal = {Langmuir},
	author = {Karapetsas, G. and Sahu, K. C. and Matar, O. K.},
	month = jul,
	year = {2016},
	note = {Publisher: American Chemical Society},
	pages = {6871--6881},
}

@article{karapetsas_convective_2012,
	title = {Convective {Rolls} and {Hydrothermal} {Waves} in {Evaporating} {Sessile} {Drops}},
	volume = {28},
	issn = {0743-7463},
	url = {https://doi.org/10.1021/la3019088},
	doi = {10.1021/la3019088},
	number = {31},
	urldate = {2024-11-20},
	journal = {Langmuir},
	author = {Karapetsas, G. and Matar, O. K. and Valluri, P. and Sefiane, K.},
	month = aug,
	year = {2012},
	note = {Publisher: American Chemical Society},
	pages = {11433--11439},
}

@article{Hu_and_Larson_2005,
	title = {Analysis of the effects of marangoni stresses on the microflow in an evaporating sessile droplet},
	volume = {21},
	issn = {07437463},
	url = {https://pubs.acs.org/doi/full/10.1021/la0475270},
	doi = {10.1021/LA0475270/ASSET/IMAGES/MEDIUM/LA0475270E00037.GIF},
	number = {9},
	urldate = {2023-08-28},
	journal = {Langmuir},
	author = {Hu, H. and Larson, R. G.},
	month = apr,
	year = {2005},
	note = {Publisher:  American Chemical Society},
	pages = {3972--3980},
}

@article{wang_role_2024,
	title = {Role of volatility and thermal properties in droplet spreading: a generalisation to {Tanner}'s law},
	volume = {987},
	issn = {0022-1120, 1469-7645},
	shorttitle = {Role of volatility and thermal properties in droplet spreading},
	url = {https://www.cambridge.org/core/product/identifier/S0022112024003859/type/journal_article},
	doi = {10.1017/jfm.2024.385},
	language = {en},
	urldate = {2024-05-22},
	journal = {Journal of Fluid Mechanics},
	author = {Wang, Z. and Karapetsas, G. and Valluri, P. and Inoue, C.},
	month = may,
	year = {2024},
	pages = {A15},
}

@article{Wang2021,
	title = {Dynamics of hygroscopic aqueous solution droplets undergoing evaporation or vapour absorption},
	volume = {912},
	issn = {0022-1120},
	url = {https://www.cambridge.org/core/product/identifier/S0022112020010733/type/journal_article},
	doi = {10.1017/jfm.2020.1073},
	urldate = {2022-12-06},
	journal = {Journal of Fluid Mechanics},
	author = {Wang, Z. and Karapetsas, G. and Valluri, P. and Sefiane, K. and Williams, A. and Takata, Y.},
	month = apr,
	year = {2021},
	note = {Publisher: Cambridge University Press},
	pages = {A2},
}

@article{WANG2022,
	title = {Wetting and evaporation of multicomponent droplets},
	volume = {960},
	issn = {0370-1573},
	doi = {10.1016/J.PHYSREP.2022.02.005},
	urldate = {2022-10-05},
	journal = {Physics Reports},
	author = {Wang, Z. and Orejon, D. and Takata, Y. and Sefiane, K.},
	month = may,
	year = {2022},
	note = {Publisher: North-Holland},
	pages = {1--37},
}

@article{Diddens2021,
	title = {Competing {Marangoni} and {Rayleigh} convection in evaporating binary droplets},
	volume = {914},
	issn = {0022-1120},
	url = {https://www.cambridge.org/core/product/identifier/S002211202000734X/type/journal_article},
	doi = {10.1017/jfm.2020.734},
	urldate = {2023-03-12},
	journal = {Journal of Fluid Mechanics},
	author = {Diddens, C. and Li, Y. and Lohse, D.},
	month = may,
	year = {2021},
	note = {Publisher: Cambridge University Press},
	pages = {A23},
}

@article{baumgartner_marangoni_2022,
	title = {Marangoni spreading and contracting three-component droplets on completely wetting surfaces},
	volume = {119},
	issn = {10916490},
	url = {https://www.pnas.org/doi/abs/10.1073/pnas.2120432119},
	doi = {10.1073/PNAS.2120432119/SUPPL_FILE/PNAS.2120432119.SM05.MP4},
	number = {19},
	urldate = {2023-07-19},
	journal = {Proceedings of the National Academy of Sciences of the United States of America},
	author = {Baumgartner, D. A. and Shiri, S. and Sinha, S. and Karpitschka, S. and Cira, N. J.},
	month = may,
	year = {2022},
	pmid = {35507868},
	note = {Publisher: National Academy of Sciences},
	pages = {e2120432119},
}

@article{Williams2021,
	title = {Spreading and retraction dynamics of sessile evaporating droplets comprising volatile binary mixtures},
	volume = {907},
	issn = {0022-1120},
	url = {https://www.cambridge.org/core/product/identifier/S002211202000840X/type/journal_article},
	doi = {10.1017/jfm.2020.840},
	urldate = {2022-10-05},
	journal = {Journal of Fluid Mechanics},
	author = {Williams, A. G. L. and Karapetsas, G. and Mamalis, D. and Sefiane, K. and Matar, O. K. and Valluri, P.},
	month = jan,
	year = {2021},
	note = {Publisher: Cambridge University Press},
	pages = {A22},
}

@article{Pradhan2016,
	title = {Influence of an adjacent droplet on fluid convection inside an evaporating droplet of binary mixture},
	volume = {500},
	issn = {0927-7757},
	doi = {10.1016/J.COLSURFA.2016.03.073},
	urldate = {2022-10-13},
	journal = {Colloids and Surfaces A: Physicochemical and Engineering Aspects},
	author = {Pradhan, T. K. and Panigrahi, P. K.},
	month = jul,
	year = {2016},
	note = {Publisher: Elsevier},
	pages = {154--165},
}

@article{Man2017,
	title = {Vapor-{Induced} {Motion} of {Liquid} {Droplets} on an {Inert} {Substrate}},
	volume = {119},
	url = {https://journals.aps.org/prl/abstract/10.1103/PhysRevLett.119.044502},
	doi = {10.1103/PhysRevLett.119.044502},
	number = {4},
	urldate = {2023-03-01},
	journal = {Physical Review Letters},
	author = {Man, X. and Doi, M.},
	month = jul,
	year = {2017},
	note = {Publisher: American Physical Society},
	pages = {044502},
}

@article{Cira2015,
	title = {Vapour-mediated sensing and motility in two-component droplets},
	doi = {10.1038/nature14272},
	author = {Cira, N. J. and Benusiglio, A. and Prakash, M.},
	year = {2015},
}

@article{charlier_waterpropylene_2022,
	title = {Water–propylene glycol sessile droplet shapes and migration: {Marangoni} mixing and separation of scales},
	volume = {933},
	issn = {0022-1120},
	url = {https://www.cambridge.org/core/journals/journal-of-fluid-mechanics/article/waterpropylene-glycol-sessile-droplet-shapes-and-migration-marangoni-mixing-and-separation-of-scales/3CBA1AA51A606D084F3D3E69B6B2BA34},
	doi = {10.1017/JFM.2021.1030},
	urldate = {2023-08-26},
	journal = {Journal of Fluid Mechanics},
	author = {Charlier, J. and Rednikov, A. Y. and Dehaeck, S. and Colinet, P. and Terwagne, D.},
	month = feb,
	year = {2022},
	note = {Publisher: Cambridge University Press},
	pages = {A45},
}

@article{sadafi_vapor-mediated_2019,
	title = {Vapor-{Mediated} versus {Substrate}-{Mediated} {Interactions} between {Volatile} {Droplets}},
	issn = {15205827},
	url = {https://pubs.acs.org/doi/full/10.1021/acs.langmuir.9b00522},
	doi = {10.1021/ACS.LANGMUIR.9B00522/ASSET/IMAGES/MEDIUM/LA-2019-005222_M011.GIF},
	urldate = {2023-07-14},
	journal = {Langmuir},
	author = {Sadafi, H. and Dehaeck, S. and Rednikov, A. and Colinet, P.},
	year = {2019},
	pmid = {31050441},
	note = {Publisher: American Chemical Society},
}

@article{wen_vapor-induced_2019,
	title = {Vapor-induced motion of two pure liquid droplets},
	volume = {15},
	issn = {1744-6848},
	url = {https://pubs.rsc.org/en/content/articlehtml/2019/sm/c8sm02584c},
	doi = {10.1039/C8SM02584C},
	number = {10},
	urldate = {2023-07-14},
	journal = {Soft Matter},
	author = {Wen, Y. and Kim, P. Y. and Shi, S. and Wang, D. and Man, X. and Doi, M. and Russell, T. P.},
	month = mar,
	year = {2019},
	pmid = {30698600},
	note = {Publisher: The Royal Society of Chemistry},
	pages = {2135--2139},
}

@article{shahidzadeh-bonn_evaporating_2006,
	title = {Evaporating droplets},
	volume = {549},
	issn = {1469-7645},
	url = {https://www.cambridge.org/core/journals/journal-of-fluid-mechanics/article/evaporating-droplets/4C1F69A8743AF186A1BBE472ADB4BC49},
	doi = {10.1017/S0022112005008190},
	urldate = {2024-02-20},
	journal = {Journal of Fluid Mechanics},
	author = {Shahidzadeh-Bonn, N. and Rafaï, S. and Azouni, A. and Bonn, D.},
	year = {2006},
	note = {Publisher: Cambridge University Press},
	pages = {307--313},
}

@article{sefiane_formation_2010,
	title = {On the {Formation} of {Regular} {Patterns} from {Drying} {Droplets} and {Their} {Potential} {Use} for {Bio}-{Medical} {Applications}},
	volume = {7},
	issn = {1672-6529},
	doi = {10.1016/S1672-6529(09)60221-3},
	number = {SUPPL.},
	urldate = {2024-01-11},
	journal = {Journal of Bionic Engineering},
	author = {Sefiane, K.},
	month = sep,
	year = {2010},
	note = {Publisher: No longer published by Elsevier},
	pages = {S82--S93},
}

@article{karpitschka_marangoni_2017,
	title = {Marangoni {Contraction} of {Evaporating} {Sessile} {Droplets} of {Binary} {Mixtures}},
	volume = {33},
	issn = {15205827},
	url = {https://pubs.acs.org/doi/full/10.1021/acs.langmuir.7b00740},
	doi = {10.1021/ACS.LANGMUIR.7B00740/ASSET/IMAGES/LA-2017-007407_M013.GIF},
	number = {19},
	urldate = {2024-02-27},
	journal = {Langmuir},
	author = {Karpitschka, S. and Liebig, F. and Riegler, H.},
	month = may,
	year = {2017},
	pmid = {28421771},
	note = {Publisher: American Chemical Society},
	pages = {4682--4687},
}

@article{AJAEV2005,
	title = {Spreading of thin volatile liquid droplets on uniformly heated surfaces},
	volume = {528},
	issn = {0022-1120},
	url = {http://www.journals.cambridge.org/abstract_S0022112005003320},
	doi = {10.1017/S0022112005003320},
	urldate = {2022-10-05},
	journal = {Journal of Fluid Mechanics},
	author = {Ajaev, V. S.},
	month = apr,
	year = {2005},
	note = {Publisher: Cambridge University Press},
	pages = {279--296},
}

@article{iqtidar_drying_2023,
	title = {Drying dynamics of sessile-droplet arrays},
	volume = {8},
	issn = {2469990X},
	url = {https://journals.aps.org/prfluids/abstract/10.1103/PhysRevFluids.8.013602},
	doi = {10.1103/PHYSREVFLUIDS.8.013602/FIGURES/7/MEDIUM},
	number = {1},
	urldate = {2024-03-04},
	journal = {Physical Review Fluids},
	author = {Iqtidar, A. and Kilbride, J. J. and Ouali, F. F. and Fairhurst, D. J. and Stone, H. A. and Masoud, H.},
	month = jan,
	year = {2023},
	note = {Publisher: American Physical Society},
	pages = {013602},
}

@article{siregar_numerical_2013,
	title = {Numerical simulation of the drying of inkjet-printed droplets},
	volume = {392},
	issn = {0021-9797},
	doi = {10.1016/J.JCIS.2012.09.063},
	number = {1},
	urldate = {2023-11-19},
	journal = {Journal of Colloid and Interface Science},
	author = {Siregar, D. P. and Kuerten, J. G. M. and van der Geld, C. W. M.},
	month = feb,
	year = {2013},
	note = {Publisher: Academic Press},
	pages = {388--395},
}

@article{deegan_contact_2000,
	title = {Contact line deposits in an evaporating drop},
	volume = {62},
	issn = {1063651X},
	url = {https://journals.aps.org/pre/abstract/10.1103/PhysRevE.62.756},
	doi = {10.1103/PhysRevE.62.756},
	number = {1},
	urldate = {2023-08-30},
	journal = {Physical Review E},
	author = {Deegan, R. D. and Bakajin, O. and Dupont, T. F. and Huber, G. and Nagel, S. R. and Witten, T. A.},
	month = jul,
	year = {2000},
	pmid = {11088531},
	note = {Publisher: American Physical Society},
	pages = {756},
}

@article{fukatani_effect_2016,
	title = {Effect of ambient temperature and relative humidity on interfacial temperature during early stages of drop evaporation},
	volume = {93},
	issn = {24700053},
	url = {https://journals.aps.org/pre/abstract/10.1103/PhysRevE.93.043103},
	doi = {10.1103/PHYSREVE.93.043103/FIGURES/17/MEDIUM},
	number = {4},
	urldate = {2024-02-20},
	journal = {Physical Review E},
	author = {Fukatani, Y. and Orejon, D. and Kita, Y. and Takata, Y. and Kim, J. and Sefiane, K.},
	month = apr,
	year = {2016},
	note = {Publisher: American Physical Society},
	pages = {043103},
}

@article{hideo_nakae_effects_1998,
	title = {Effects of surface roughness on wettability},
	volume = {46},
	issn = {1359-6454},
	doi = {10.1016/S1359-6454(98)80012-8},
	number = {7},
	urldate = {2024-02-20},
	journal = {Acta Materialia},
	author = {{Hideo Nakae} and {Ryuichi Inui} and {Yosuke Hirata} and {Hiroyuki Saito}},
	month = apr,
	year = {1998},
	note = {Publisher: Pergamon},
	pages = {2313--2318},
}

@article{starov_evaporation_2009,
	title = {On evaporation rate and interfacial temperature of volatile sessile drops},
	volume = {333},
	issn = {0927-7757},
	doi = {10.1016/J.COLSURFA.2008.09.047},
	number = {1-3},
	urldate = {2023-11-19},
	journal = {Colloids and Surfaces A: Physicochemical and Engineering Aspects},
	author = {Starov, V. and Sefiane, K.},
	month = feb,
	year = {2009},
	note = {Publisher: Elsevier},
	pages = {170--174},
}

@article{sefiane_self-excited_2008,
	title = {Self-excited hydrothermal waves in evaporating sessile drops},
	volume = {93},
	issn = {00036951},
	url = {/aip/apl/article/93/7/074103/336088/Self-excited-hydrothermal-waves-in-evaporating},
	doi = {10.1063/1.2969072/336088},
	number = {7},
	urldate = {2023-11-19},
	journal = {Applied Physics Letters},
	author = {Sefiane, K. and Moffat, J. R. and Matar, O. K. and Craster, R. V.},
	month = aug,
	year = {2008},
	note = {Publisher: AIP Publishing},
	pages = {74103},
}

@article{Bennacer2014,
	title = {Vortices, dissipation and flow transition in volatile binary drops},
	volume = {749},
	url = {https://doi.org/10.1017/jfm.2014.220},
	doi = {10.1017/jfm.2014.220},
	journal = {J. Fluid Mech},
	author = {Bennacer, R. and Sefiane, K.},
	year = {2014},
	note = {Publisher: Cambridge University Press},
	pages = {649--665},
}

@article{dugas_droplet_2005,
	title = {Droplet evaporation study applied to {DNA} chip manufacturing},
	volume = {21},
	issn = {07437463},
	url = {https://pubs.acs.org/doi/full/10.1021/la050764y},
	doi = {10.1021/LA050764Y/ASSET/IMAGES/LARGE/LA050764YF00006.JPEG},
	number = {20},
	urldate = {2023-11-19},
	journal = {Langmuir},
	author = {Dugas, V. and Broutin, J. and Souteyrand, E.},
	month = sep,
	year = {2005},
	pmid = {16171342},
	note = {Publisher:  American Chemical Society},
	pages = {9130--9136},
}

@article{jing_automated_1998,
	title = {Automated high resolution optical mapping using arrayed, fluid-fixed {DNA} molecules},
	volume = {95},
	issn = {00278424},
	url = {https://www.pnas.org/doi/abs/10.1073/pnas.95.14.8046},
	doi = {10.1073/PNAS.95.14.8046/ASSET/D89FF7C0-77B1-4E72-BD13-B52DBE615DA7/ASSETS/GRAPHIC/PQ1381472004.JPEG},
	number = {14},
	urldate = {2023-11-19},
	journal = {Proceedings of the National Academy of Sciences of the United States of America},
	author = {Jing, J. and Reed, J. and Huang, J. and Hu, X. and Clarke, V. and Edington, J. and Housman, D. and Anantharaman, T. S. and Huff, E. J. and Mishra, B. and Porter, B. and Shenker, A. and Wolfson, E. and Hiort, C. and Kantor, R. and Aston, C. and Schwartz, D. C.},
	month = jul,
	year = {1998},
	pmid = {9653137},
	note = {Publisher: 
        The National Academy of Sciences},
	pages = {8046--8051},
}

@article{brutin_pattern_2011,
	title = {Pattern formation in drying drops of blood},
	volume = {667},
	issn = {1469-7645},
	url = {https://www.cambridge.org/core/journals/journal-of-fluid-mechanics/article/pattern-formation-in-drying-drops-of-blood/F79775CAE43EC2731B5141B40C86A5DE},
	doi = {10.1017/S0022112010005070},
	urldate = {2023-11-19},
	journal = {Journal of Fluid Mechanics},
	author = {Brutin, D. and Sobac, B. and Loquet, B. and Sampol, J.},
	month = jan,
	year = {2011},
	note = {arXiv: 1010.2510
Publisher: Cambridge University Press},
	pages = {85--95},
}

@article{kim_spray_2007,
	title = {Spray cooling heat transfer: {The} state of the art},
	volume = {28},
	issn = {0142-727X},
	doi = {10.1016/J.IJHEATFLUIDFLOW.2006.09.003},
	number = {4},
	urldate = {2023-11-19},
	journal = {International Journal of Heat and Fluid Flow},
	author = {Kim, J.},
	month = aug,
	year = {2007},
	note = {Publisher: Elsevier},
	pages = {753--767},
}

@article{bar-cohen_direct_2006,
	title = {Direct liquid cooling of high flux micro and nano electronic components},
	volume = {94},
	issn = {00189219},
	doi = {10.1109/JPROC.2006.879791},
	number = {8},
	urldate = {2023-11-19},
	journal = {Proceedings of the IEEE},
	author = {Bar-Cohen, A. and Arik, M. and Ohadi, M.},
	year = {2006},
	note = {Publisher: Institute of Electrical and Electronics Engineers Inc.},
	pages = {1549--1570},
}

@article{singh_inkjet_2010,
	title = {Inkjet {Printing}—{Process} and {Its} {Applications}},
	volume = {22},
	issn = {1521-4095},
	url = {https://onlinelibrary.wiley.com/doi/full/10.1002/adma.200901141},
	doi = {10.1002/ADMA.200901141},
	number = {6},
	urldate = {2023-11-19},
	journal = {Advanced Materials},
	author = {Singh, M. and Haverinen, H. M. and Dhagat, P. and Jabbour, G. E.},
	month = feb,
	year = {2010},
	pmid = {20217769},
	note = {Publisher: John Wiley \& Sons, Ltd},
	pages = {673--685},
}

@article{Matar2002,
	title = {A simple derivation of the time-dependent convective-diffusion equation for surfactant transport along a deforming interface},
	volume = {14},
	url = {https://doi.org/10.1063/1.1516597},
	doi = {10.1063/1.1516597},
	journal = {Physics of Fluids},
	author = {Matar, O. K.},
	year = {2002},
	pages = {111},
}

@article{Saenz2017,
	title = {Dynamics and universal scaling law in geometrically-controlled sessile drop evaporation},
	volume = {8},
	issn = {2041-1723},
	url = {https://www.nature.com/articles/ncomms14783},
	doi = {10.1038/ncomms14783},
	number = {1},
	urldate = {2023-03-03},
	journal = {Nature Communications 2017 8:1},
	author = {Sáenz, P. J. and Wray, A. W. and Che, Z. and Matar, O. K. and Valluri, P. and Kim, J. and Sefiane, K.},
	month = mar,
	year = {2017},
	pmid = {28294114},
	note = {Publisher: Nature Publishing Group},
	pages = {1--9},
}

@article{Charitatos2021,
	title = {Droplet evaporation on soft solid substrates},
	volume = {17},
	issn = {1744-6848},
	url = {https://pubs.rsc.org/en/content/articlehtml/2021/sm/d1sm00828e},
	doi = {10.1039/D1SM00828E},
	number = {41},
	urldate = {2023-04-19},
	journal = {Soft Matter},
	author = {Charitatos, V. and Kumar, S.},
	month = oct,
	year = {2021},
	pmid = {34596647},
	note = {Publisher: The Royal Society of Chemistry},
	pages = {9339--9352},
}

@article{Diddens2017,
	title = {Modeling the evaporation of sessile multi-component droplets},
	volume = {487},
	issn = {10957103},
	doi = {10.1016/j.jcis.2016.10.030},
	urldate = {2022-12-02},
	journal = {Journal of Colloid and Interface Science},
	author = {Diddens, C. and Kuerten, J. G. M. and van der Geld, C. W. M. and Wijshoff, H. M. A.},
	month = feb,
	year = {2017},
	pmid = {27810511},
	note = {Publisher: Academic Press},
	pages = {426--436},
}

@article{Diddens2017a,
	title = {Evaporating pure, binary and ternary droplets: thermal effects and axial symmetry breaking},
	volume = {823},
	url = {https://doi.org/10.1017/jfm.2017.312},
	doi = {10.1017/jfm.2017.312},
	journal = {J. Fluid Mech},
	author = {Diddens, C. and Tan, H. and Lv, P. and Versluis, M. and Kuerten, J. G. M. and Zhang, X. and Lohse, D.},
	year = {2017},
	note = {Publisher: Cambridge University Press},
	pages = {470--497},
}

@article{Hatte2019,
	title = {Universal evaporation dynamics of ordered arrays of sessile droplets},
	volume = {866},
	issn = {0022-1120},
	url = {https://www.cambridge.org/core/journals/journal-of-fluid-mechanics/article/universal-evaporation-dynamics-of-ordered-arrays-of-sessile-droplets/2AC01D10DF4623DA32FA60EB5035B250},
	doi = {10.1017/JFM.2019.105},
	urldate = {2023-05-10},
	journal = {Journal of Fluid Mechanics},
	author = {Hatte, S. and Pandey, K. and Pandey, K. and Chakraborty, S. and Basu, S.},
	month = may,
	year = {2019},
	note = {Publisher: Cambridge University Press},
	pages = {61--81},
}

@article{Craster2009,
	title = {Dynamics and stability of thin liquid films},
	volume = {81},
	issn = {0034-6861},
	url = {https://link.aps.org/doi/10.1103/RevModPhys.81.1131},
	doi = {10.1103/RevModPhys.81.1131},
	number = {3},
	urldate = {2022-10-05},
	journal = {Reviews of Modern Physics},
	author = {Craster, R. V. and Matar, O. K.},
	month = aug,
	year = {2009},
	note = {Publisher: American Physical Society},
	pages = {1131--1198},
}

@article{barrio2024droplet,
  title={Droplet motion driven by humidity gradients during evaporation and condensation},
  author={Barrio-Zhang, Hern{\'a}n and Ruiz-Guti{\'e}rrez, {\'E}lfego and Orejon, Daniel and Wells, Gary G and Ledesma-Aguilar, Rodrigo},
  journal={The European Physical Journal E},
  volume={47},
  number={5},
  pages={32},
  year={2024},
  publisher={Springer}
}

@article{whitaker1992species,
  title={The species mass jump condition at a singular surface},
  author={Whitaker, Stephen},
  journal={Chemical Engineering Science},
  volume={47},
  number={7},
  pages={1677--1685},
  year={1992},
  publisher={Elsevier}
}

@article{Ristenpart2007,
	title = {Influence of {Substrate} {Conductivity} on {Circulation} {Reversal} in {Evaporating} {Drops}},
	volume = {99},
	issn = {0031-9007},
	url = {https://link.aps.org/doi/10.1103/PhysRevLett.99.234502},
	doi = {10.1103/PhysRevLett.99.234502},
	number = {23},
	urldate = {2023-03-01},
	journal = {Physical Review Letters},
	author = {Ristenpart, W. D. and Kim, P. G. and Domingues, C. and Wan, J. and Stone, H. A.},
	month = dec,
	year = {2007},
	note = {Publisher: American Physical Society},
	pages = {234502},
}

@article{deegan_capillary_1997,
	title = {Capillary flow as the cause of ring stains from dried liquid drops},
	volume = {389},
	issn = {1476-4687},
	url = {https://www.nature.com/articles/39827},
	doi = {10.1038/39827},
	number = {6653},
	urldate = {2023-08-28},
	journal = {Nature 1997 389:6653},
	author = {Deegan, R. D. and Bakajin, O. and Dupont, T. F. and Huber, G. and Nagel, S. R. and Witten, T. A.},
	year = {1997},
	note = {Publisher: Nature Publishing Group},
	pages = {827--829},
}

@article{Saenz2015,
	title = {Evaporation of sessile drops: a three-dimensional approach},
	volume = {772},
	issn = {0022-1120},
	url = {https://www.cambridge.org/core/product/identifier/S0022112015002244/type/journal_article},
	doi = {10.1017/jfm.2015.224},
	urldate = {2022-10-05},
	journal = {Journal of Fluid Mechanics},
	author = {Sáenz, P. J. and Sefiane, K. and Kim, J. and Matar, O. K. and Valluri, P.},
	month = jun,
	year = {2015},
	note = {Publisher: Cambridge University Press},
	pages = {705--739},
}

@article{Majhy2020,
	title = {Evaporation-induced transport of a pure aqueous droplet by an aqueous mixture droplet},
	volume = {32},
	url = {https://doi.org/10.1063/1.5139002},
	doi = {10.1063/1.5139002},
	journal = {Phys. Fluids},
	author = {Majhy, B. and Sen, A. K.},
	year = {2020},
	pages = {32003},
}

@book{colinet2001nonlinear,
  title={Nonlinear dynamics of surface-tension-driven instabilities},
  author={Colinet, P. and Legros, J. C. and Velarde, M. G.},
  volume={522},
  year={2001},
  publisher={Wiley Online Library}
}

@article{cazabat2010evaporation,
  title={Evaporation of macroscopic sessile droplets},
  author={Cazabat, A.-M. and Guena, G.},
  journal={Soft Matter},
  volume={6},
  number={12},
  pages={2591--2612},
  year={2010},
  publisher={Royal Society of Chemistry}
}

@article{MURISIC2011,
	title = {On evaporation of sessile drops with moving contact lines},
	volume = {679},
	issn = {0022-1120},
	url = {https://www.cambridge.org/core/product/identifier/S0022112011001339/type/journal_article},
	doi = {10.1017/jfm.2011.133},
	urldate = {2022-10-05},
	journal = {Journal of Fluid Mechanics},
	author = {Murisic, N. and Kondic, L.},
	month = jul,
	year = {2011},
	pages = {219--246},
}

@article{fabrikant_potential_1985,
	title = {On the potential flow through membranes},
	volume = {36},
	issn = {00442275},
	url = {https://link.springer.com/article/10.1007/BF00945301},
	doi = {10.1007/BF00945301/METRICS},
	number = {4},
	urldate = {2023-07-26},
	journal = {ZAMP Zeitschrift für angewandte Mathematik und Physik},
	author = {Fabrikant, V. I.},
	month = jul,
	year = {1985},
	note = {Publisher: Birkhäuser-Verlag},
	pages = {616--623},
}

@article{Wray2020,
	title = {Competitive evaporation of multiple sessile droplets},
	volume = {884},
	issn = {0022-1120},
	url = {https://www.cambridge.org/core/product/identifier/S0022112019009194/type/journal_article},
	doi = {10.1017/jfm.2019.919},
	urldate = {2023-03-01},
	journal = {Journal of Fluid Mechanics},
	author = {Wray, A. W. and Duffy, B. R. and Wilson, S. K.},
	month = feb,
	year = {2020},
	note = {Publisher: Cambridge University Press},
	pages = {A45},
}

@article{Wray2021,
	title = {Contact-line deposits from multiple evaporating droplets},
	volume = {6},
	issn = {2469-990X},
	url = {https://link.aps.org/doi/10.1103/PhysRevFluids.6.073604},
	doi = {10.1103/PhysRevFluids.6.073604},
	number = {7},
	urldate = {2023-03-01},
	journal = {Physical Review Fluids},
	author = {Wray, A. W. and Wray, P. S. and Duffy, B. R. and Wilson, S. K.},
	month = jul,
	year = {2021},
	note = {Publisher: American Physical Society},
	pages = {073604},
}

@article{Plesset1976,
	title = {Flow of vapour in a liquid enclosure},
	volume = {78},
	issn = {0022-1120},
	url = {https://www.cambridge.org/core/product/identifier/S002211207600253X/type/journal_article},
	doi = {10.1017/S002211207600253X},
	number = {3},
	urldate = {2022-10-05},
	journal = {Journal of Fluid Mechanics},
	author = {Plesset, M. S. and Prosperetti, A.},
	month = dec,
	year = {1976},
	note = {Publisher: Cambridge University Press},
	pages = {433--444},
}

@article{Masoud2021,
	title = {Evaporation of multiple droplets},
	volume = {927},
	issn = {0022-1120},
	url = {https://www.cambridge.org/core/journals/journal-of-fluid-mechanics/article/evaporation-of-multiple-droplets/2BF278F6273E587A8BDBC4D91A5D6F40},
	doi = {10.1017/JFM.2021.785},
	urldate = {2023-04-12},
	journal = {Journal of Fluid Mechanics},
	author = {Masoud, H. and Howell, P. D. and Stone, H. A.},
	month = nov,
	year = {2021},
	note = {Publisher: Cambridge University Press},
	pages = {R4},
}

@article{Christy2011,
	title = {Flow {Transition} within an {Evaporating} {Binary} {Mixture} {Sessile} {Drop}},
	volume = {106},
	issn = {0031-9007},
	url = {https://link.aps.org/doi/10.1103/PhysRevLett.106.205701},
	doi = {10.1103/PhysRevLett.106.205701},
	number = {20},
	urldate = {2022-10-05},
	journal = {Physical Review Letters},
	author = {Christy, J. R. E. and Hamamoto, Y. and Sefiane, K.},
	month = may,
	year = {2011},
	note = {Publisher: American Physical Society},
	pages = {205701},
}

@article{Dunn2009,
	title = {The strong influence of substrate conductivity on droplet evaporation},
	volume = {623},
	issn = {1469-7645},
	url = {https://www.cambridge.org/core/journals/journal-of-fluid-mechanics/article/strong-influence-of-substrate-conductivity-on-droplet-evaporation/CDCA592888292CC74D22DC0473433C5E},
	doi = {10.1017/S0022112008005004},
	urldate = {2023-03-01},
	journal = {Journal of Fluid Mechanics},
	author = {Dunn, G. J. and Wilson, S. K. and Duffy, B. R. and David, S. and Sefiane, K.},
	year = {2009},
	note = {Publisher: Cambridge University Press},
	pages = {329--351},
}

@article{Moosman1980,
	title = {Evaporating menisci of wetting fluids},
	volume = {73},
	issn = {0021-9797},
	doi = {10.1016/0021-9797(80)90138-1},
	number = {1},
	urldate = {2022-10-05},
	journal = {Journal of Colloid and Interface Science},
	author = {Moosman, S. and Homsy, G. M.},
	month = jan,
	year = {1980},
	note = {Publisher: Academic Press},
	pages = {212--223},
}

\end{document}